%% file: main.tex
\documentclass[reprint,amsmath,amssymb,aps,prb,]{revtex4-2}

\usepackage[utf8]{inputenc}
\usepackage{graphicx}
\usepackage{overpic}
 \usepackage{rotating}
 \usepackage{mwe, tikz}
 \usepackage{amsmath,amssymb}
\usepackage{pgfplots}
\usetikzlibrary{calc,arrows,decorations.markings,shapes.geometric} 
\usepackage{amssymb}
\usepackage[finalnew]{trackchanges}
\usepackage{lipsum}
\usepackage{romannum}
\usepackage{float}
\usepackage{romannum}
\usepackage{xfp}
\usepackage{graphicx}
\usepackage{amsmath}
\usepackage{mathtools} 
\usepackage{color}
\usepackage{overpic}
\usepackage[export]{adjustbox}
\usepackage{mwe, tikz}
\usepackage{rotating}
\usepackage{latexsym}
\usepackage{csquotes}
\usepackage{tikz} 
\usetikzlibrary{calc,arrows,decorations.markings} 
\usepackage{pict2e}
\usepackage{pgfplots}
\usepackage{float}
\usepackage{hyperref}
\usepackage{xr}
\usepackage{rotating}
\usepackage{amsmath,amssymb}
\usepackage{fp}
\usetikzlibrary{calc} 
\usepackage{booktabs, tabularx}
\usepackage{romannum}
\usepackage{bm}
\usepackage{float}
\usepackage{adjustbox}
\usepackage{booktabs, tabularx}
\usepackage{longtable}
\usepackage{romannum}
\usepackage{footnote}
\usepackage{hyperref}

\input{macros}
\soulregister\ref7
\soulregister\cite7
\soulregister\citeA7
\soulregister\romannum7
\soulregister\section7
\newcommand*{\Labelxy}[4]{\put(#1,#2) {\setlength{\fboxsep}{0pt}\colorbox{white}{\strut\textcolor{black}{\begin{turn}{#3}{#4}\end{turn}}}}}
\newcommand*{\LabelFig}[3]{\put(#1,#2) {\setlength{\fboxsep}{0pt}\colorbox{white}{\textcolor{black}{#3}}} }
\newcommand*{\LabelFigg}[3]{\put(#1,#2) {\setlength{\fboxsep}{1pt}\colorbox{white}{\textcolor{black}{#3}}} }
\newcommand*{\romn}[1]{\emph{\textbf{\romannum{#1}}}}
\definecolor{darkolivegreen}{rgb}{0.33, 0.42, 0.18}
\definecolor{darkspringgreen}{rgb}{0.09, 0.45, 0.27}
\definecolor{darkslategray}{rgb}{0.18, 0.31, 0.31}
\definecolor{darkred}{rgb}{0.55, 0.0, 0.0}
\newcommand{\comp}{NiCoCr~}

\newcommand{\compFarkas}{FeNiCoCrCu~}
\newcommand{\potOne}{Li-Sheng-Ma~}
\newcommand{\potTwo}{Farkas-Caro~}
\addeditor{KK}
\addeditor{AN}
\begin{document}
\title{Dislocation plasticity in equiatomic NiCoCr alloys: The effect of short-range order
}

\author{Amir H. Naghdi$^{1}$}
\email{These authors made equal contributions.}
\author{Kamran Karimi$^{1*}$}
\email{Corresponding author: kamran.karimi@ncbj.gov.pl}
\author{Axel E. Poisvert$^1$}
\author{Amin Esfandiarpour$^1$}
\author{Rene Alvarez$^1$}
\author{Pawel Sobkowicz$^1$}
\author{Mikko Alava$^{1,2}$}
\author{Stefanos Papanikolaou$^1$}

\affiliation{%
 $^1$ NOMATEN Centre of Excellence, National Center for Nuclear Research, ul. A. Sołtana 7, 05-400 Swierk/Otwock, Poland\\
 $^{2}$ Aalto University, Department of Applied Physics, PO Box 11000, 00076 Aalto, Espoo, Finland
}%

\begin{abstract}

Equiatomic NiCoCr solid solutions have been shown by recent experiments and atomistic simulations to display exceptional mechanical properties, that have been suggested to be linked to nanostructural short-range order (SRO) features that may arise from thermal treatments, such as annealing or/and aging. 
Here, we use hybrid Monte Carlo-Molecular Dynamics (MC-MD) simulations to gain further insights of thermal effects on the SRO formation as well as the edge dislocation plasticity mechanisms of equiatomic NiCoCr face-centered cubic solid solution. For that purpose, we utilize two well-known NiCoCr interatomic potentials, one of which displays well documented SRO, believed to be linked to experimental evidence and labeled as the Li-Sheng-Ma potential, while the other (Farkas-Caro) does not. We use these two potentials to discern short range ordering (from inherent randomness in random solid solutions) and understand how SROs influence dislocation depinning dynamics in various thermal annealing scenarios. In this context, we used robust, scale-dependent metrics to infer a characteristic SRO size in the Li-Sheng-Ma case by probing local concentration fluctuations which otherwise indicate uncorrelated patterns in the Farkas-Caro case {in a close agreement with random alloys}. 
Our Voronoi-based analysis shows meaningful variations of local misfit properties owing to the presence of SROs. Using relevant order parameters, we also report on the drastic increase of chemical ordering within the stacking fault region.
More importantly, we find that the Li-Sheng-Ma potential leads to excellent edge dislocation depinning strength with low stacking fault width.
Our findings indicate an enhanced roughening mechanism due to the SROs-misfit synergy that leads to significant improvements in dislocation glide resistance. We argue that the improvements in alloy strength have their atomistic origins in the interplay between nanoscopic SROs and atomic-level misfit properties.

\end{abstract}
\maketitle

\input{sections/introduction}

\input{sections/Methods}

\input{sections/Results}
\input{sections/discussions}

\input{sections/conclusions}
\clearpage
\bibliography{Ref}
\renewcommand{\thefigure}{S\arabic{figure}}
\renewcommand{\thesection}{S~\Roman{section}}
\setcounter{figure}{0}    
\setcounter{section}{0}    
\input{sections/sm}
\end{document}

%% file: macros.tex
\newcommand{\legSq}[5]{
                \coordinate (a) at (0,0); 
                \node[white] at (a) {\tiny.};               %
			     \coordinate (center1) at (#1,#2); 
                \coordinate (b) at ($ (center1) - (.075,0) $);
                \coordinate (c) at ($ (b) - (.1,0) $); 
                \coordinate (d) at ($ (center1) + (.075,0) $);
                \coordinate (e) at ($ (d) + (.1,0) $); 
				 \draw[#3,fill] ($(b)-(0.0,0.075)$) rectangle ($ (d) + (0.0,0.075) $); 
                \draw[#3][line width=0.1mm] (b) -- (c); 
                \draw[#3][line width=0.1mm] (d) -- (e); 
		    \node at ($(center1)+(#5,0)$) {{\color{black}#4}}; 
}

\newcommand{\legCirc}[5]{
                \coordinate (a) at (0,0); 
                \node[white] at (a) {\tiny.};               %
                \coordinate (center1) at (#1,#2); 
				 \draw[#3,fill=#3] (center1) circle (2pt);
                \coordinate (b) at ($ (center1) - (.075,0) $);
                \coordinate (c) at ($ (b) - (.1,0) $); 
                \coordinate (d) at ($ (center1) + (.075,0) $);
                \coordinate (e) at ($ (d) + (.1,0) $); 
                \draw[#3][line width=0.1mm] (b) -- (c); 
                \draw[#3][line width=0.1mm] (d) -- (e); 
		    \node at ($(center1)+(#5,0)$) {{\color{black}#4}}; 
}

\definecolor{light-gray}{gray}{0.85}

\newcommand{\drawSlope}[6]{ 
				\coordinate (center1) at (#1,#2); 
				 \FPeval{\nx}{cos(#4*pi/180)}%
				 \FPeval{\ny}{sin(#4*pi/180)}%
				 \FPeval{\absNx}{abs(\nx)}%
				 \FPeval{\absNy}{abs(\ny)}%
				\coordinate (b) at ($ (center1) + #3*(-\ny,+\nx) $);
				\coordinate (c) at ($ (center1) - #3*(-\ny,+\nx) $); 
				 \FPeval{\xx}{(-\nx*\ny)}%
				\ifdim\xx pt < 0pt 
				\coordinate (d) at ($ (c) -2*#3*\ny*(1,0) $);
				\node at ($(d) +(0,#3*\nx)-0.2*(\nx/\absNx,0)$) {{\color{#5}$#6$}}; 
				\node at ($(d)+(#3*\ny,0)-0.2*(0,\ny/\absNy)$) {{\color{#5}$\scriptstyle 1$}}; 
				\else
				\coordinate (d) at ($ (c) +2*#3*\nx*(0,1) $);
				\node at ($(d)-#3*\nx*(0,1)-0.2*\nx/\absNx*(1,0)$) {{\color{#5}$#6$}}; 
				\node at ($(d)-#3*\ny*(1,0)-0.2*\ny/\absNy*(0,1)$) {{\color{#5}$\scriptstyle 1$}}; 
				\fi
				\draw[#5,line width=0.1mm] (d) -- (b); 
				\draw[#5,line width=0.1mm] (d) -- (c); 
				\draw[#5,line width=0.1mm] (b) -- (c); 
} 

\newcommand{\scaleBarr}[8]{
			     \coordinate (center1) at (#1,#2); 
                \coordinate (b) at ($ (center1) - (#3,#4/2) $); 
                \coordinate (d) at ($ (center1) + (#3,#4/2) $);
				 \draw[black,fill] ($(b)$) rectangle ($(d)$); 

				\node at ($(b)-(0,#4)$){{\color{black} $#5$}};
				\node at ($(b)+(2*#3,-#4)$) {{\color{black} $ #6$}};
				\node at ($(b)+(2*#3,+2.6*#4)$) {{\color{black}$\text{#8}$}}; 
			     \coordinate (center2) at ($ (center1) + 2*(#3,0.0)$);  
                \coordinate (b) at ($ (center2) - (#3,#4/2) $);
                \coordinate (d) at ($ (center2) + (#3,#4/2) $);
				 \draw[black] ($(b)$) rectangle ($(d)$); 
				\node at ($(d)-(0,2*#4)$) {{\color{black} $#7$}};		
				}	



%% file: sections/Introduction.tex
\begin{figure*}[t]
  \centering
  \includegraphics[width=0.75\textwidth]{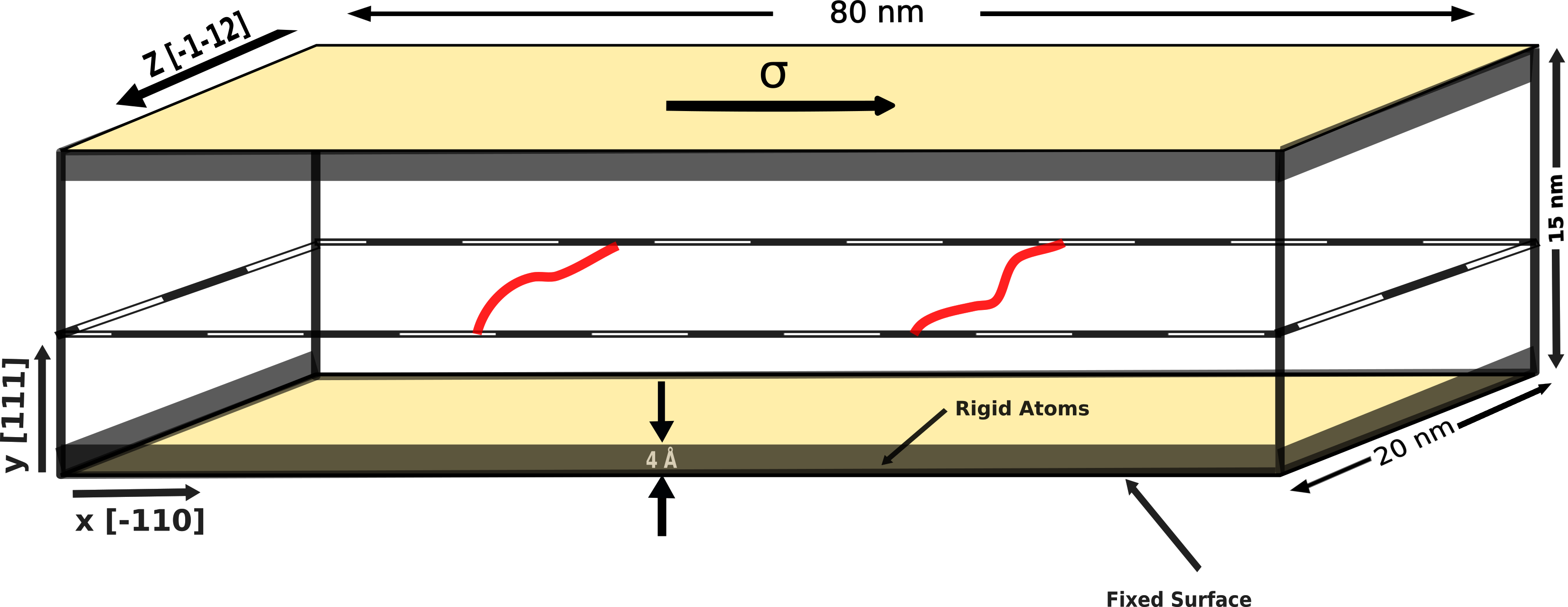}
  \caption{A sketch of the shear setup including the partial dislocations. Periodic boundary conditions are applied parallel to the dislocation lines and the glide direction ($x$ and $z$ dimensions, respectively) and fixed boundaries are implemented along the $y$ direction (rigid slabs in gray). To shear, rigid atoms within the bottom surface are held fixed (i.e. zero displacements) while the shear stress $\sigma$ is applied on the rigid top plane.} 
  \label{fig:sketchLoadSetup}
\end{figure*}

\section{\label{sec:introduction}Introduction}

Metallurgy of alloys is at the core of technological progress. 
Concentrated solid solution alloys (CSAs) have recently emerged as major candidates for novel alloys for extreme conditions applications \cite{li2019mechanical,shang2021mechanical}. 
Out of all, the equiatomic NiCoCr CSA represents a simple enough composition that has been consistently reported to show exceptional mechanical properties \cite{li2019mechanical}. 
These include (among others) a relatively high (tensile) strength and ductility, fracture toughness, as well as (micro-)hardness, that often exceed those of the “Cantor” alloy \cite{Gludovatz2016}, yet with a fewer number of principal components. 
This is most likely rooted in the chemical composition and underlying sub-structure.
However, the microstructural origin of the exceptional mechanical properties has been heavily debated, with  a possible explanation being the presence of nm-level (chemical/structural) short-range order (SRO) \cite{zhang2020short,wu2021short,zhou2022atomic,chen2021direct} that {arises from particular thermal processing and} influences dislocation pinning and stacking fault widths. In addition, lattice distortions and local crystalline misfits, due to atomic size differences~\cite{Yin2020,noehring2019correlation}, have been shown to be correlated to the exceptional mechanical behavior of this alloy. Given the apparent importance of local misfit volumes~\cite{Yin2020}, it remains a challenge to identify the role of SRO for exceptional mechanical properties.
In this paper, we focus on extensive molecular simulations to understand how SRO influences dislocation plasticity and how it might depend on processing parameters (i.e. annealing temperature), the properties of the ordered phase, as well as the role of the MD atomic potentials in the ordering process. 
{
In this framework, we investigated two case studies involving commonly-used NiCoCr interatomic potentials i) Li-Sheng-Ma and ii) Farkas-Caro potential. 
While i) leads to the formation of short range ordering upon aging, ii) displays chemical/structural features, under the exact same thermal treatment, that are almost indistinguishable from random solid solutions.
Our multi-scale characterization of local ordering were based upon the use of novel descriptors that exhibit distinct structural/chemical signatures owing to the presence of nanoscopic SROs.
}
We study the effects of the SRO on mechanical properties, showing that it significantly influences them, via the interplay of dislocations and SRO structures.
{Such an interplay was quantified via a detailed analysis of the dislocation substructure indicating enhanced roughening properties due to combined SRO-misfit effects.}

Short range order has been at the core of studies in CSAs across the board \cite{liu2021nanoprecipitate,he2021understanding,wolverton2000short}. 
Thermodynamically speaking, SROs' ubiquity at low-temperature alloys has been mainly attributed to dominant enthalpic effects that, in the absence of entropy-driven mechanisms, do not favor idealistic perfect mixtures of equimolar elements \cite{Li2019}.
In this context, SROs typically refer to coherent compositional deviations apart from (statistically) random distributions of atoms within the solution matrix as in random solid solution alloys (RSAs).
More importantly, (thermal) processing parameters associated with annealing and homogenization procedures (i.e. temperature and time) or irradiation may have a drastic effect on the nucleation of SROs and associated substructural features \cite{Yin2020,walsh2021magnetically,zhang2017local}.
Owing to their nanoscopic scales, laboratory-based observations of SROs are quite nontrivial involving intensive use of advanced characterization techniques such as high-resolution electron microscopy and atomic-resolution energy dispersive spectrometry mapping \cite{wang2022chemical,chen2021direct}.
The latter are strongly tied to underlying physical mechanisms that govern fundamental alloy properties. 
For instance, the formation of Ni-rich nano-precipitates and associated inhomogeneities within the annealed NiCoCr matrix has been recently suggested to tune the stacking fault width with evident consequences in terms of the alloy strengthening \cite{ding2018tunable,Li2019}. 
Similar conclusions were drawn experimentally by Ritchie et al. \cite{zhang2020short} who reported on the emergence of SROs in aged NiCoCr CSAs with significant impacts on the dislocation activation energy and hardness. 
In studies of NiCoCr-based alloys, stacking fault energy, hardness, and fracture toughness, as bulk properties, were recently shown to strongly correlate with the degree of Ni-rich SROs and corresponding structural features \cite{yang2022chemical,jian2020effects,liu2022exceptional,miao2021ordering}.

\begin{figure*}[t]
    \centering
    \begin{overpic}[width=0.45\textwidth]{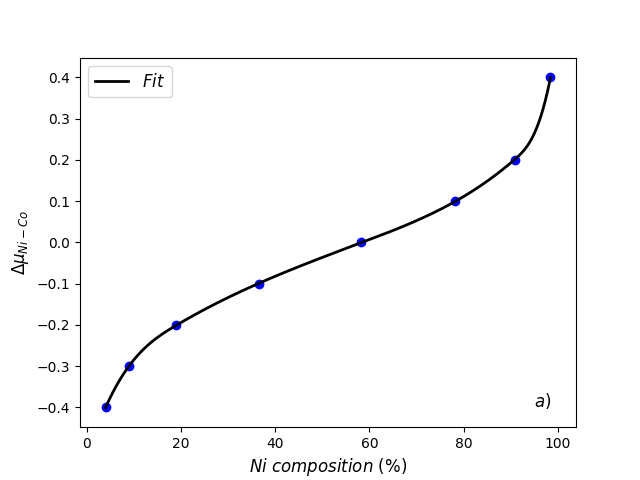}
        \LabelFig{83.5}{12}{$a)$}
    \end{overpic}
    \begin{overpic}[width=0.45\textwidth]{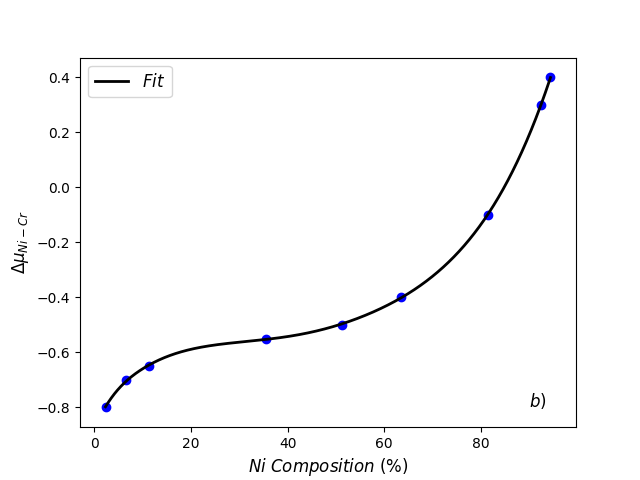}
        \LabelFig{83.0}{12}{$b)$}
    \end{overpic}
    \caption{Chemical potential differences for \textbf{a)} Ni-Co and \textbf{b)} Ni-Cr pairs used for variance constrained semi-grand canonical (VCSGC) ensemble.} 
    \label{fig:Chemical Poyentials}
\end{figure*}

In contrast, local misfit properties have been conventionally viewed as a key solid solution strengthening mechanism \cite{sohn2019ultrastrong}.
{In this framework, the inherent yield strength of alloys (or Peierls stress) associates to the stress threshold required for dislocation depinning and in the context of multi-component high-entropy alloys, their intrinsic hetrogeneity and randomness gives rise to somewhat uncorrelated perturbations to the local thresholds. }
This concept was theoretically put forward in the seminal work by Labusch \cite{Labusch} who hypothesized that the motion of dislocations within a random set of solute obstacles leads to significant hardening effects in dilute solutions. 
In a series of relevant papers, Varvenne and Curtin (VC) \cite{VARVENNE2016164,VARVENNE2017660, VARVENNE201892} further investigated the RSA context in terms of elastic-type long-range interactions between dislocation lines and residual strain fields resulting from atomic size misfits. 
Along these lines, a mean-field theoretical framework was proposed to make fairly accurate predictions of yield strengths solely based on the effective medium elastic properties and, more importantly, local misfit fluctuations. 
{The proposed theory accounts for local compositional fluctuations described by spatial distributions of misfit volumes which are accessible through atomistic simulations and experimentation.
The VC framework was further generalized to additionally account for thermal and strain-rate effects on the alloys’ strength and checked in the context of random alloys \cite{VARVENNE2017660, Yin2020}.
In the case of high-entropy alloys (HEAs), however, there may often be a considerable degree of short range ordering and, therefore, HEAs cannot be simply treated as fully random.}

Short-range order and local misfits are not one-way streets but they combine and interplay in ways that are somewhat unpredictable in advance and inseparable \cite{jian2020effects}.
Moreover, almost all crucial properties related to alloy strengths are strongly dependent on the underlying microstructure and processing methods used to synthesize CSAs.
{In an aged multicomponent alloy, the effective Peierls stress will be influenced by both randomness in the local composition distribution (giving rise to misfit fluctuations) and short ranged (but still finite) spatial correlations introduced by SROs.
Naturally, this combined effect leads to an effective yield stress that typically exceeds that of a random solid solution, lacking this finite range ordering component. 
A naive picture in this context is that dislocations will move by locally bending between pinning sites to overcome locally-fluctuating Peierls stresses leading to extra strengthening \cite{utt2022origin}.
Nevertheless, a systematic study accounting for dislocations' glide resistance and their substructure discerning the separate roles of SROs and misfits seems necessary.  
}

\begin{figure*}[t]
   \centering
   \begin{overpic}[width=0.75\textwidth]{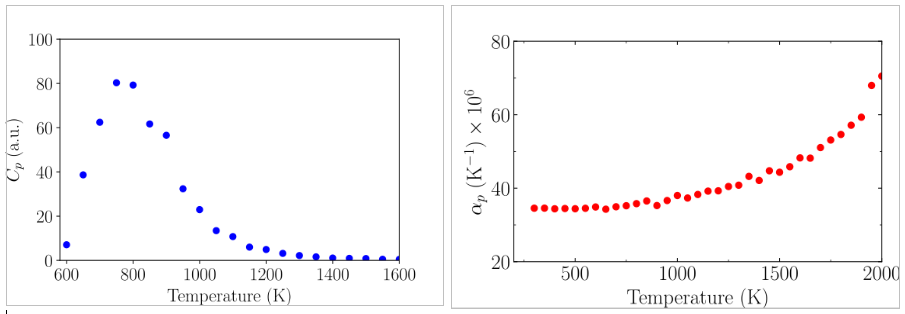}
        \LabelFig{8}{27}{$a)$}
        \LabelFig{58}{27}{$b)$}
        \Labelxy{0.5}{13}{90}{$~~~~C_p(J$K$^{-1}$)}
        \Labelxy{51.2}{10}{90}{$\alpha_p(\times 10^{-6}$K$^{-1})$}
        \put(22,13) {\includegraphics[width=.2\textwidth]{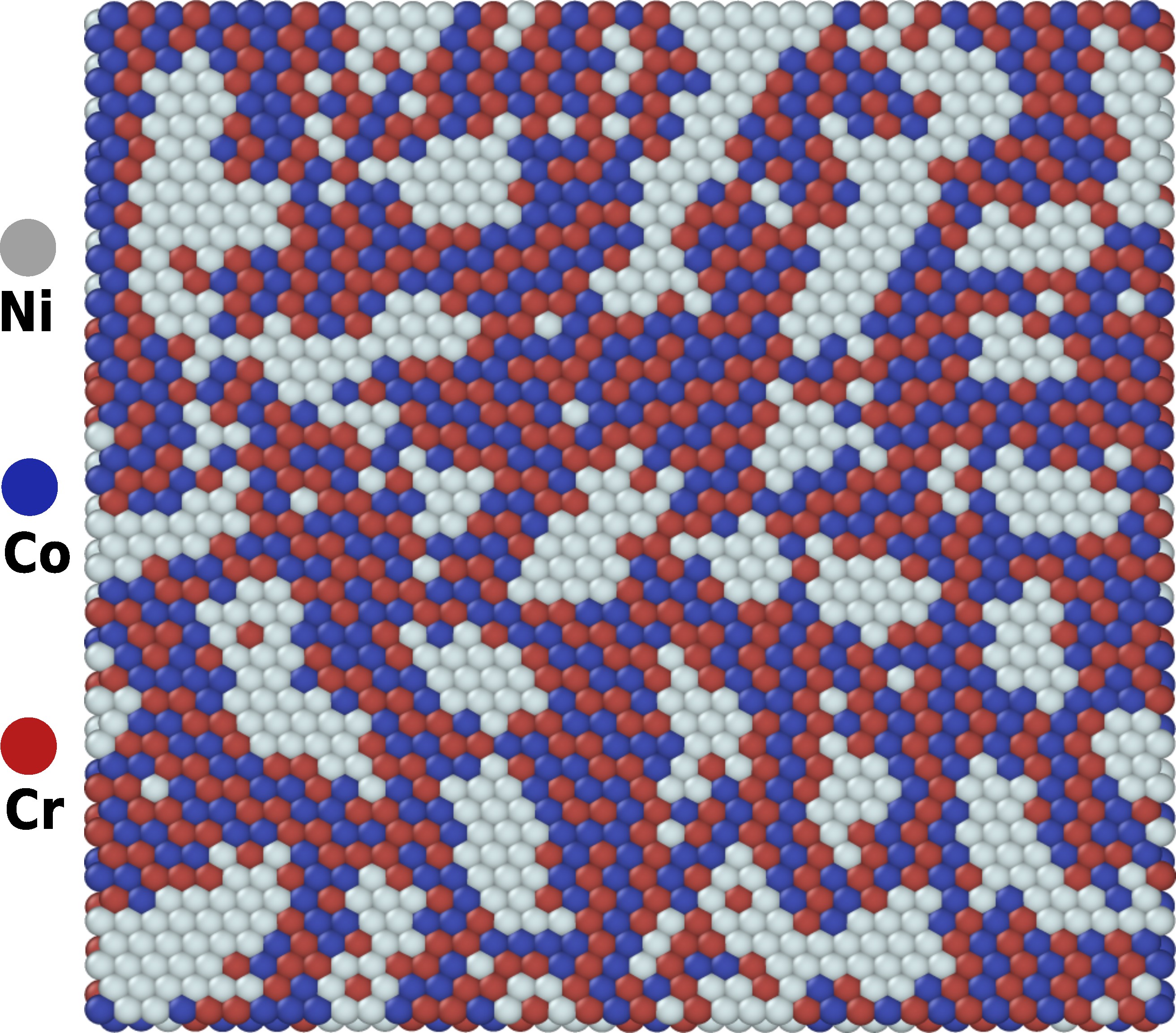}}
        \begin{tikzpicture} 
                \coordinate (a) at (0,0);
                \node[white] at (a) {\tiny.};
            \scaleBarr{4.3}{1.2}{0.35}{0.15}{$\tiny 0$}{$\tiny 15$}{$\tiny 30$}{\footnotesize\r{A}}
        \end{tikzpicture}
    \end{overpic}
   \caption{Annealing temperature effects on NiCoCr based on the \potOne potential. \textbf{a)} Heat capacity $C_p$ \textbf{b)} thermal expansion coefficient $\alpha_p$ versus anealing temperature $T_a$. The inset represents a $(111)$ cross-section of Ni (grey), Co (blue), and Cr (red) atoms at $T_a=800$ K.} 
   \label{fig:thermo}
\end{figure*}
\begin{figure*}[t]
    \centering
    \begin{overpic}[width=0.24\textwidth]{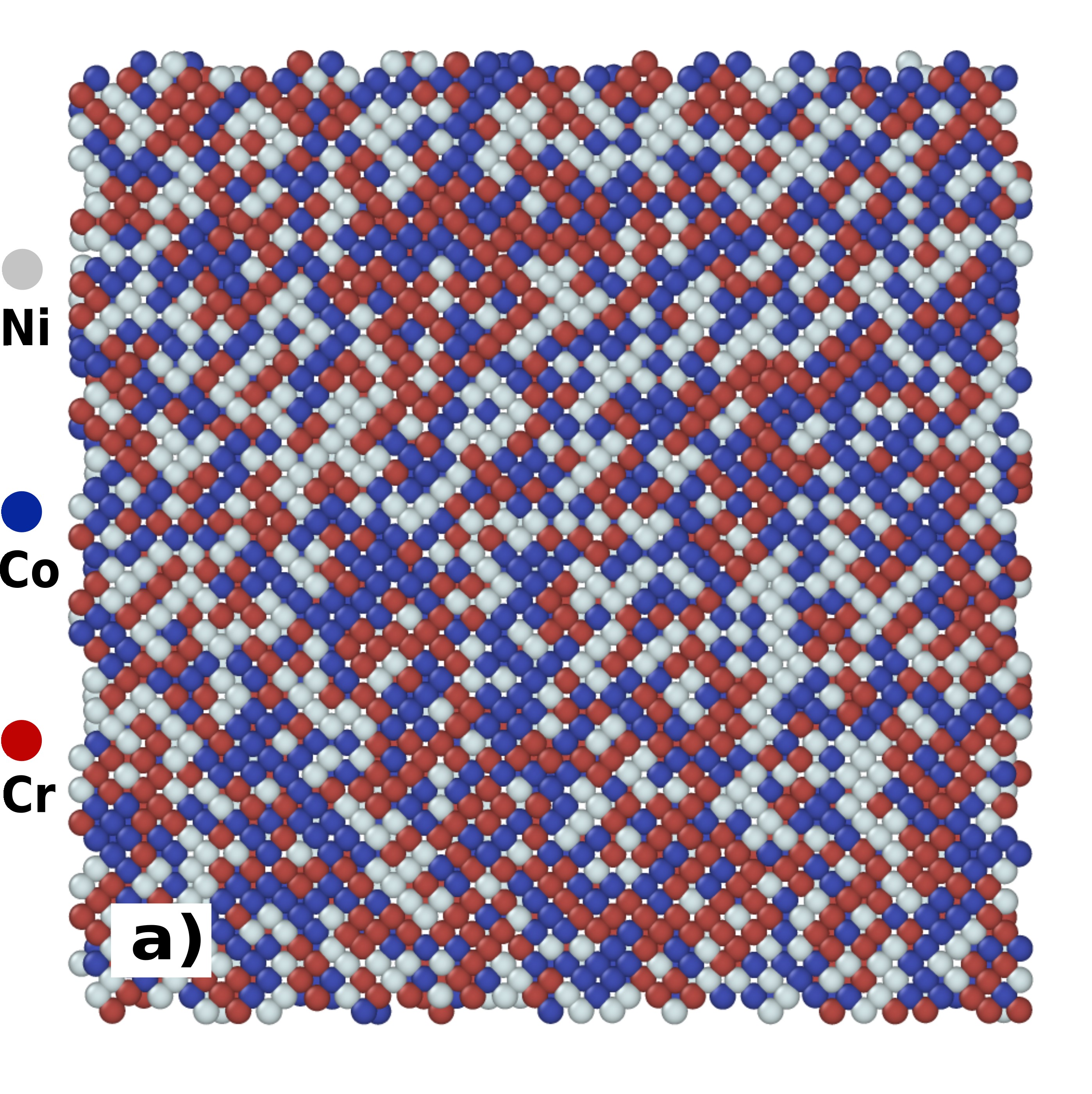}
        \LabelFig{12}{12}{$a)$}
    \end{overpic}
    \begin{overpic}[width=0.24\textwidth]{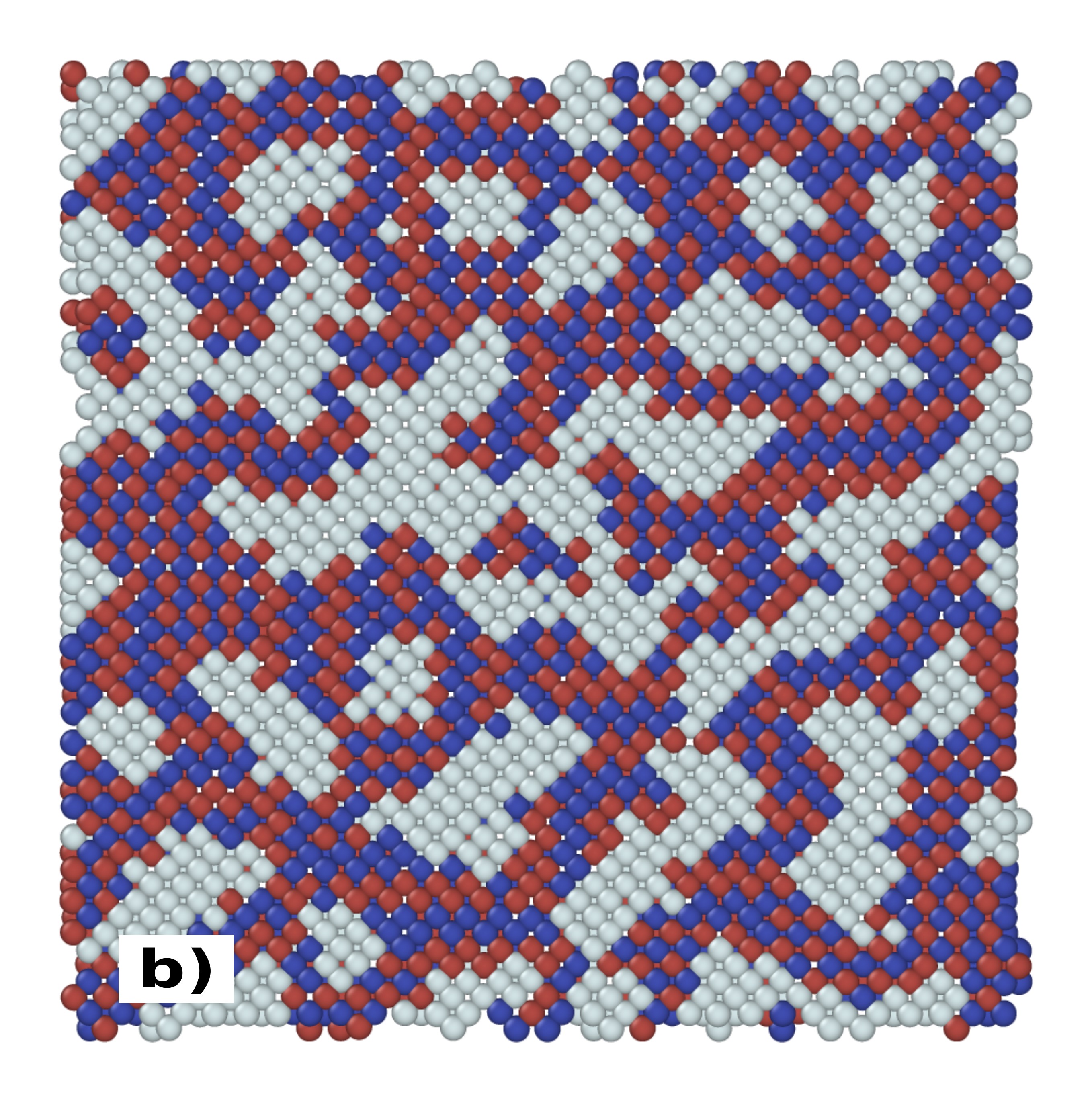}
        \LabelFig{13}{10}{$b)$}
    \end{overpic}
    \begin{overpic}[width=0.24\textwidth]{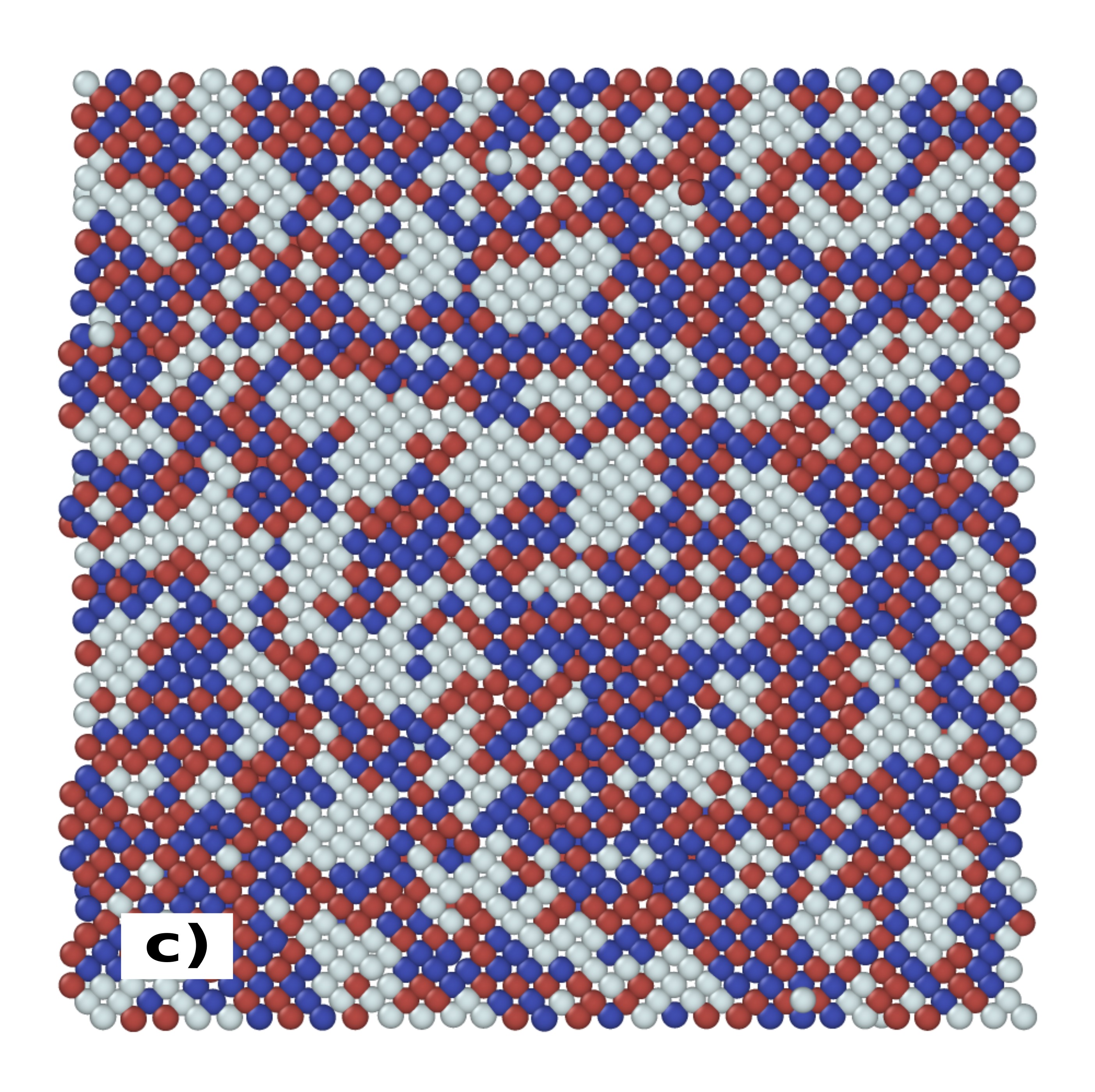}
        \LabelFig{13}{10.5}{$c)$}
    \end{overpic}
    \begin{overpic}[width=0.24\textwidth]{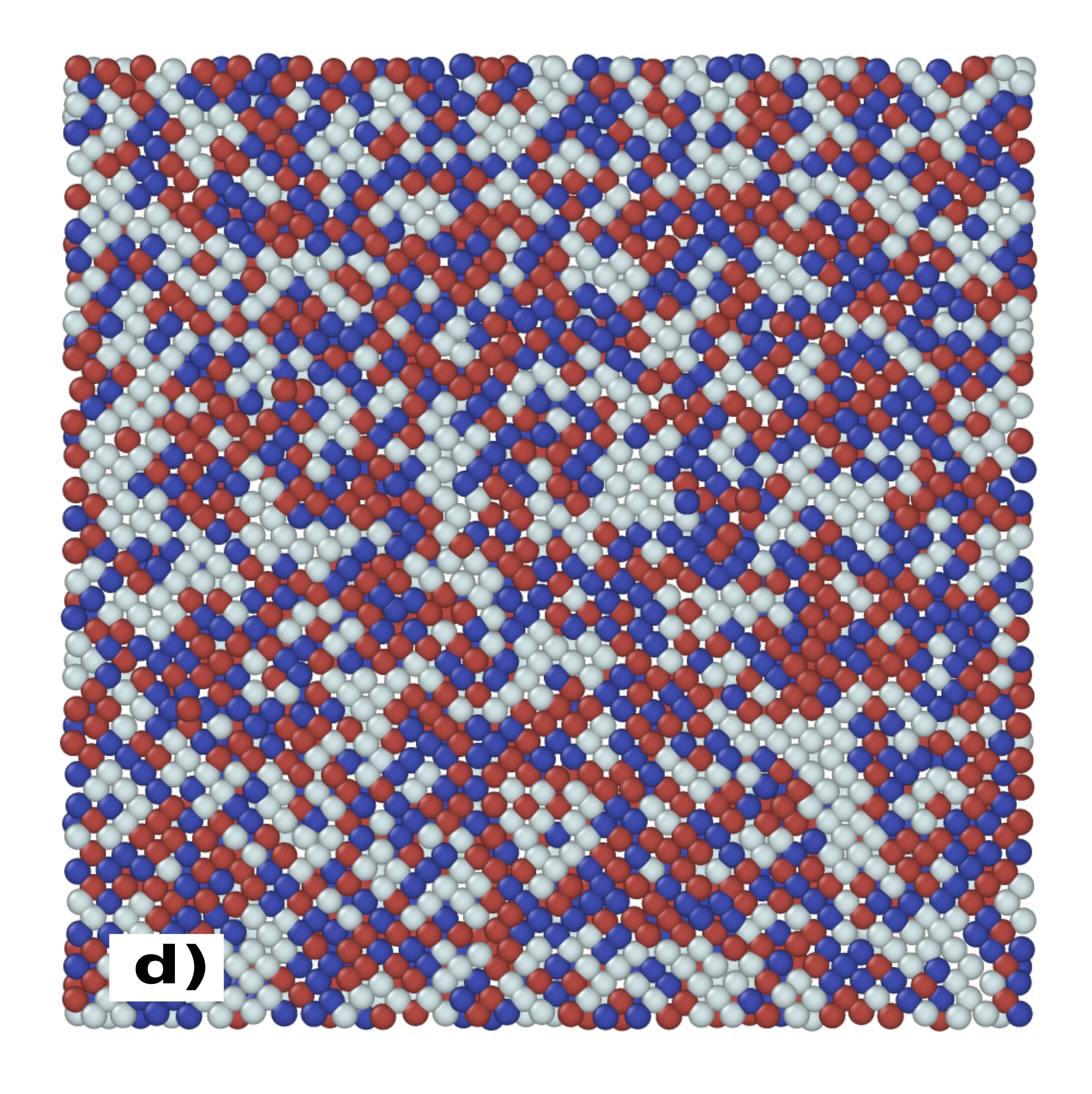}
        \LabelFig{12}{10}{$d)$}
    \end{overpic}
    \caption{Snapshots of NiCoCr samples \textbf{a)} RSA equilibrated at $T=400$ K, \textbf{b)} annealed at $T_a=400$ K, \textbf{c)} annealed at $T_a=800$ K and \textbf{d)} annealed at $T_a=1400$ K.} 
    \label{fig:Annealed}
\end{figure*}

\begin{figure*}[t]
   \centering
     \begin{overpic}[width=0.24\textwidth]{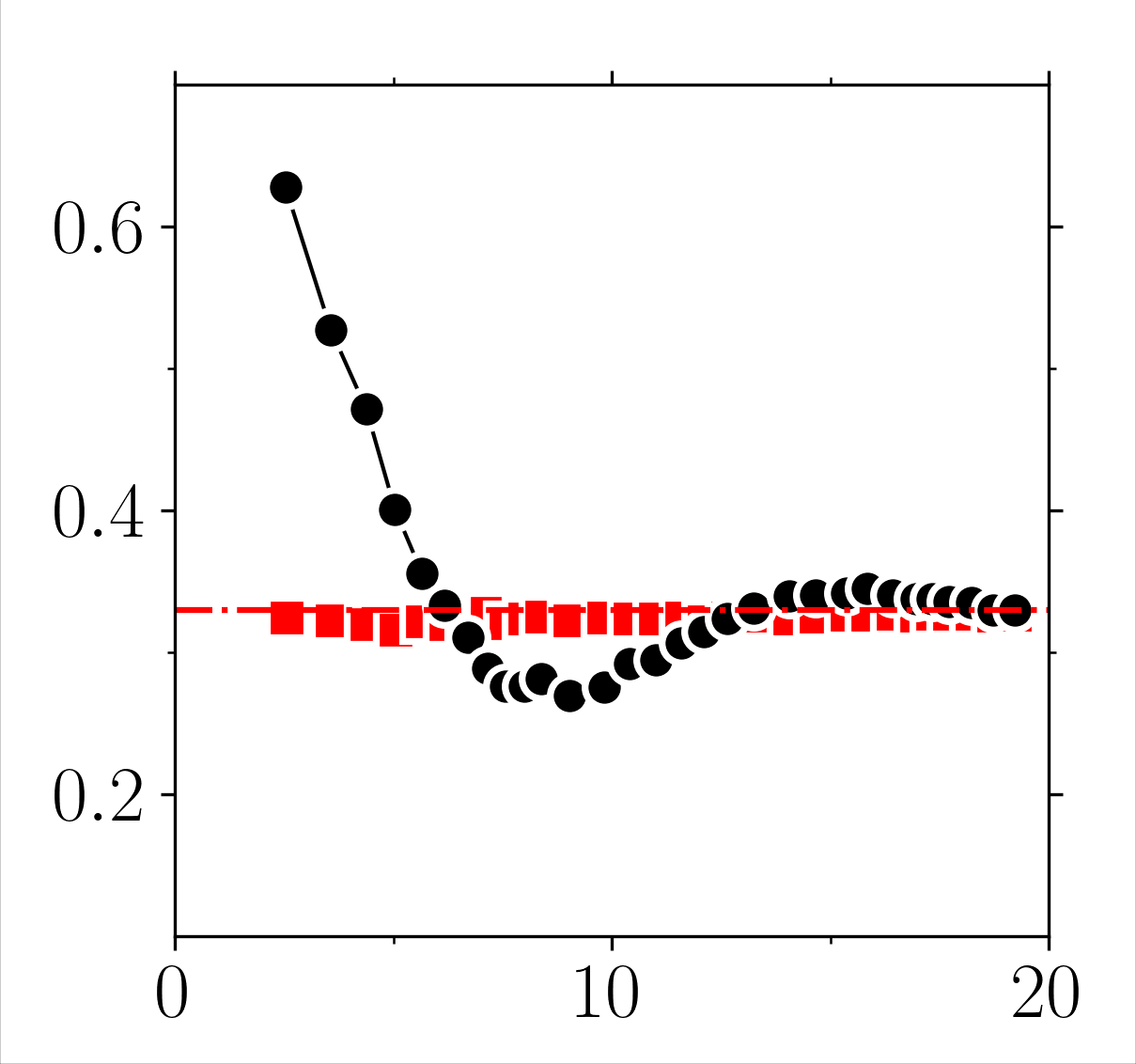}
        \Labelxy{50}{-3}{0}{$r$(\r{A})}
        \Labelxy{-6}{35}{90}{$p_\text{NiNi}$}
        \LabelFig{19}{76}{$a)$}
    \end{overpic}
    \begin{overpic}[width=0.24\textwidth]{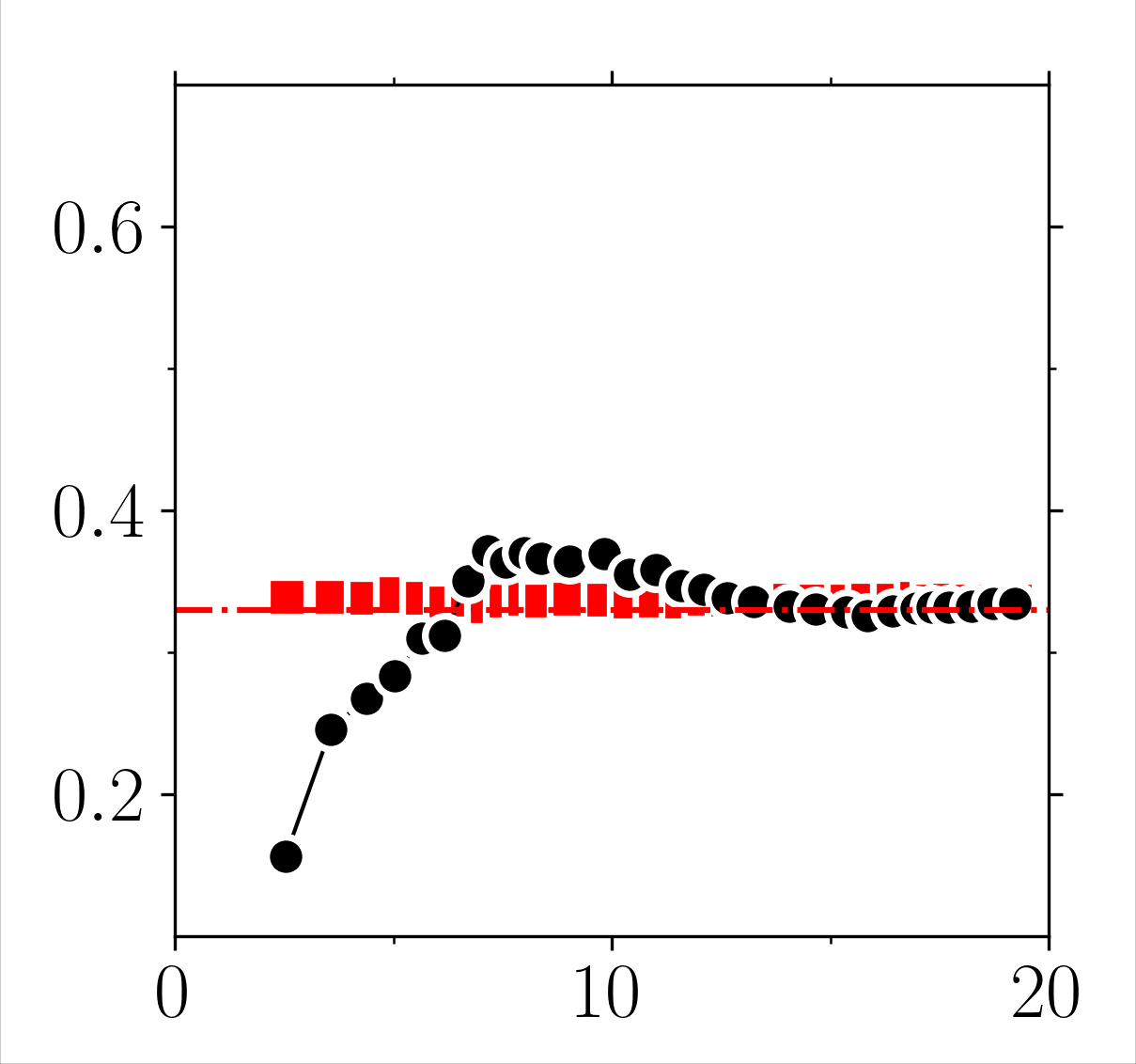}
        \Labelxy{50}{-3}{0}{$r$(\r{A})}
        \Labelxy{-6}{35}{90}{$p_\text{NiCo}$}
        \LabelFig{19}{76}{$b)$}
        \begin{tikzpicture}
            \legCirc{1.6}{3}{black}{\footnotesize annealed}{0.9}
            \legSq{1.6}{2.6}{red}{\footnotesize random CSA}{1.2}
        \end{tikzpicture}
    \end{overpic}
    \begin{overpic}[width=0.24\textwidth]{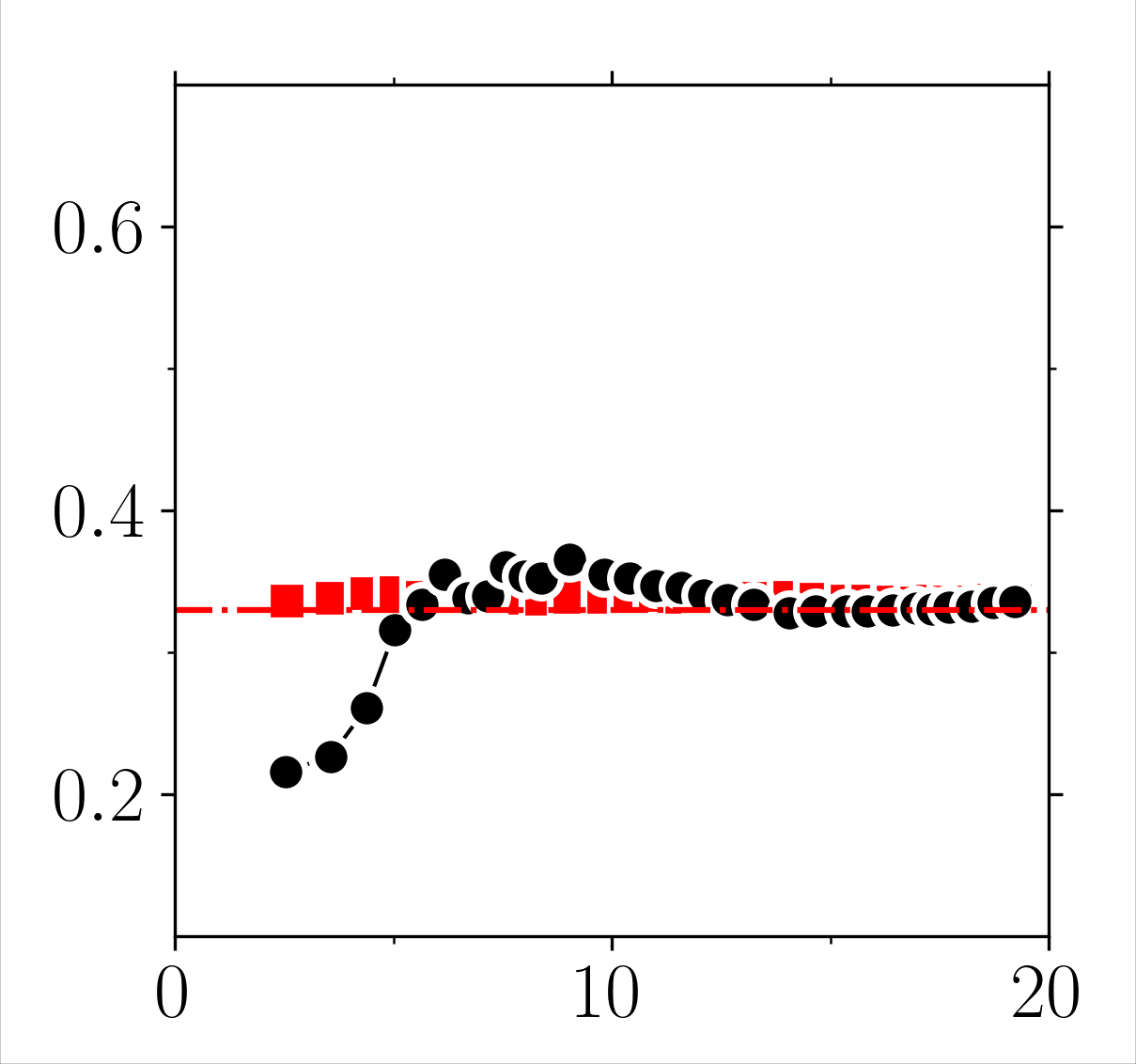}
        \Labelxy{50}{-3}{0}{$r$(\r{A})}
        \Labelxy{-6}{35}{90}{$p_\text{NiCr}$}
        \LabelFig{19}{76}{$c)$}
        \put(53,50) {\includegraphics[width=.15\textwidth]{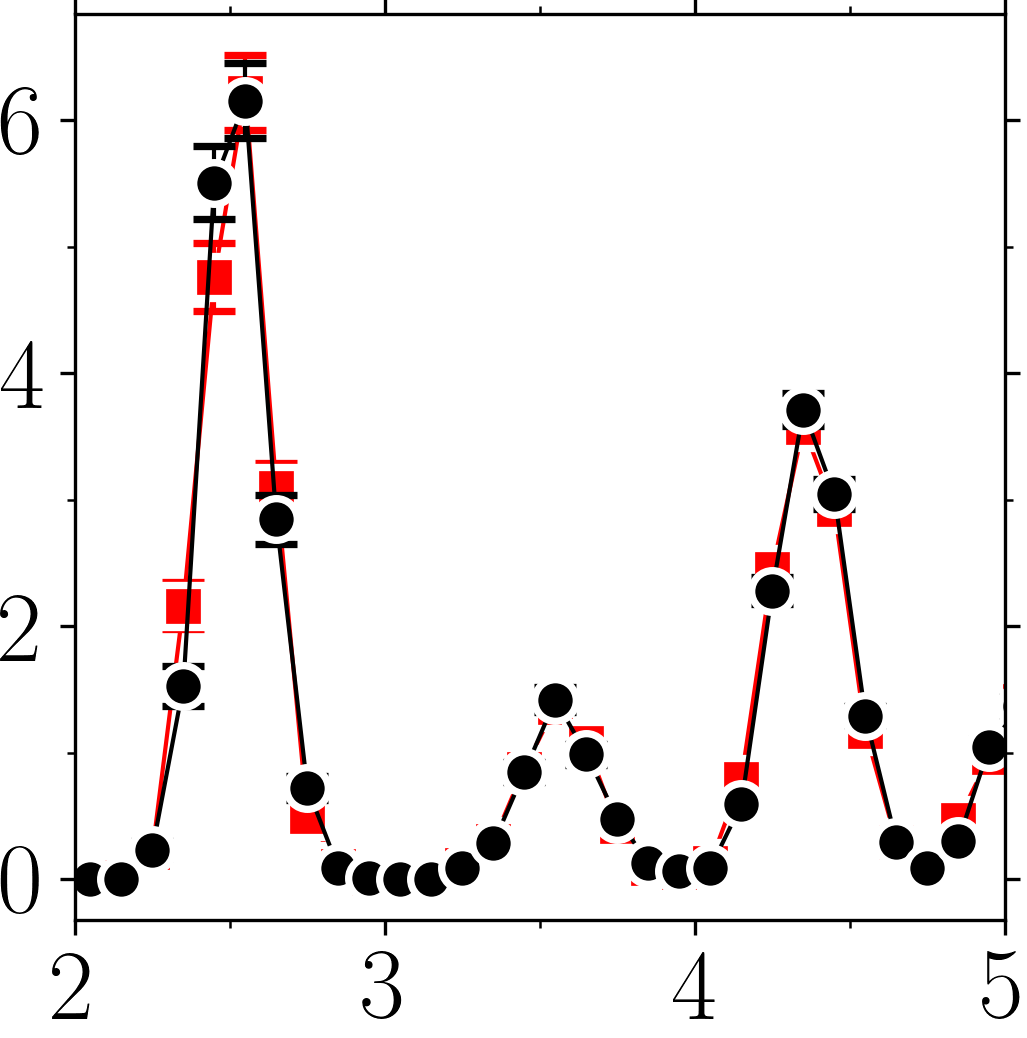}}
        \Labelxy{78}{48}{0}{$\scriptstyle r$\scriptsize (\r{A})}
        \Labelxy{116}{80}{90}{$\scriptstyle g(r)$}
        \LabelFig{104}{105}{$g)$}
    \end{overpic}

    \begin{overpic}[width=0.24\textwidth]{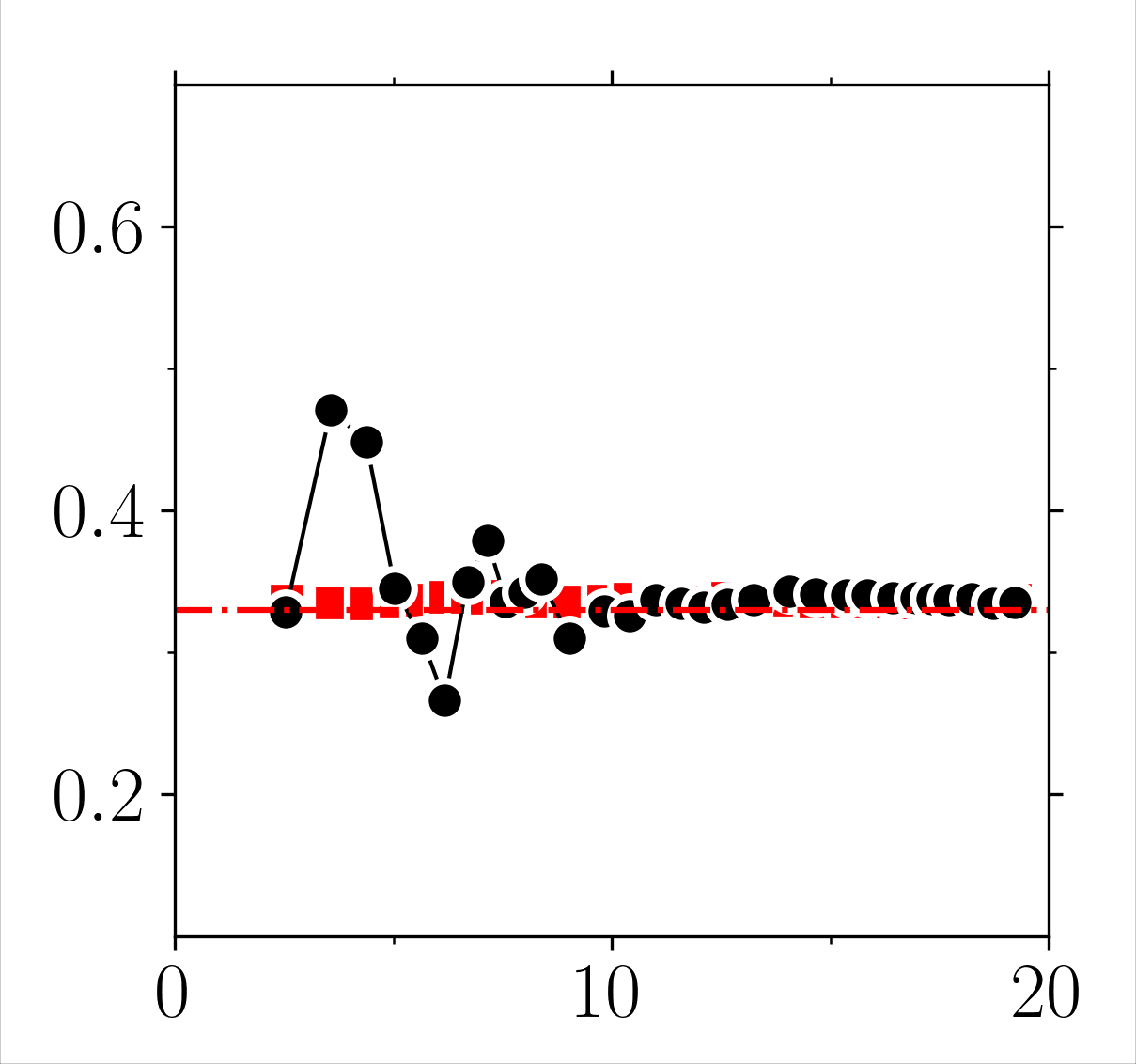}
        \Labelxy{50}{-3}{0}{$r$(\r{A})}
        \Labelxy{-6}{35}{90}{$p_\text{CoCo}$}
        \LabelFig{19}{76}{$d)$}
    \end{overpic}
    \begin{overpic}[width=0.24\textwidth]{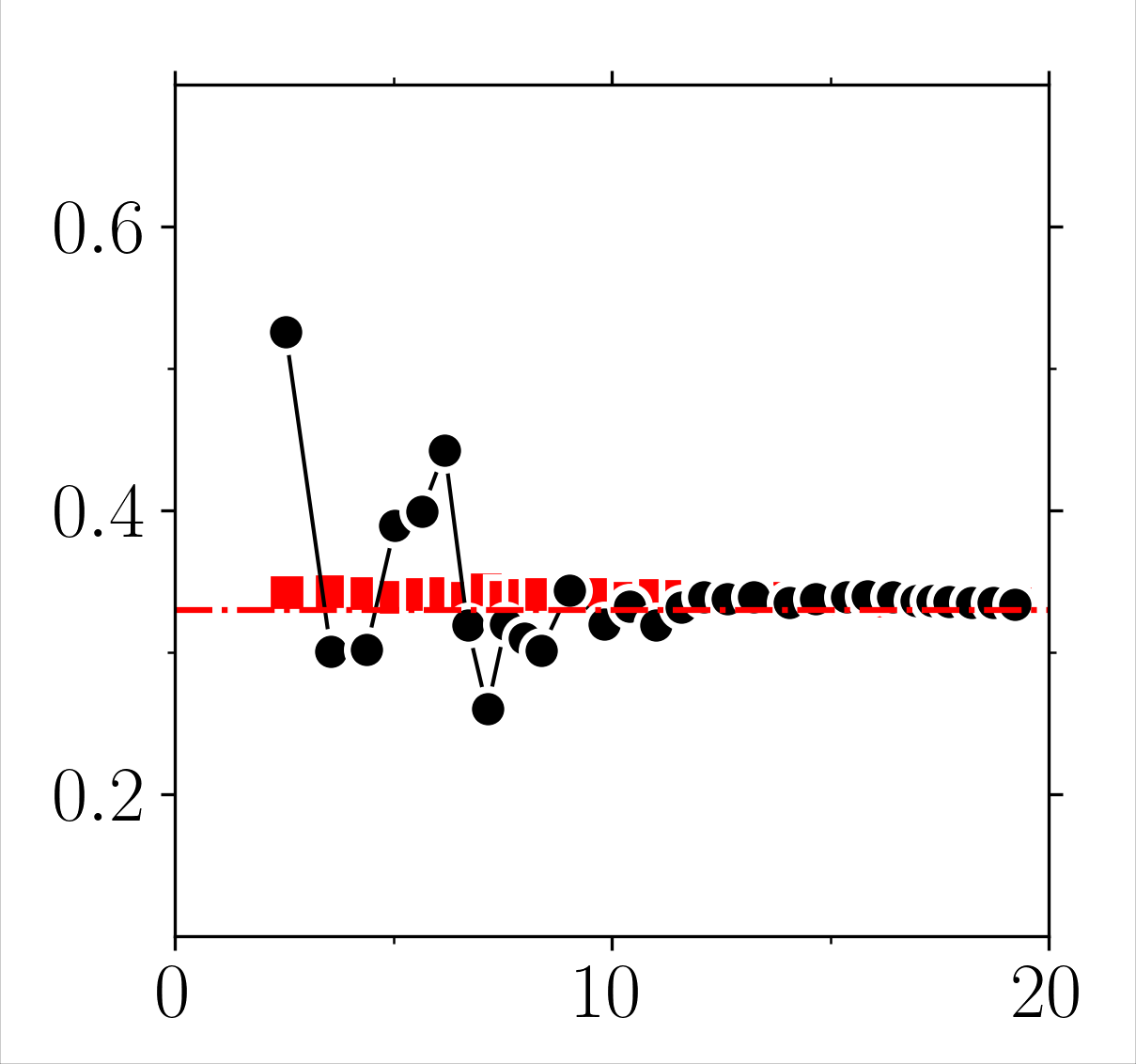}
        \Labelxy{50}{-3}{0}{$r$(\r{A})}
        \Labelxy{-6}{35}{90}{$p_\text{CoCr}$}
        \LabelFig{19}{76}{$e)$}
    \end{overpic}
    \begin{overpic}[width=0.24\textwidth]{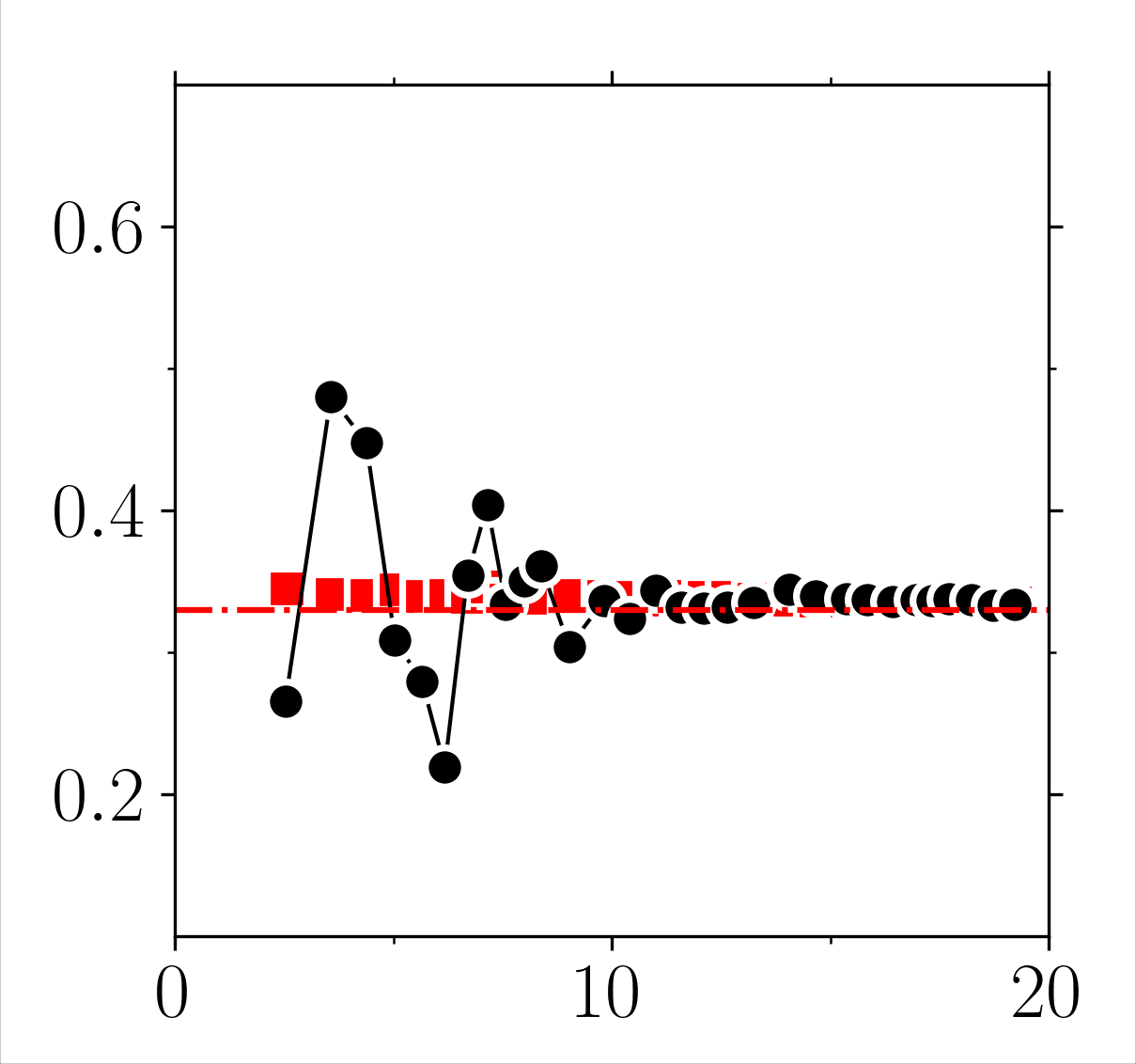}
        \Labelxy{50}{-3}{0}{$r$(\r{A})}
        \Labelxy{-6}{35}{90}{$p_\text{CrCr}$}
        \LabelFig{19}{76}{$f)$}
    \end{overpic}
   \caption{Short range ordering in annealed NiCoCr CSAs based on the \potOne potential. Warren–Cowley SRO parameters including \textbf{a)} $p_\text{NiNi}$ \textbf{b)} $p_\text{NiCo}$ \textbf{c)} $p_\text{NiCr}$ \textbf{d)} $p_\text{CoCo}$ \textbf{e)} $p_\text{CoCr}$ \textbf{f)} $p_\text{CrCr}$ plotted against distance $r$ at $T_a=400$ K. \textbf{g)} Pair correlation function $g(r)$ at $T_a=400$ K. The base (red) dashdotted line indicates the random concentration.} 
   \label{fig:sroSheng}
\end{figure*}

\begin{figure}[b]
    \begin{overpic}[width=0.28\textwidth]{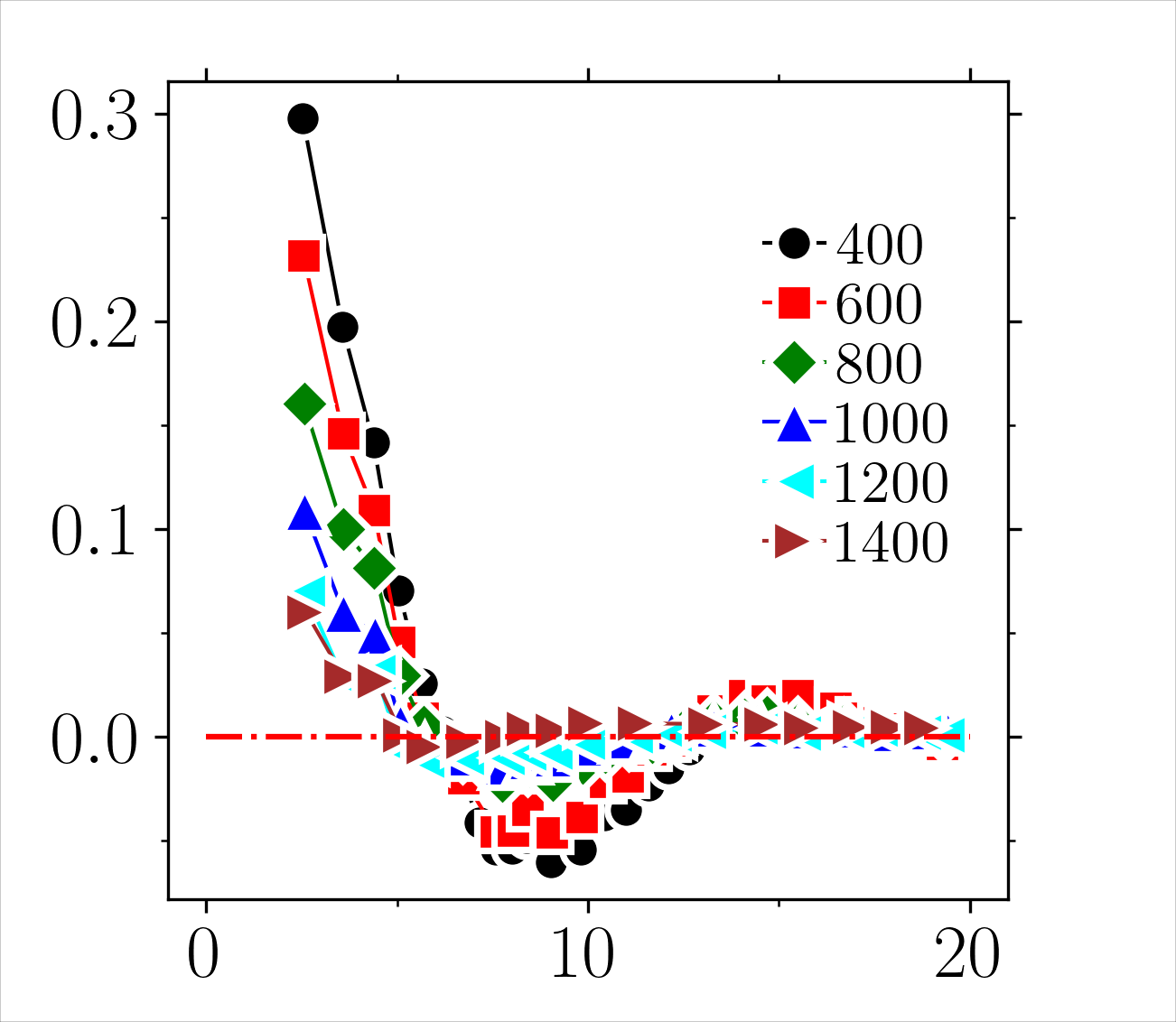}
        \LabelFig{17}{15}{$a)$}
        \Labelxy{44}{-3}{0}{$r$(\r{A})}
        \Labelxy{62}{71}{0}{$\scriptstyle T_a$\scriptsize (K)}
        \Labelxy{-6}{35}{90}{$p_\text{NiNi}-p^\text{rsa}_\text{NiNi}$}
    \end{overpic}
    \begin{overpic}[width=0.28\textwidth]{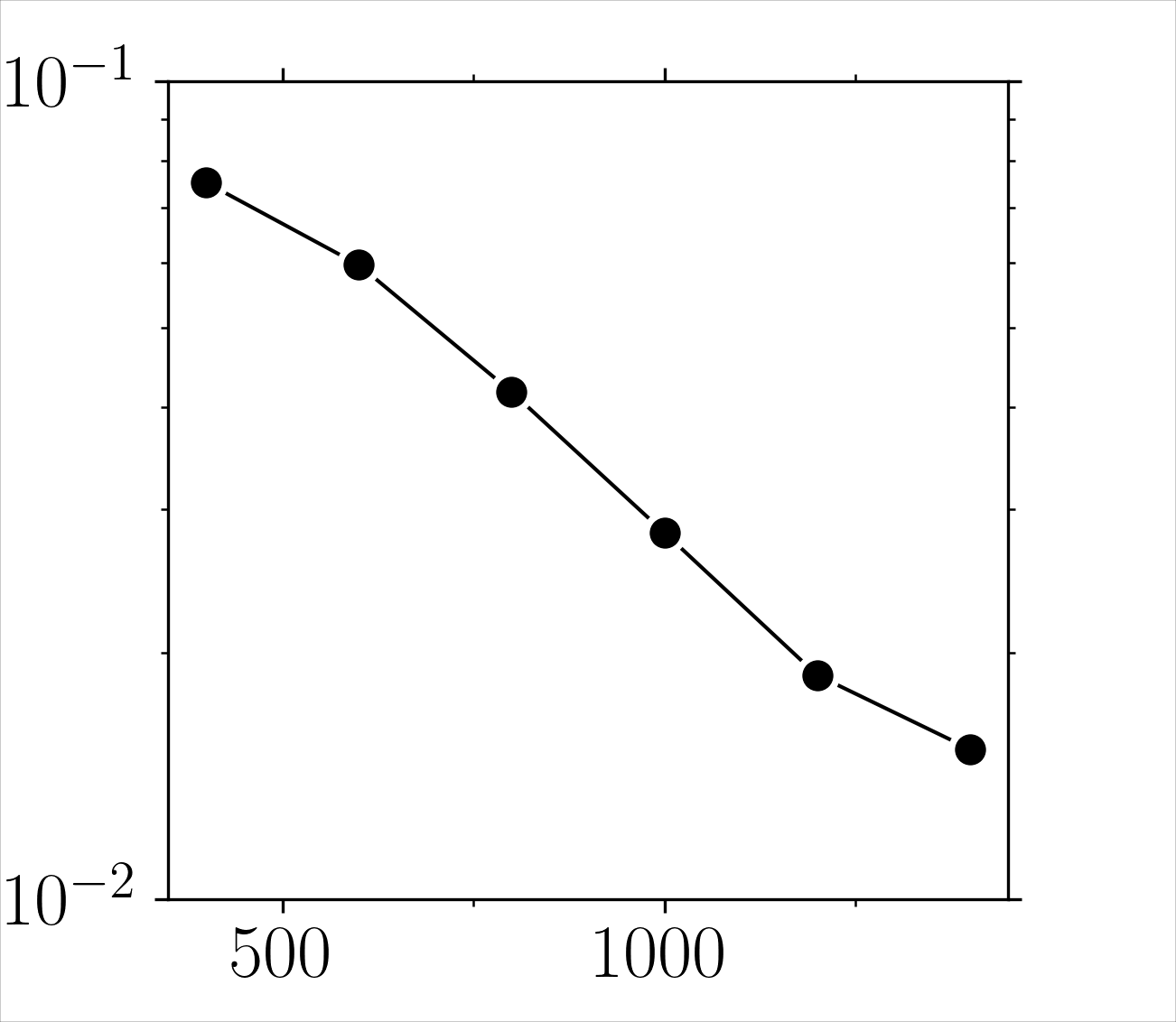}
        %
        \Labelxy{44}{-3}{0}{$T_a$\scriptsize (K)}
        \Labelxy{-6}{35}{90}{$p^\text{rms}_\text{NiNi}$}
        \LabelFig{17}{15}{$b)$}

    \end{overpic}
    %

    %
   \caption{SRO variations with annealing temperature $T_a$. \textbf{a)} $p_\text{NiNi}-p^\text{rss}_\text{NiNi}$ as a function of pair distance $r$ \textbf{b)} root-mean-squared fluctuations $p^\text{rms}_\text{NiNi}$ plotted against $T_a$. The (red) base line indicate zero correlations associated with RSAs. The results are based on the \potOne interatomic potential.
   \note{PS: consider splitting this figure in two. The first should contain only the upper panel (make it larger), plus an attempt to some scaling of the $p_\text{NiNi}-p^\text{rss}_\text{NiNi}$ as a function of pair distance $r$ for various annealing temperatures. Can the functions be scaled (e.g. linearly) so that that they "look the same"? If yes then the second picture (after Fig 6) could provide the scaling from figure 4 and fig 6, giving quantitative view on the effects of the annealing temperature.}
   \note[KK]{tried a simple rescaling by dividing $p_\text{NiNi}-p^\text{rsa}_\text{NiNi}$ by $p^\text{rms}_\text{NiNi}$ but didn't get a master curve. The curves show quite complex nonmonotonic features with a couple of characteristic lengthscales that i) are hard to measure ii) scale non-trivially with temperature.} \note[PS]{Thanks for checking. Maybe we should openly point out this nontrivial differences in the text, any non-triviality is "sexy". }
   } 
   \label{fig:sroTemp}
\end{figure}

To this end, we use a versatile approach to investigate the micro-structural/chemical origin of strengthening in NiCoCr including enthalpy-driven ordering effects and local distortions. 
We perform hybrid Monte Carlo/Molecular Dynamics (MC/MD) simulations at a range of annealing temperatures based on two commonly-used NiCoCr interatomic potentials. 
We find that the emergence of SROs is not a robust feature of annealed model NiCoCr CSAs but, to a great extent, depends on the chosen potential energy.
More specifically, the two models generate microstructurally different alloys (with/out SROs) with the exact same thermal processing.
Following the numerical framework in \cite{Li2019}, we probed effects of SROs in terms of local concentration fluctuations, stacking fault widths, dislocation glide resistance, and misfit volumes of NiCoCr as well as thermodynamic properties such as the specific heat and thermal expansion coefficient. 
Our analysis indicates meaningful correlations of the above observables with varying degrees of SROs in aged alloys, making them easily distinguishable from RSAs. 
{We further observe a marked growth in the population of SROs inside the stacking fault region and remarkable strengthening behavior against dislocation glide with the latter rooted in the interplay between short range ordering and local misfit properties.}

The organization of this paper is the following.
In Sec.~\ref{sec:methods}, we describe the numerical setup, sample preparation (including aging/annealing), loading protocols, and relevant simulation details {including the hybrid MD/MC model, interatomic forces, and shear test description}. 
Section~\ref{sec:results} presents our simulation results relevant to the chemical/microstructural characterization of SROs and their potential effects on dynamics of dislocations.
In this context, Sec.~\ref{sec:sro_temp} introduces robust structural/compositional metrics {associated with local elemental fluctuations } to characterize the temperature-dependence of SROs and {distribution of misfit volumes}.
Lattice distortions in the presence of short range ordering will be discussed in Sec.~\ref{sec:LatticeDistortions}.
In Sec.~\ref{sec:sro_disl}, we provide an in-depth analysis of partial dislocations and their depinning mechanism in the presence of SROs.
{This includes auto-correlation analyses associated with the dislocation line fluctuations as well as local variations of the dislocation velocity .}
Section~\ref{sec:discussions} presents relevant discussions and conclusions.

%% file: sections/Methods.tex
\section{\label{sec:methods}Methods and Materials}
Molecular dynamics simulations were carried out in LAMMPS \cite{LAMMPS} by implementing atomistic samples of size $N=500,000$ and $1,700,000$ within a three-dimensional periodic cell. 
In order to study SRO properties (in the absence of dislocations), we prepared cubic samples with length $10$~nm along
the $x[100]$, $y[010]$, and $z[001]$ directions.
The NPT ensembles were implemented via a Nose-Hoover thermostat and barostat with relaxation time scales $\tau_d^\text{therm}=10$ fs and $\tau_d^\text{bar}=100$ fs.
We also set the discretization time to $\Delta t\simeq 1.0$ fs.
Samples were initially prepared via an energy minimization at $T=0$ K (at a fixed volume) and subsequently thermalized at different temperatures \add[KK]{and constant pressure $P=0$ bar} for $100$ ps prior to annealing. 

The interatomic forces are based on two commonly-used embedded-atom method (EAM) potentials in the context of \comp solid solution alloys: \romn{1}) the Li-Sheng-Ma potential proposed in \cite{Li2019} which has been utilized in recent SRO studies, modeling dislocation nucleation and glide dynamics \cite{cao2020novel,jian2020effects}, and nanoindentation tests \cite{yang2022chemical} 
\romn{2}) the EAM Farkas–Caro potential \cite{farkas2018model} originally developed to model equimolar high-entropy \compFarkas alloys but used here to validate the SRO formation and its robustness against different potentials.

Annealed configurations were obtained performing hybrid MC/MD simulations based on the variance constrained semi-grand canonical (VCSGC) ensemble \cite{PhysRevB.85.184203} within the annealing temperature range $T_a=400-1300$ K. 
In order to determine the values of $\Delta\mu_{X_1X_2} = \mu_{X_1} - \mu_{X_2}$ which minimizes the composition errors we perform a set of semi-grand canonical simulations varying the chemical composition at $T = 1500$ K, and fitted the MC data using the equation: \(\Delta\mu(X_1,P,T)=T\ln(X_1/[1-X_1])+\sum_{i=0}^{n}A_iX_{1}^{i}\) (Fig.~\ref{fig:Chemical Poyentials}), where $X_1$ is the reference element (Ni in our work) and $A_i$ are the fitting parameters \cite{Becker}. This allow us to perform hybrid MD/VCSGC-MC with a fixed target composition. During the annealing process, we perform $1$ MC cycle consisting of $N/2$ attempts of transmutation every $20$ MD steps \remove[KK]{ and swapping}. 
We carried out a total number of $800,000$ MC cycles at all annealing temperatures $T_a$ to ensure that the configurations reach thermal equilibrium and that the structure of SROs are statistically indifferent.

We also studied dynamics of a $\frac{1}{2}[\bar{1}10](111)$ edge dislocation which, under an external perturbation, dissociates into two separate partials with a stacking fault in face-centered cubic (fcc) crystals.
To this end, we constructed a simulation cell with dimensions $L_x\simeq80$~nm, $L_y\simeq 20$~nm, and $L_z\simeq 15$~nm (see Fig.~\ref{fig:sketchLoadSetup}) and performed annealing at a desired temperature $T_a$.
We subsequently equilibrated the annealed alloy at a low temperature $T=5$~K and pressure $P=0$ for the duration of $100$ ps.
To create an edge dislocation within the aged sample, we used the periodic array of dislocation (PAD) model proposed in \cite{osetsky2003} to ensure a periodic setup along the Burgers vector ($x[\bar{1}10]$) and dislocation line ($z[\bar{1}\bar{1}2]$). 
The dislocated sample was further relaxed using the NPT framework with $P_{xx}=P_{zz}=0$ and $T=5$ K (for 100 ps) leading to the dislocation dissociation into two Shockley partials.
A stress-controlled framework was employed within the NVT ensemble at $T=5$ K by applying additional forces on the top plane (normal to $y$) with the bottom layer held fixed (with a thickness $\simeq 4$ \r{A}) during the course of shear simulations. 
The applied stress was gradually increased from $\sigma=50$~MPa (in a quasi-static fashion) above a critical (depinning) stress in order to move the partial dislocations.

%% file: sections/Results.tex
\section{\label{sec:results}Results}
The above methodology to prepare and anneal model CSAs indicates that aged NiCoCr alloys may exhibit varying degrees of chemical ordering under different annealing temperatures.
Using robust composition-based metrics in Sec.~\ref{sec:sro_temp}, we further confirm the formation of SROs and their strong dependence on the thermal processing.
Section~~\ref{sec:LatticeDistortions} presents the misfit volume analysis indicating meaningful correlations with the degree of short range ordering. 
In Sec.~\ref{sec:sro_disl}, we analyze dynamics of partial dislocations in the presence of SROs and discuss potential implications in terms of the SRO-induced CSA strengthening.

\begin{figure*}[t]
   \centering
    \begin{overpic}[width=0.24\textwidth]{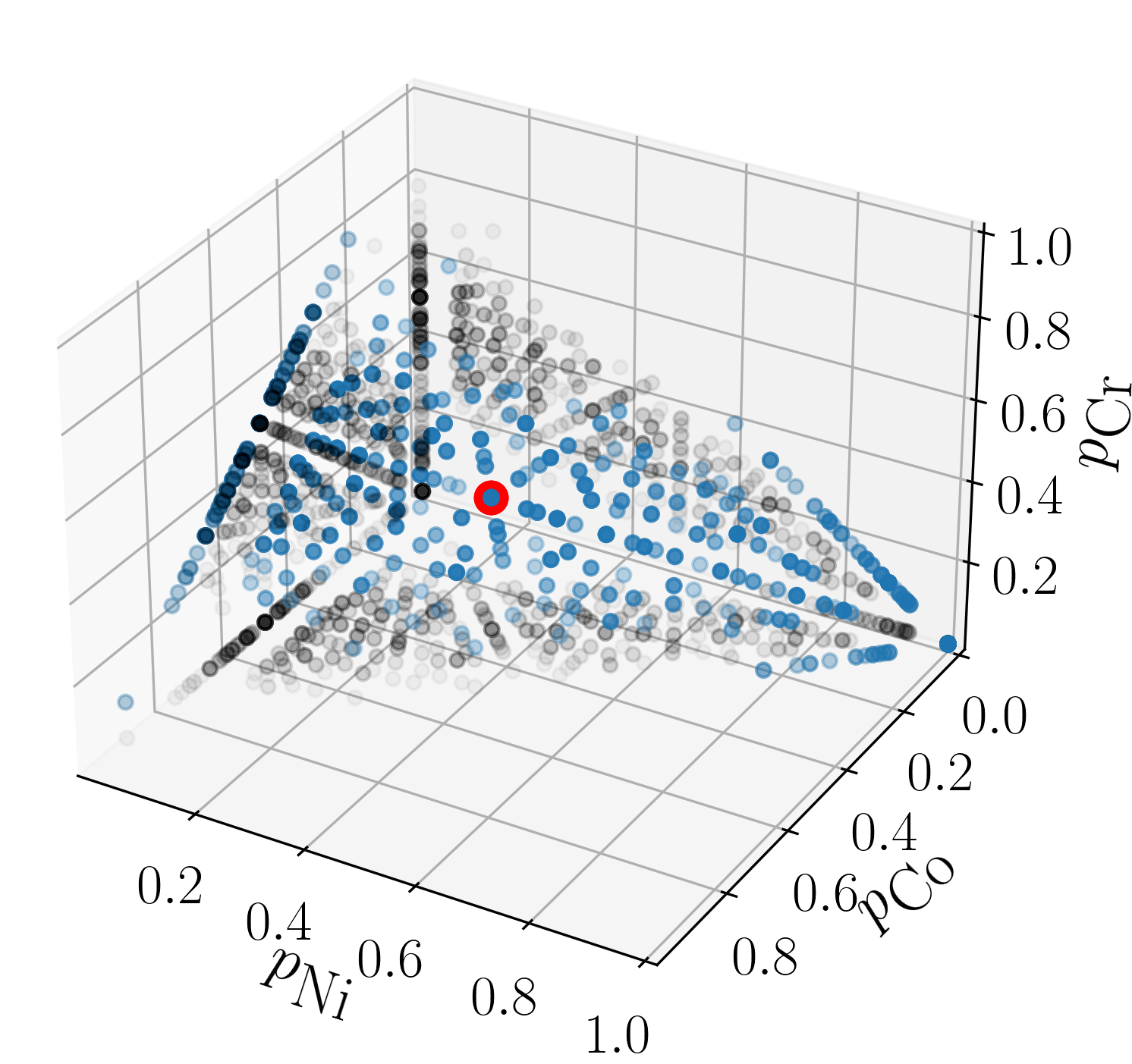}
        \LabelFig{6}{82}{$a)~r_c=5~$\r{A}}
    \end{overpic}
    \begin{overpic}[width=0.24\textwidth]{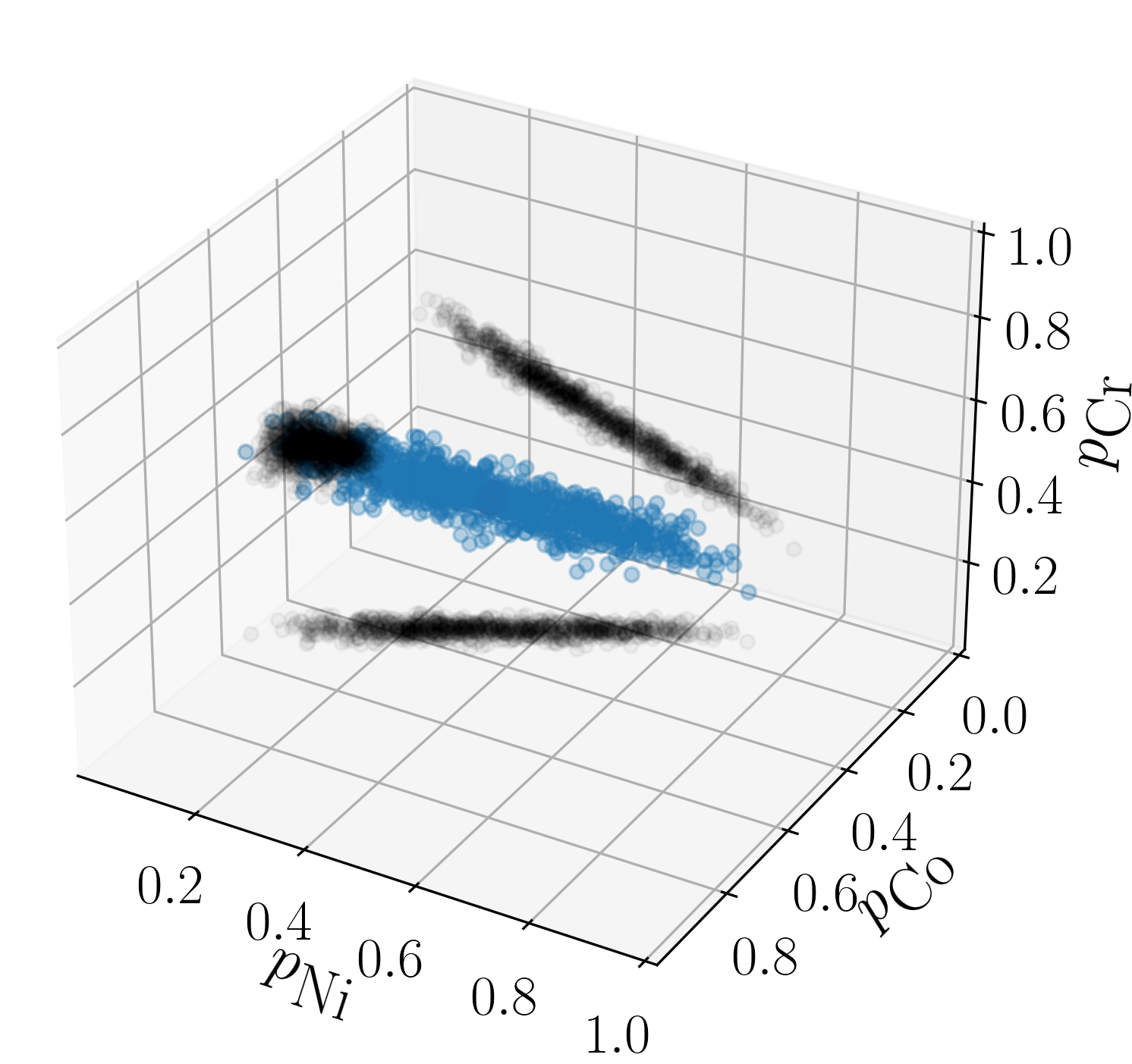}
        \LabelFig{6}{82}{$b)~r_c=10~$\r{A}}
    \end{overpic}
    \begin{overpic}[width=0.24\textwidth]{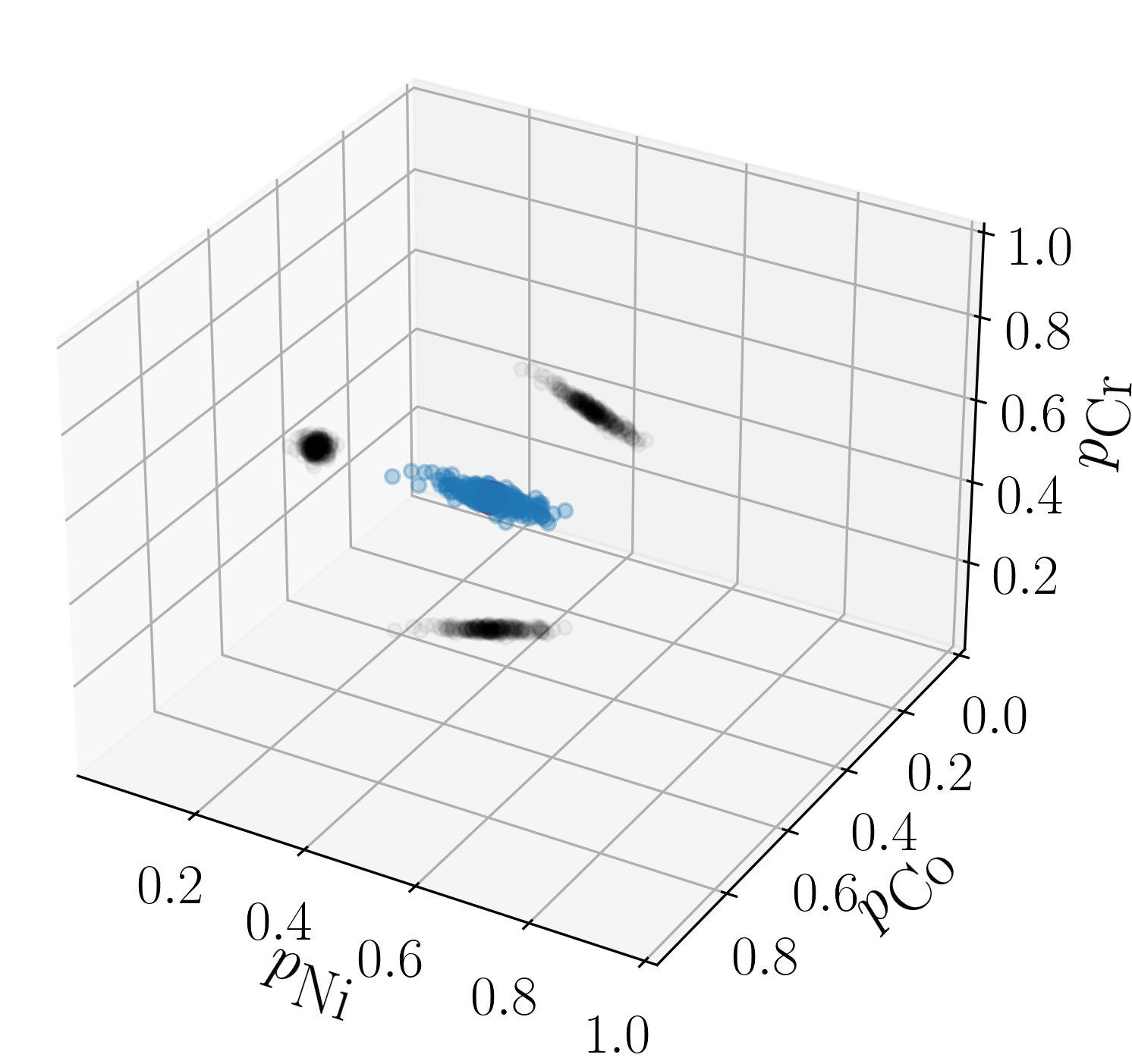}
        \LabelFig{6}{82}{$c)~r_c=18~$\r{A}}
    \end{overpic}
   \caption{Local concentration fluctuations $p_b$ with $b$ being Ni, Co, and Cr in an annealed CSA (based on the \potOne potential) at $T_a=400$ K and multiple lengthscales \textbf{a)} $r_c=5~$\r{A} \textbf{b)} $r_c=10~$\r{A} \textbf{c)} $r_c=18~$\r{A}. \add[KK]{We note that at low distance scales $r_c=5$ and $10$ \r{A}, the segregation of Ni and Co/Cr domains is very strong.} The black scatter points represent the two dimensional projections. The red dot denotes the equimolar concentration. \note{PS: comparing fluctuations for large radii -- much larger than the size of segregation does not give much physical information. decrease of the fluctuations with increasing radius is pure statistics. Consider three panels with $r_c$ equal to 5 and 10 and 18 \AA} \note[KK]{done.} \note[PS]{Great. But I would add a comment in the caption signalling the meaining: that at low distance scales $r_c$ of 10, 5 \AA the segregation of Ni and Co/Cr domains is very strong.}} 
   \label{fig:concentrationFluctuations}
\end{figure*}

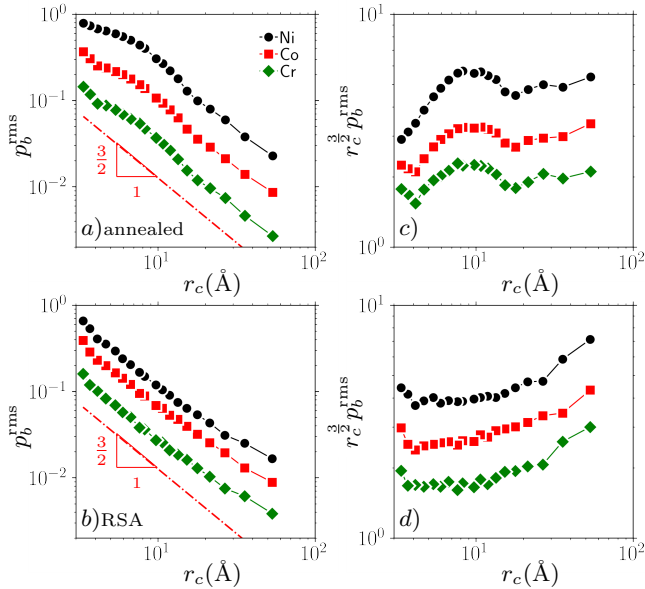
\begin{figure}[b]
    \begin{overpic}[width=0.23\textwidth]{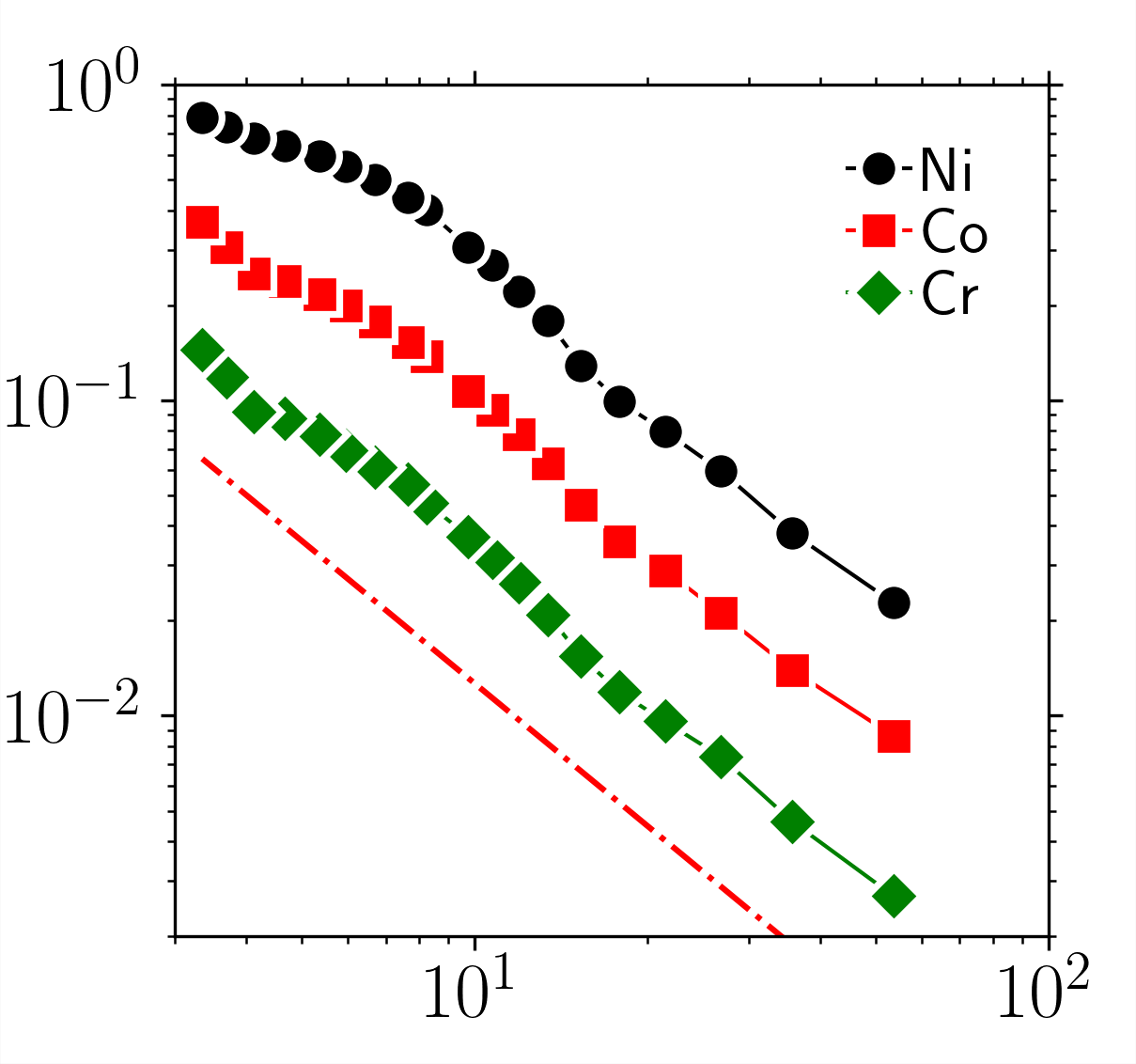}
        \LabelFig{17}{15}{$a)$\scriptsize annealed}
        \Labelxy{50}{-3}{0}{$r_c$(\r{A})}
        \Labelxy{-6}{40}{90}{$p^\text{rms}_b$}
       \begin{tikzpicture}
            \coordinate (a) at (0,0); 
            \node[white] at (a) {\tiny.};               %
                 \drawSlope{1.3}{1.5}{0.35}{50}{red}{\frac{3}{2}}
 		\end{tikzpicture}
 	\end{overpic}
    \begin{overpic}[width=0.23\textwidth]{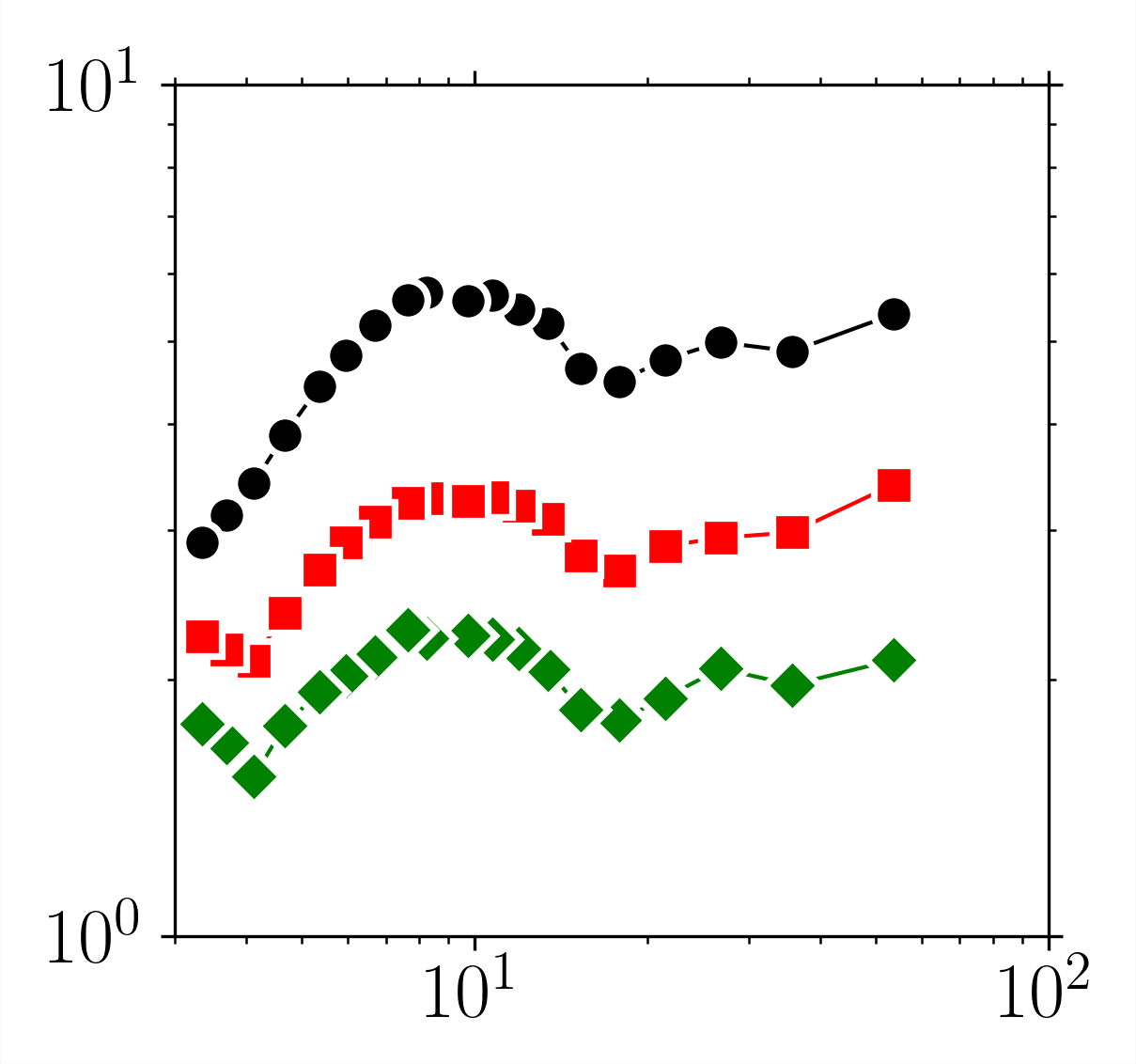}
        \Labelxy{-6}{40}{90}{$r_c^{\frac{3}{2}}p^\text{rms}_b$}
        \Labelxy{50}{-3}{0}{$r_c$(\r{A})}
        \LabelFig{17}{15}{$c)$}
 	\end{overpic}
    \begin{overpic}[width=0.23\textwidth]{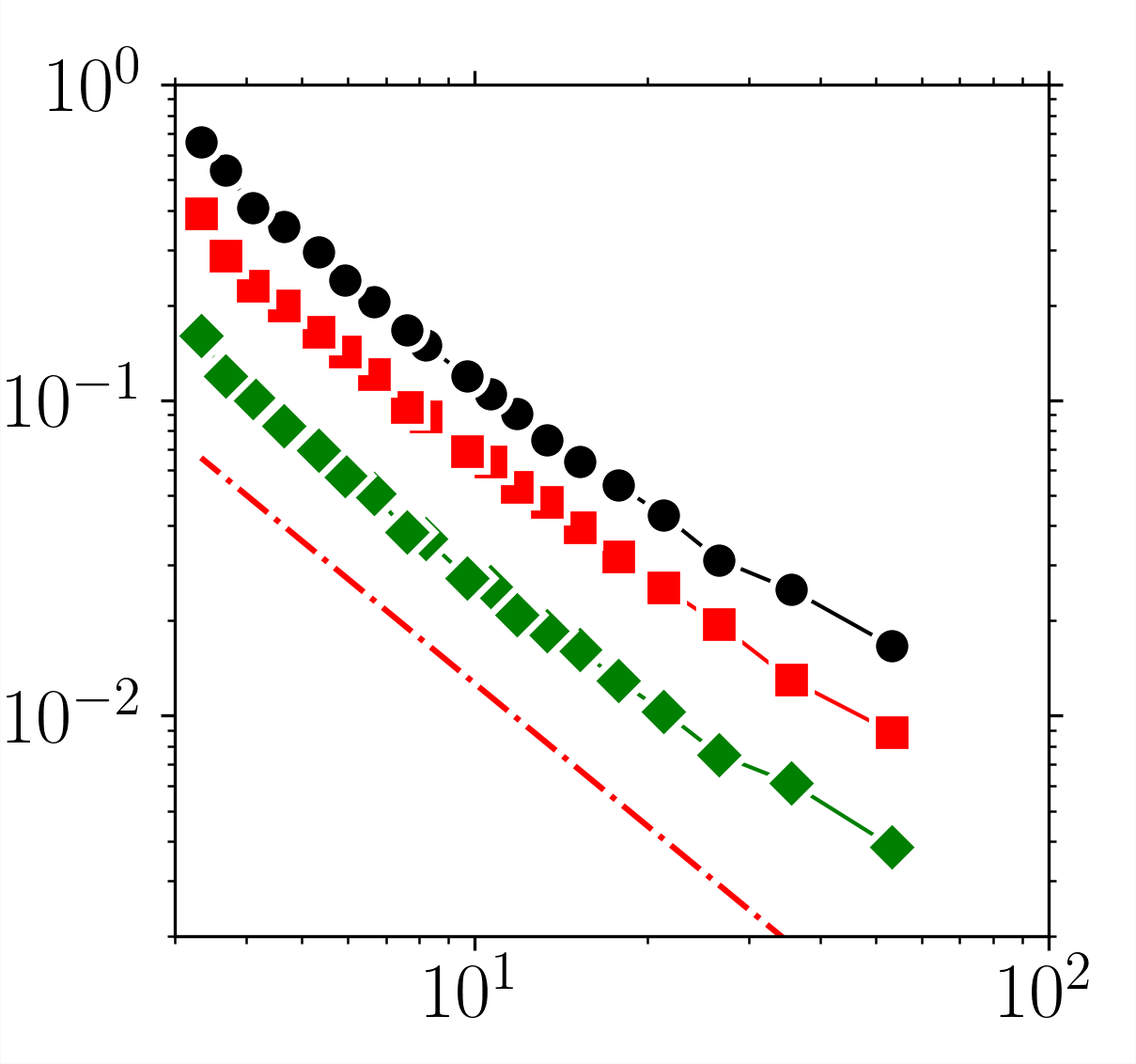}
        \LabelFig{17}{15}{$b)$\scriptsize RSA}
        \Labelxy{-6}{40}{90}{$p^\text{rms}_b$}
        \Labelxy{50}{-3}{0}{$r_c$(\r{A})}
       \begin{tikzpicture}
            \coordinate (a) at (0,0); 
            \node[white] at (a) {\tiny.};               %
                 \drawSlope{1.3}{1.5}{0.35}{50}{red}{\frac{3}{2}}
 		\end{tikzpicture} 
 	\end{overpic}
    \begin{overpic}[width=0.23\textwidth]{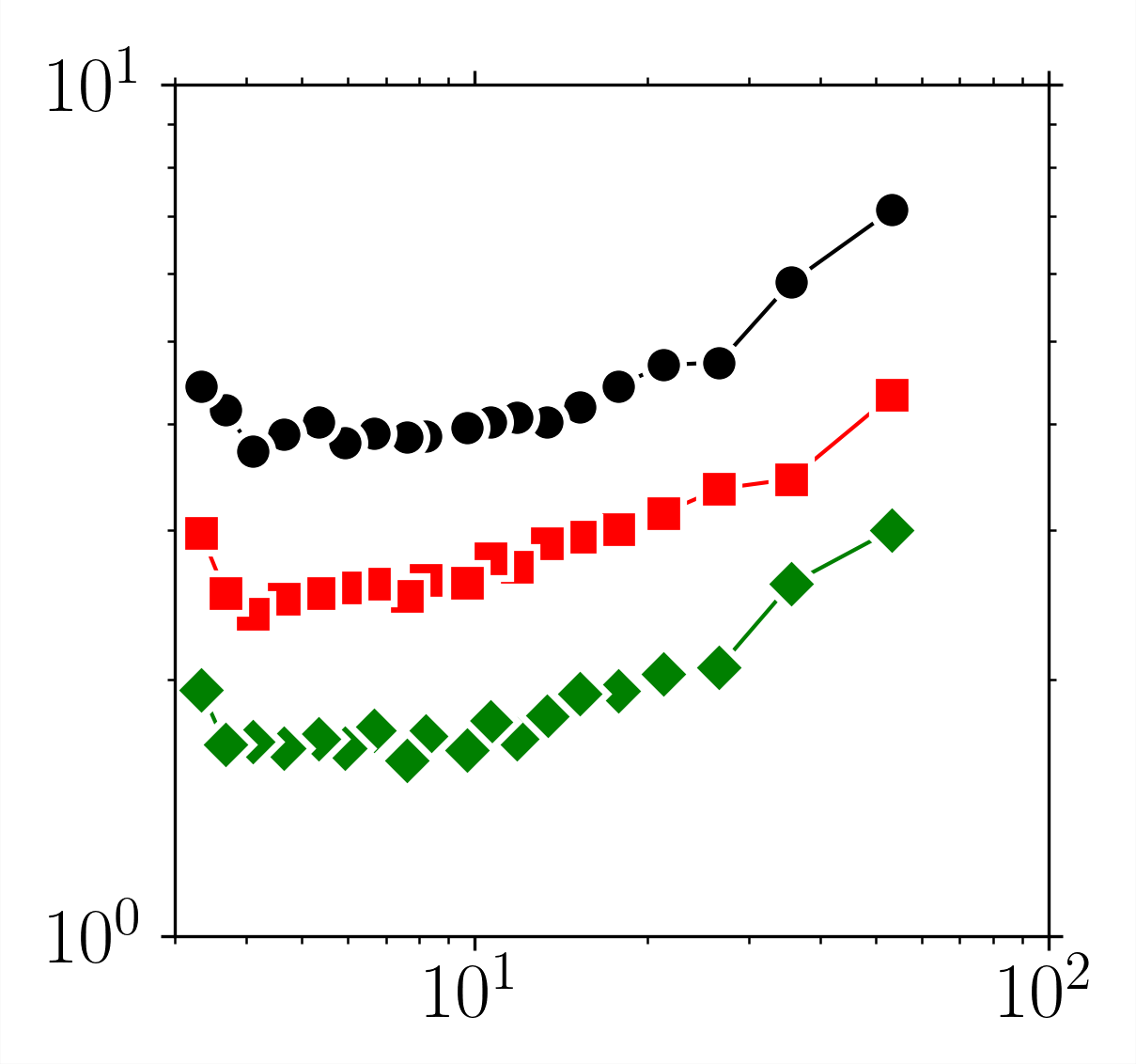}
        \Labelxy{-6}{40}{90}{$r_c^{\frac{3}{2}}p^\text{rms}_b$}
        \Labelxy{50}{-3}{0}{$r_c$(\r{A})}
        \LabelFig{17}{15}{$d)$}
 	\end{overpic}
   \caption{Root-mean-squared (rms) fluctuations $p^\text{rms}_b$ of local Ni, Co, and Cr concentrations as a function of window size $r_c$ in \textbf{a)} annealed NiCoCr CSAs at $T_a=400$ K \textbf{b)} NiCoCr RSAs at $T=400$ K based on the \potOne potential. The panels in \textbf{c)} and \textbf{d)} are the same as \textbf{a)} and \textbf{b)} but with the $y$-axis rescaled by $r_c^{-\frac{3}{2}}$. The curves are shifted for the sake of clarity.}  
   \label{fig:std_concentrationFluctuations}
\end{figure}

\subsection{Temperature-dependence of SROs}\label{sec:sro_temp}
Figure~\ref{fig:thermo} illustrates that the formation of SROs within the solid solutions may strongly depend on the annealing temperature $T_a$.
The color map of atoms, the inset of Fig.~\ref{fig:thermo}(a), indicates a clear segregation of Ni-rich phases (gray circles) in the NiCoCr CSA modeled via the \potOne potential at $T_a=800$ K.
Overall, the observed clustering features tend to become less pronounced at higher annealing temperatures as shown in Fig.~\ref{fig:Annealed}.

The presence of SROs appears to have a direct relevance on the (constant-pressure) heat capacity $C_p=\partial_T H$ in Fig.~\ref{fig:thermo}(a) featuring a maximum around $T_a\simeq 800$~K.
Here $H$ denotes the enthalpy \footnote{We only report the (excess) configurational heat capacity (and thermal expansion coefficient) neglecting (ideal) kinetic contributions.}.
The data presented in Fig.~\ref{fig:thermo} correspond to a sample equilibrated at $T=300$ K and subsequently aged at multiple annealing temperatures.
The emerging peak in $C_p$ may suggest a dominant role of enthalpic interactions over entropic effects that rule out the formation of an ideal random solid solution \cite{Li2019}.   
We note that the (ideal) heat capacity associated with the latter rises monotonically within the temperature range of interest.
Such deviations seem to be less pronounced in terms of the thermal expansion coefficient $\alpha_p=\frac{1}{V}\partial_T V$, with $V$ being the system volume, as shown in Fig.~\ref{fig:thermo}(b).

The notion of SROs typically refers to coherent compositional deviations from (statistically) random distributions of atoms within the solution matrix.
Along these lines, we investigated the Warren–Cowley SRO parameters $p_{ab}(r)$ \cite{wolverton2000short}
probing the concentration variations of type-$b$ atoms at a distance $r$ from a center type-$a$ element. 
For an equi-molar random \comp solid solution, one obtains $p^{\text{rsa}}_{ab}=0.33$ (on average) at any $r$ ---more precisely, between the successive valleys of the pair correlation function $g(r)$ as in Fig.~\ref{fig:sroSheng}(g).
The SRO parameters could be also determined locally for individual atoms which will presumably show strong fluctuations in the presence of SROs.
Nevertheless, the ``averaged" Warren–Cowley parameters should be relevant as the system tends to be statistically homogeneous beyond the mean SRO size.

Figure~\ref{fig:sroSheng}(a-f) illustrates deviations of $p_{ab}$ associated with the annealed CSAs from $p_{ab}^{\text{rsa}}$ including the six (distinct) elemental pairs at $T_a=400$ K.
The order parameters reveal several interesting features describing the SRO microstructure.
The abundance of the Ni-Ni pairs beyond random concentrations is remarkable and appears to persist up to $r\simeq 5$~\r{A} in Fig.~\ref{fig:sroSheng}(a).
It should be noted that twice this lengthscale is in a rough agreement with the visual impression one gets from the segregation map, the inset of Fig.~\ref{fig:thermo}(a), regarding the mean SRO size.
Below the base line, the dip in $p_{ab}$ corresponding to Ni-Ni pairs should indicate their scarcity above the mean size.
The order parameter re-crosses the horizontal line beyond which it features a fairly broad peak at $r\simeq 15$~\r{A} before going asymptotically to $p_{ab}^{\text{rsa}}$.
We remark that the inferred lengthscale could potentially relate to the average spacing between adjacent SROs.
Figure~\ref{fig:sroSheng}(b-c) associated with $p_\text{NiCo}$ and $p_\text{NiCr}$ feature fairly identical properties but with opposite trends as seen in $p_\text{NiNi}$ since they must all add up to unity.

As opposed to Ni-Ni ordering, we observe less coherent patterns associated with the identical (same-element) pairs Co-Co and Cr-Cr as in Figure~\ref{fig:sroSheng}(d) and (f).  
In particular, $p_\text{CoCo}$ and $p_\text{CrCr}$ seem to indicate ordering as well as anti-ordering (potentially due to repulsion) at the first and next nearest neighbor distances.
The strong bonding between Co-Cr in Fig.~\ref{fig:sroSheng}(e) at the first nearest neighbor is also remarkable (see also \cite{Li2019,jian2020effects,yang2022chemical}).
We further note that, unlike $p_{ab}$, the pair correlation function $g(r)$ does not suggest any \emph{structural} differences between annealed and random solid solutions as shown in Fig.~\ref{fig:sroSheng}(g).

\begin{figure}[t]
    \centering
    \begin{overpic}[width=0.32\textwidth]{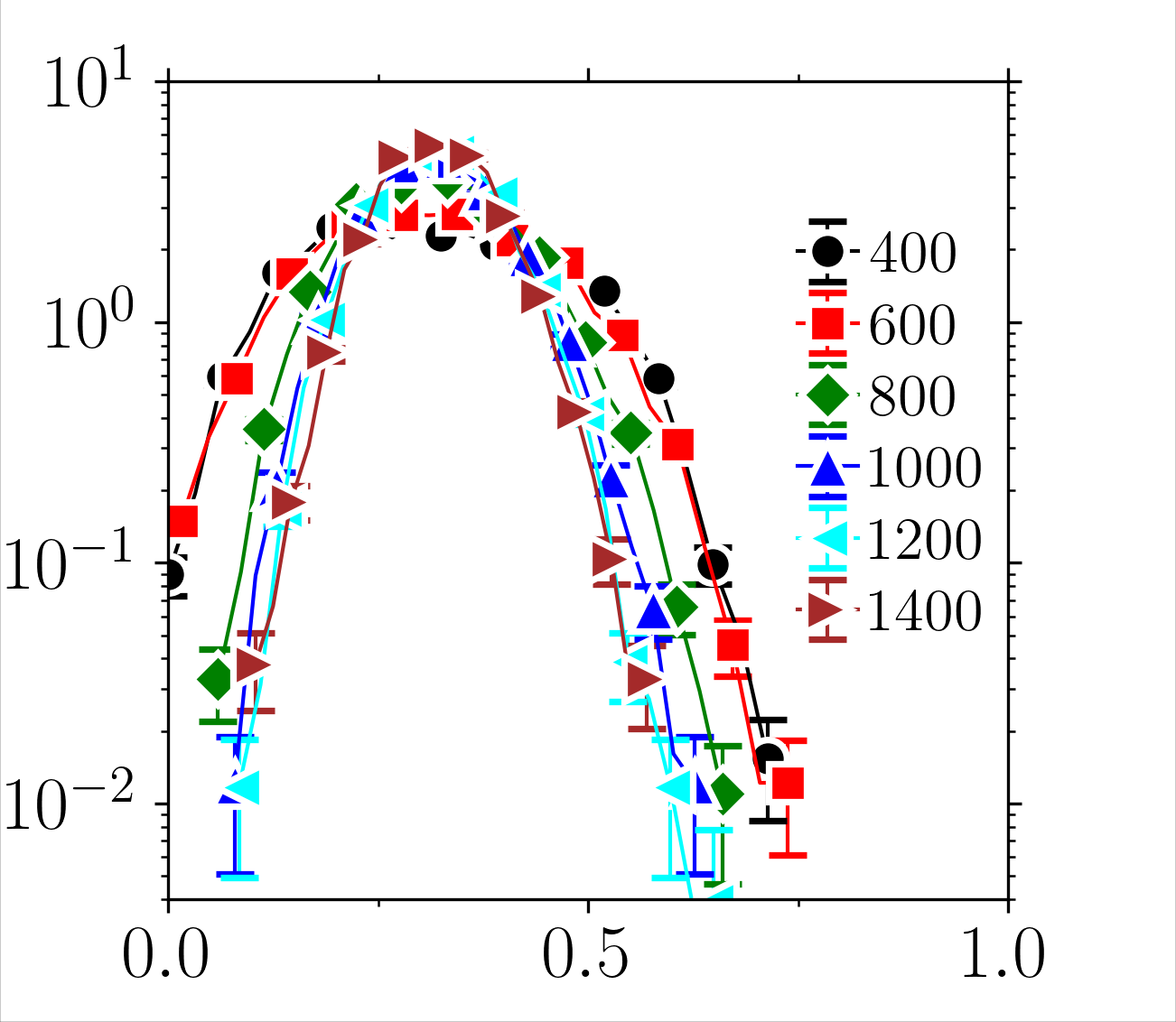}
        \Labelxy{68}{71}{0}{$\scriptstyle T_a$\scriptsize (K)}
        \Labelxy{50}{-3}{0}{$p_\text{Ni}$}
        \Labelxy{-6}{40}{90}{PDF}
    \end{overpic}
    %
    %
   \caption{Probability distribution functions of the local concentrations $p_\text{Ni}$ in an annealed CSA based on the \potOne potential at the window size $r_c=10$ \r{A} and various $T_a$. \note{PS: rather than comparing window sizes, compare the distributions for different annealing temperatures for a given $r_c$. Then look for a scaling factor that could transform all these PDF shapes into a single one: this would provide quantitative measure of the effects of annealing.}\note[KK]{please see the updated figures and relevant discussions in the text}. \note[PS]{Can we normalize the PDF histograms, so that the CDF sums to 1? I am asking because then one can state which fraction of environments has $p_{Ni}$ equal to zero (or close to it). for $T_a$ of 400 and 600 it is a non-negligible fraction, which docuents strong segregation of Ni and Co/Cr phases.}\note[KK]{these are probability density functions, normalized such that the *integral* over the entire range is 1. In other words, associated CDFs should be between 0 and 1 by construction. I am adding relevant discussions on the relative abundance of regions with very low and high Ni content in the text.} }
   \label{fig:pdf_concentrationFluctuations}
\end{figure}

Figure~\ref{fig:sroTemp} quantifies the abundance of Ni-Ni elemental pairs upon annealing at multiple temperatures between $T_a=400-1400$ K.
\add[KK]{As shown in Fig.~\ref{fig:sroTemp}(a), the curves show quite complex nonmonotonic features with certain characteristic lengthscales that seem to scale non-trivially with temperature.}
Nevertheless, the SRO-related features in $p_\text{NiNi}-p^\text{rsa}_\text{NiNi}$ become less and less pronounced with an increase in $T_a$ continually approaching their asymptote at the zero-valued base line.
This is mathematically quantified by the metric $p^\text{rms}_\text{NiNi}=\langle (p_\text{NiNi}-p^\text{rsa}_\text{NiNi})^2\rangle^{\frac{1}{2}}$ as a root-mean-squared (rms) measure of deviations from RSAs.
Figure.~\ref{fig:sroTemp}(b) features a monotonic growth of $p^\text{rms}_\text{NiNi}$ upon decreasing $T_a$.
We also note the tendency for the saturation of $p^\text{rms}_\text{NiNi}$ at $T_a\simeq 1400$ K or above potentially due to residual SRO distributions at atomistic levels that preclude an ideal RSA formation.
The above analysis was repeated for NiCoCr alloys simulated based on the \potTwo potential. 
Interestingly, we found no clear signature of clustering in these samples as opposed to those generated via the \potOne potential (see Fig.~\ref{fig:sroFarkas}).

We probed fluctuations in \emph{local} concentrations of the constituent elements in annealed NiCoCr alloys that, in the presence of SROs, should deviate from those of random solid solution alloys.
In this context, the entire space was partitioned using sub-volumes of varying size $r_c$ and local molar compositions $p_\text{Ni}$, $p_\text{Co}$, and $p_\text{Cr}$ were determined within each cube.
As illustrated in the scatter plot of Fig.~\ref{fig:concentrationFluctuations} associated with NiCoCr CSAs annealed at $T_a=400$ K, the fluctuations tend to self-average at larger $r_c$ which could be also understood in terms of counting statistics.

We also investigated \emph{local} fluctuations in CSA elemental concentrations in space that, in the presence of SROs, show marked deviations from those of RSAs. 
Figure~\ref{fig:std_concentrationFluctuations} shows rms fluctuations of local concentrations $p_\text{Ni}$, $p_\text{Co}$, and $p_\text{Cr}$ and their decay with distance $r_c$. 
In Fig.~\ref{fig:std_concentrationFluctuations}(a) and (c), rms fluctuations in annealed NiCoCr CSAs seem to self-average but only above some certain lengthscale above which the decay is well-predicted by the expected $r_c^{-3/2}$ power-law behavior. 
The latter is the relevant scaling in purely random atomic configurations as illustrated in Fig.~\ref{fig:std_concentrationFluctuations}(b) and (d).
We take the characteristic peak in Fig.~\ref{fig:std_concentrationFluctuations}(c) as a signature of spatial correlations which may be interpretted as the average SRO size $\xi^\text{sro}\simeq 10$ \r{A}. 
Furthermore, the inferred lengthscale closely agrees with the one extracted from the SRO order parameters in the preceding paragraphs which is within the typical range of SRO size ($0.5-1.5$ nm) seen experimentally \cite{zhang2020short,wang2022chemical}.
\add[KK]{Figure~\ref{fig:pdf_concentrationFluctuations} illustrates the probability distribution functions (PDFs) of the local Ni concentrations $p_\text{Ni}$ at $r_c=10$ \r{A} for different annealing temperatures. 
We note the marked deviation of the low-$T_a$ PDFs from a standard Gaussian distribution which, otherwise, seems to be the asymptotic limit for the $p_\text{Ni}$ distributions at higher annealing temperatures.
One could also observe a marked abundance of low and high (local) Ni concentrations away from the average at $p_\text{Ni}=0.33$ for $T_a=400$ and $600$ K which is indicative of the strong segregation of Ni phases.}

\begin{figure}[b]
  \centering
  \begin{overpic}[width=0.24\textwidth]{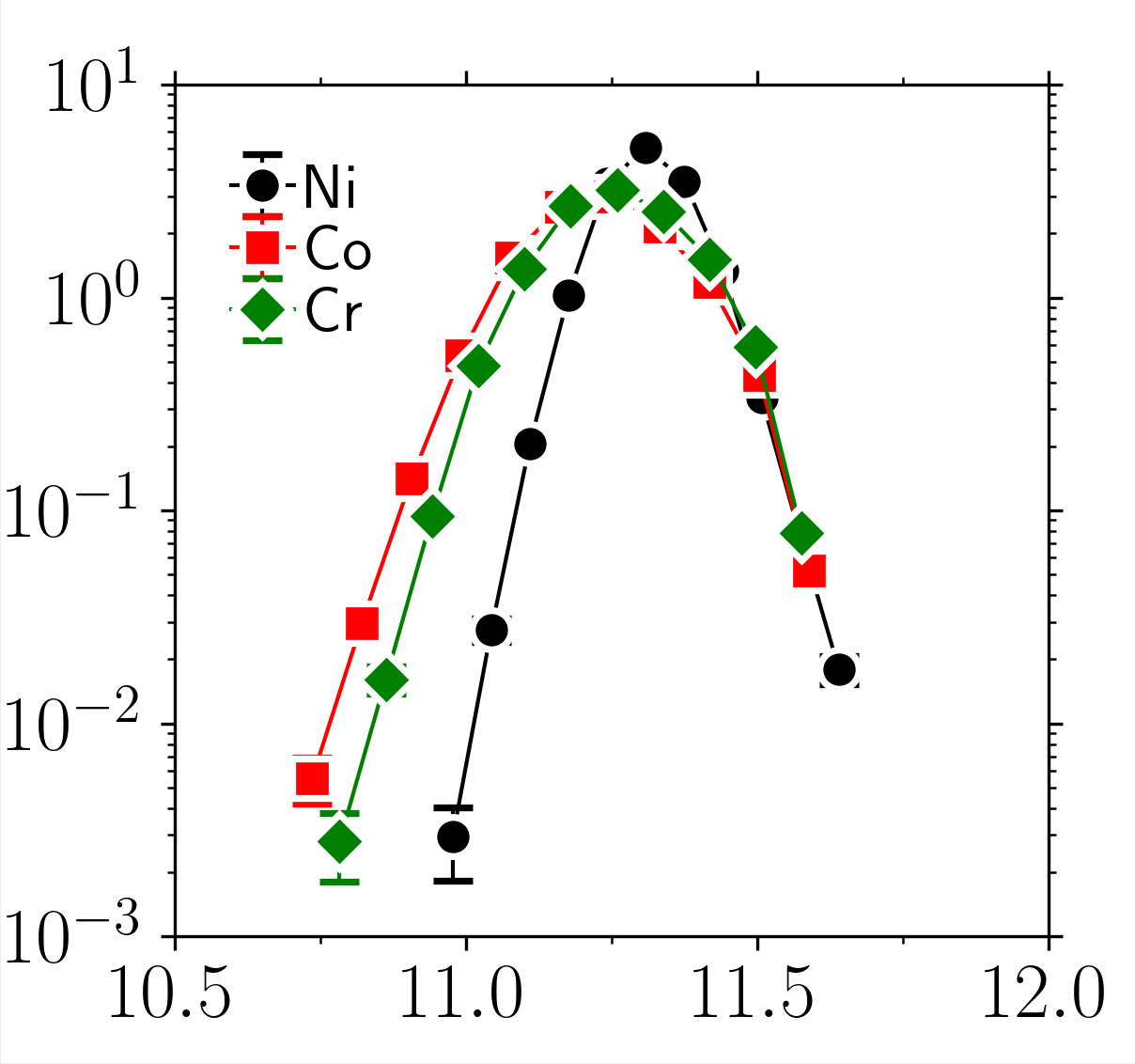}
        \LabelFig{19}{89}{$a)$ annealed}
        \Labelxy{50}{-4}{0}{$V_b$(\r{A}$^3$)}
        \Labelxy{-6}{40}{90}{PDF}
  \end{overpic}
  
  \begin{overpic}[width=0.24\textwidth]{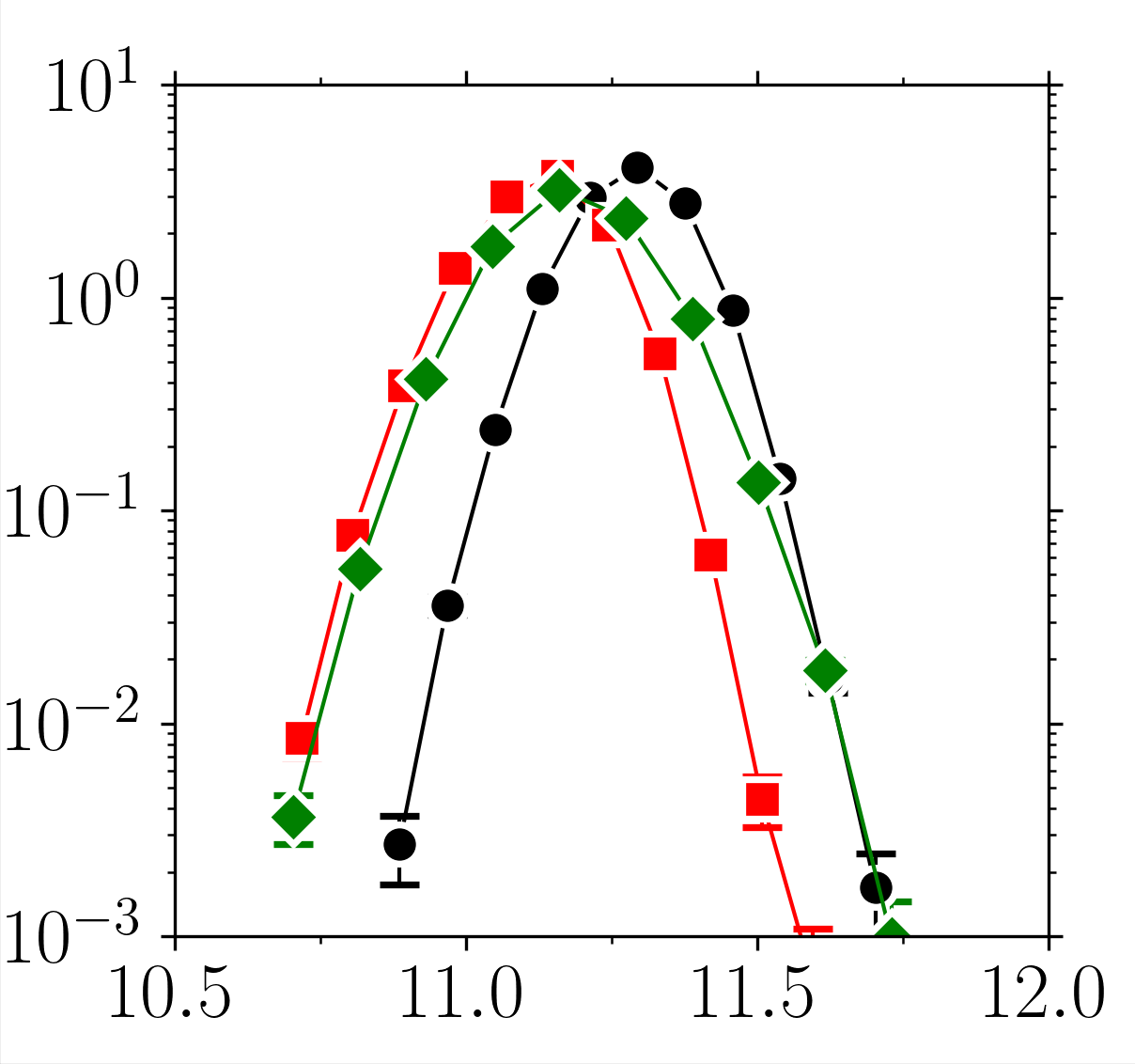}
        \LabelFig{19}{89}{$b)$ RSA}
        \Labelxy{50}{-4}{0}{$V_b$(\r{A}$^3$)}
        \Labelxy{-6}{40}{90}{PDF}
  \end{overpic}
  \caption{Probability distributions of alloy atomic (Voronoi) volumes $V_b$ with $b$ denoting the Ni, Co, and Cr atoms in \textbf{a)} the \comp alloy aged at $T_a=600$ K and \textbf{b)} \comp RSA. The volume measurements were carried out at $5$ K.
  \note{PS: graphics: please make the figure graphically consistent, all Y axis descriptions on one side. }\note[KK]{done.}} 
  \label{fig:voronoi}
\end{figure}

\begin{figure}[t]
  \centering
  \begin{overpic}[width=0.24\textwidth]{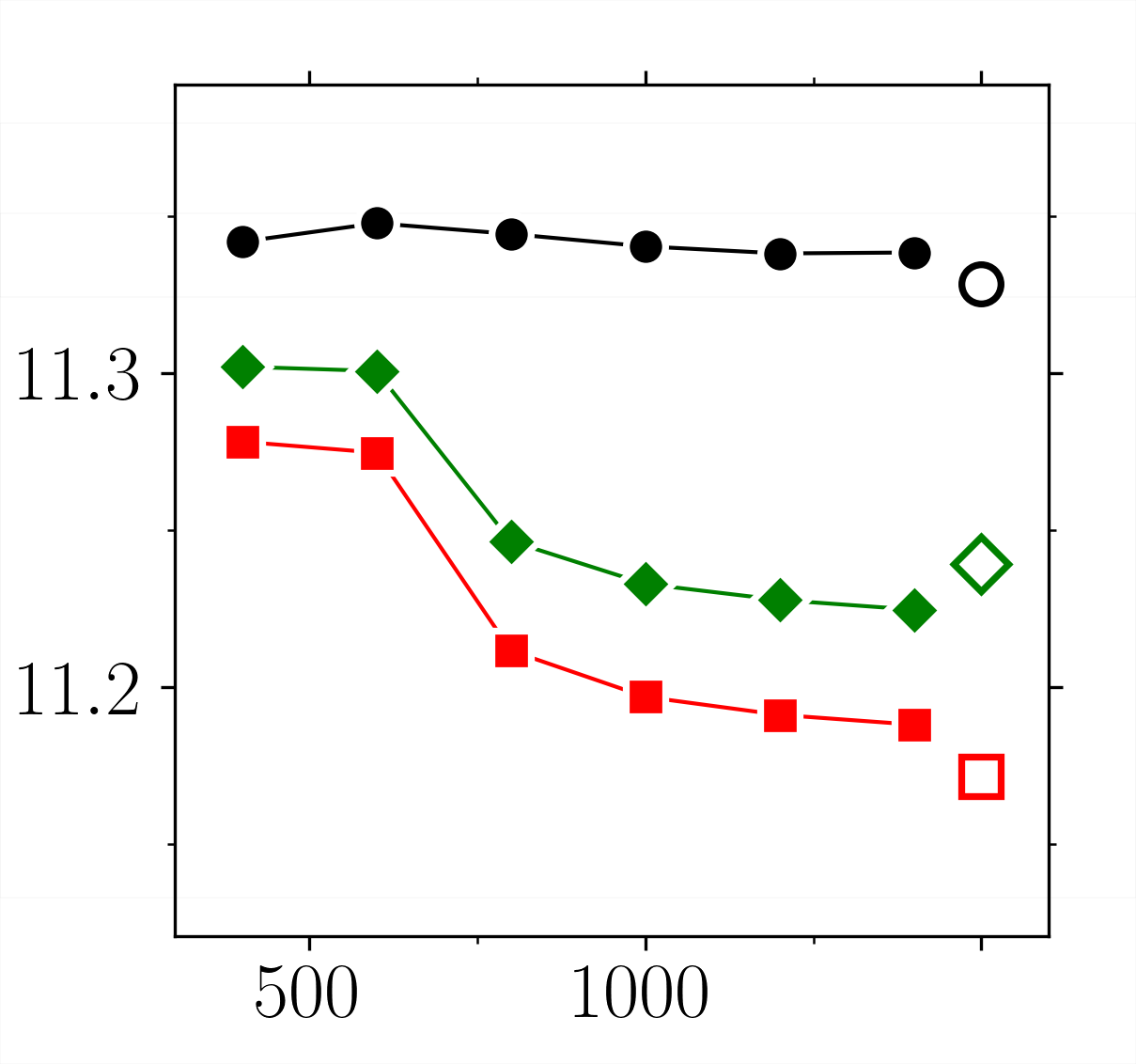}
        \LabelFig{19}{76}{$a)$}
        \Labelxy{50}{-4}{0}{$T_a$(K)}
        \Labelxy{83}{4}{0}{\scriptsize RSA}
        \Labelxy{-8}{40}{90}{$\langle V_b \rangle$}
  \end{overpic}
  
  \begin{overpic}[width=0.24\textwidth]{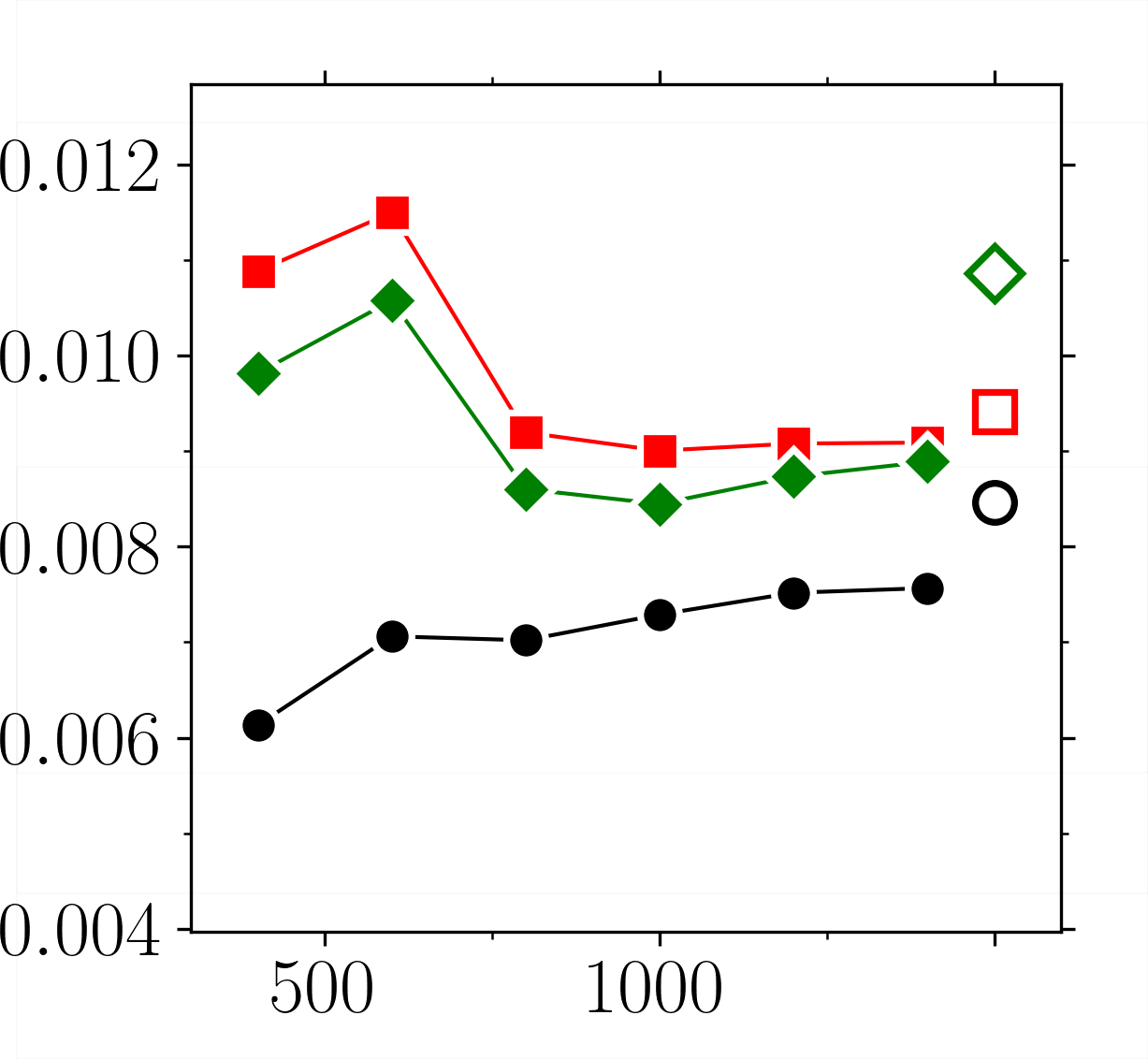}
        \LabelFig{19}{76}{$b)$}
        \Labelxy{50}{-4}{0}{$T_a$(K)}
        \Labelxy{83}{4}{0}{\scriptsize RSA}
        \Labelxy{-12}{40}{90}{$\text{var}^{\frac{1}{2}}(V_b)/\langle V_b \rangle$}
  \end{overpic}
  \caption{Dependence of \textbf{a)} average Voronoi volume $\langle V_b \rangle$ and \textbf{b)} rms volume fluctuations scaled by the average value $\text{var}^{\frac{1}{2}}(V_b)/\langle V_b \rangle$ on the annealing temperature $T_a$. The empty symbols in correspond to a \comp RSA.} 
  \label{fig:voronoiRMS}
\end{figure}

\begin{figure}[b]
  \centering
  \begin{overpic}[width=0.5\textwidth]{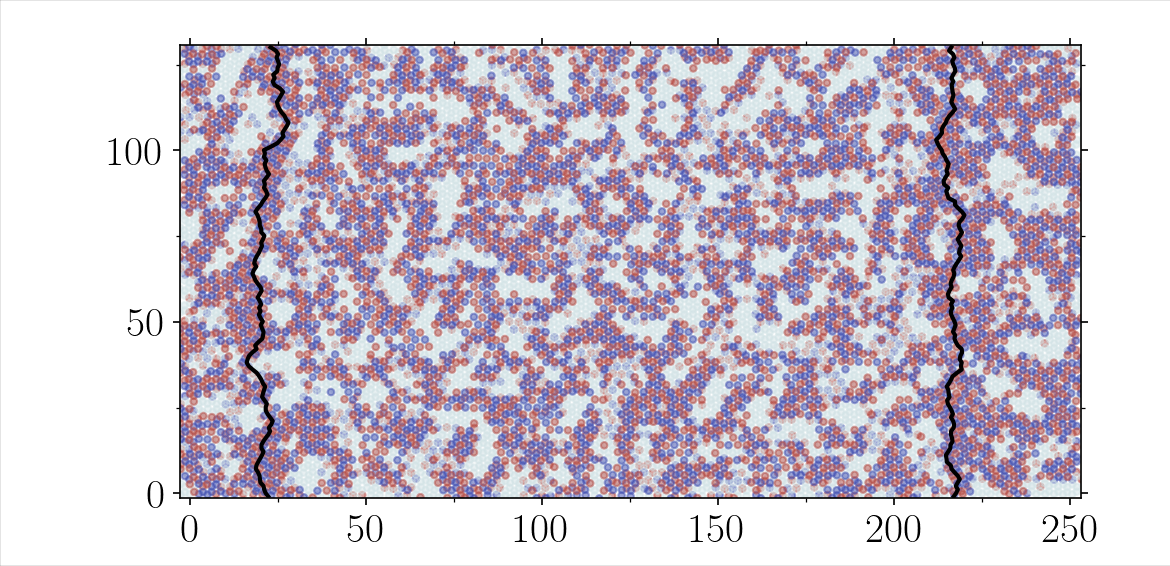}
        %
        \LabelFig{16}{8}{$a)~y=90$ \scriptsize \r{A}}
        \Labelxy{50}{-1}{0}{$x$(\r{A})}
        \Labelxy{4}{20}{90}{$z$(\r{A})}
    \end{overpic}

   \begin{overpic}[width=.5\textwidth]{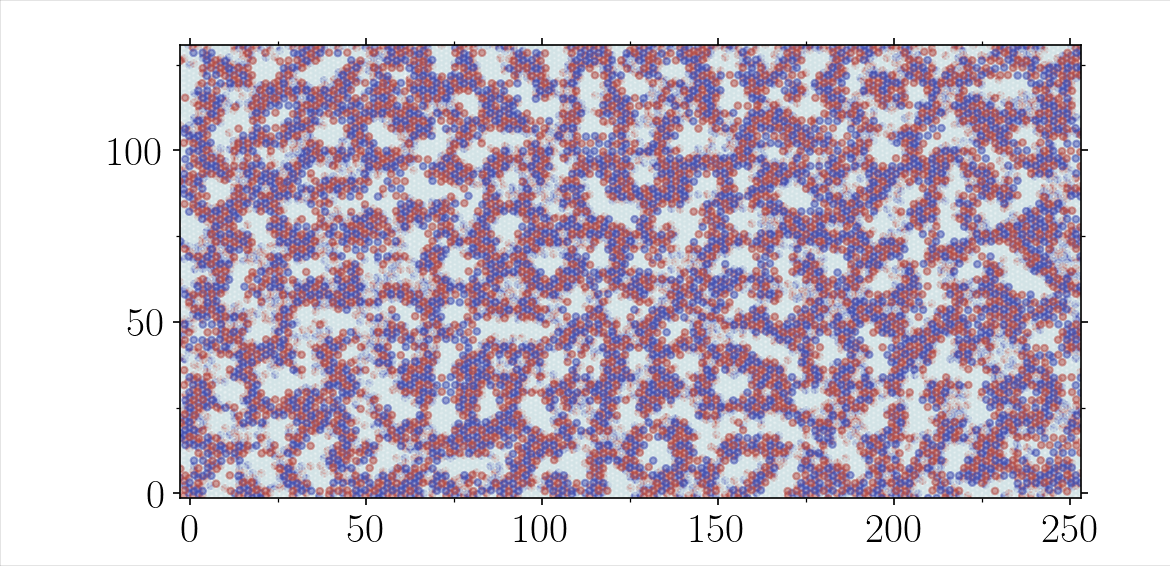}
        \LabelFig{16}{8}{$b)~y=10$ \scriptsize \r{A}}
        \Labelxy{50}{-1}{0}{$x$(\r{A})}
        \Labelxy{4}{20}{90}{$z$(\r{A})}
   \end{overpic}
     \begin{overpic}[width=0.27\textwidth]{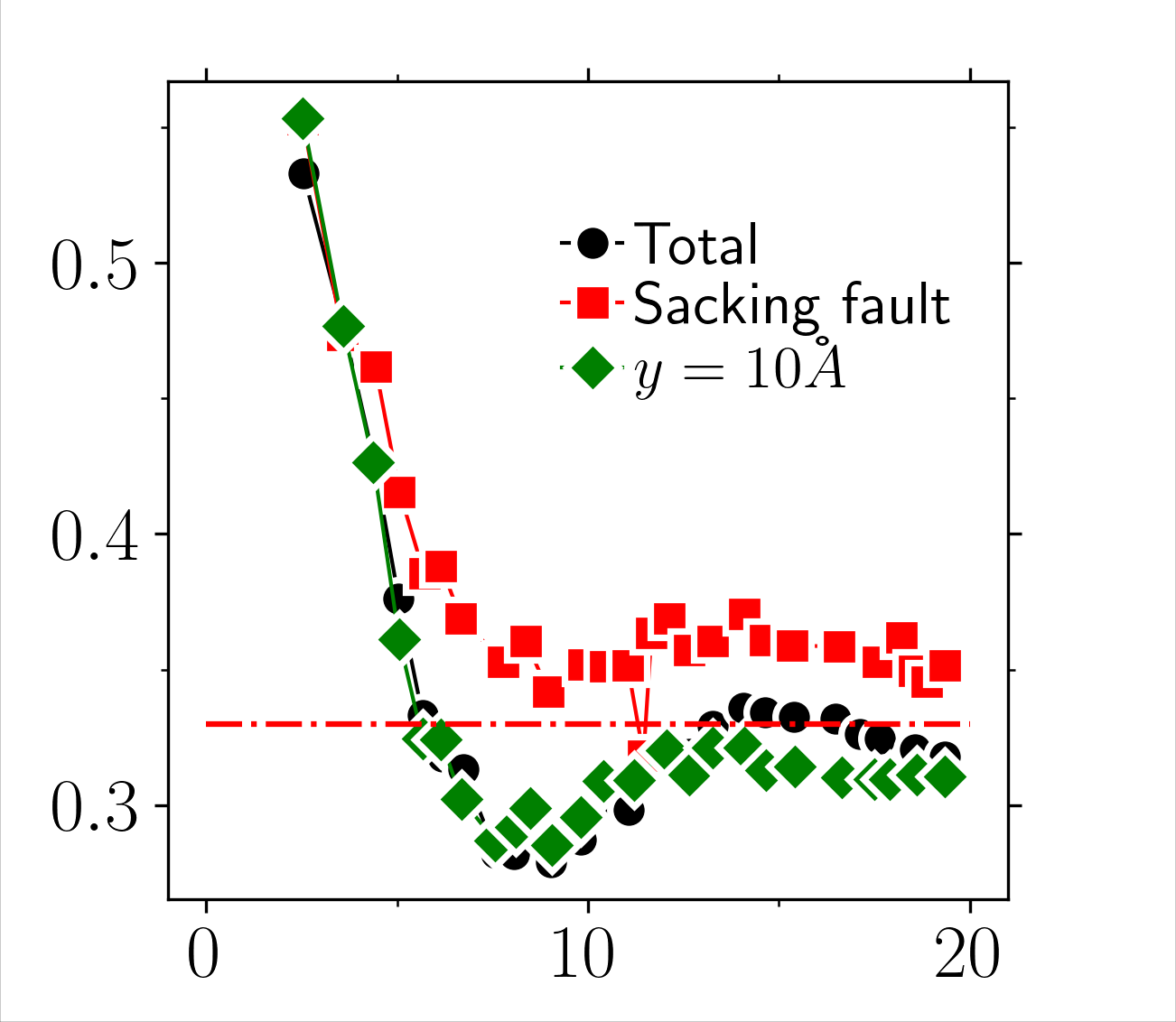}
        \Labelxy{46}{-5}{0}{$r$(\r{A})}
        \Labelxy{-5}{36}{90}{$p_\text{NiNi}$}
        \LabelFig{16}{14}{$c)$}
        \LabelFigg{52}{60}{\tiny Stacking fault}
    \end{overpic}

  \caption{SRO microstructure in the presence of (partial) dislocations in aging NiCoCr at $T_a=600$ K. \textbf{a}) cross section containing the stacking fault \textbf{b}) cross section at $y=10$~\r{A} \textbf{c}) the SRO parameter $p_\text{NiNi}$ as a function of pairwise distance $r$. The different curves in \textbf{c}) correspond to the entire configuration as well as the two dimensional stacks depicted in \textbf{a}) and \textbf{b}). The black segments denote dislocation lines. The dashdotted (red) line indicates RSAs. {The stacking fault region in \textbf{a}) lies between $x\simeq 25$ and $220$ \r{A}.} \note[PS]{Expend the caption by direct explanation of the meaning: near the stacking fault Ni is over-represented. What does it mean?} \note[KK]{could you elaborate what you are asking?}}
  \label{fig:stackingFault}
\end{figure}

\begin{figure*}
    \centering
    \begin{overpic}[width=\textwidth]{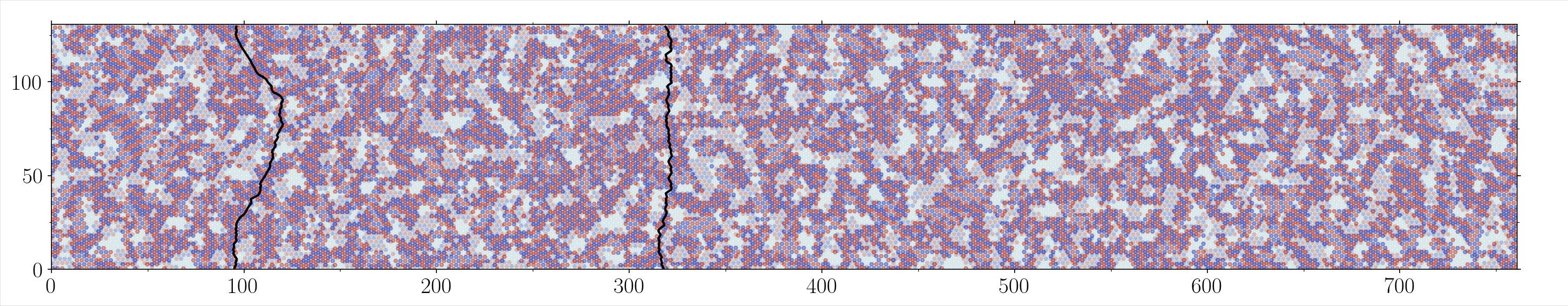}
        \LabelFig{4}{16}{$a)$ \scriptsize annealed}
        \Labelxy{-2}{6}{90}{$z$(\r{A})}
    \end{overpic}
    \begin{overpic}[width=\textwidth]{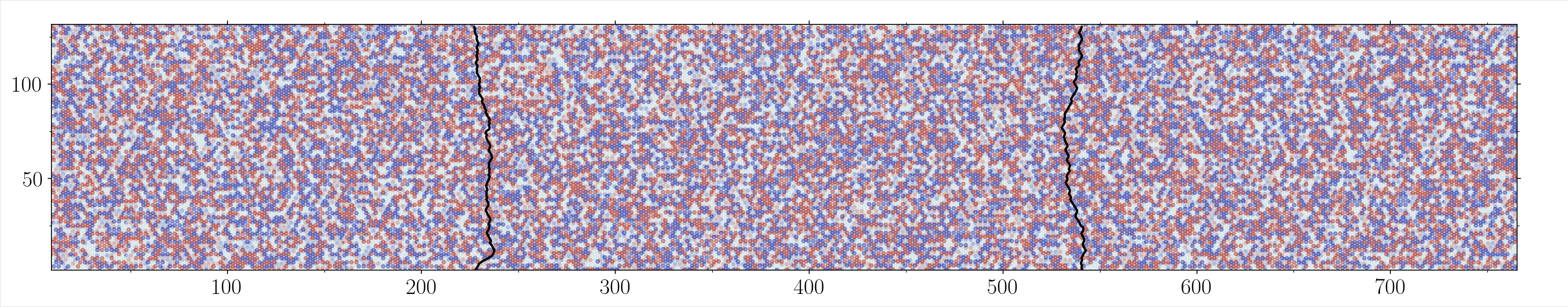}
        \LabelFig{4}{16}{$b)$ \scriptsize RSA}
        \Labelxy{50}{-2}{0}{$x$(\r{A})}
        \Labelxy{-2}{6}{90}{$z$(\r{A})}
    \end{overpic}
   \caption{Realizations of partial dislocations in \textbf{a)} NiCoCr annealed at $T_a=600$ K \textbf{b)} RSA.} 
   \label{fig:dislStackingFault}
\end{figure*}

\subsection{Lattice distortions}\label{sec:LatticeDistortions}
To characterize local distortion properties, we analyzed fluctuations in atomic Voronoi cell volumes and associated temperature-dependence in annealed alloys.
The aged solid solutions were equilibrated at $5$ K upon annealing in order to suppress thermal fluctuations.
Figure~\ref{fig:voronoi}(a) and (b) shows alloy atomic volumes and associated PDFs for the Ni, Co, and Cr atoms in annealed and random solid solutions. 
Both alloys feature quite narrow distributions with well-defined mean values $\langle V_\text{Ni} \rangle$, $\langle V_\text{Co} \rangle$, and $\langle V_\text{Cr} \rangle$ that show slight variations with $T_a$ as in Fig.~\ref{fig:voronoiRMS}(a).
The measured mean atomic volume in annealed \comp samples is $\langle V\rangle \simeq 11.3$ \r{A}$^3$ ---which is equivalent to the average lattice constant of $a=3.56$ \r{A}--- in very close agreement with experimental observations reported in \cite{Yin2020}.
Figure~\ref{fig:voronoiRMS}(a) indicates features near a characteristic annealing temperature $T_a\simeq 800$ K below which the mean atomic volumes seem to accelerate, potentially a signature of remarkable enthalpy-driven ordering \cite{Li2019}. 
This observation is in accordance with the heat capacity $C_p$ developing a characteristic peak in Fig.~\ref{fig:thermo}(a).

\begin{figure*}
    \centering
    \begin{overpic}[width=0.24\textwidth]{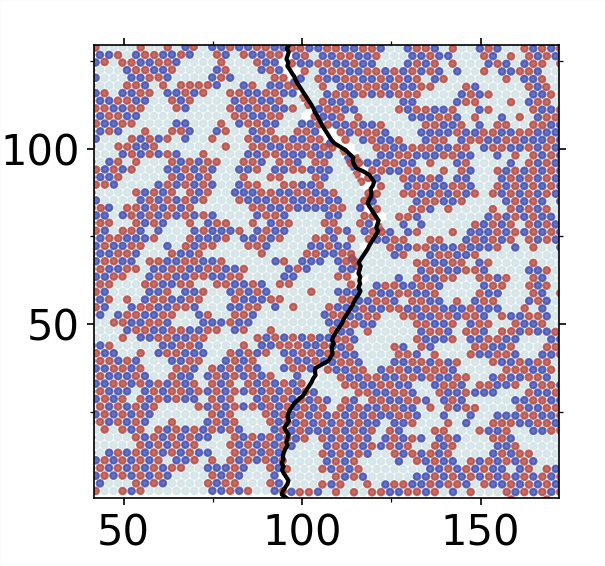}
        \LabelFig{19}{76}{$a)\scriptscriptstyle\sigma=500$\tiny(Mpa)}
        \Labelxy{50}{-6}{0}{$x$(\r{A})}
        \Labelxy{-6}{40}{90}{$z$(\r{A})}
    \end{overpic}
    \begin{overpic}[width=0.24\textwidth]{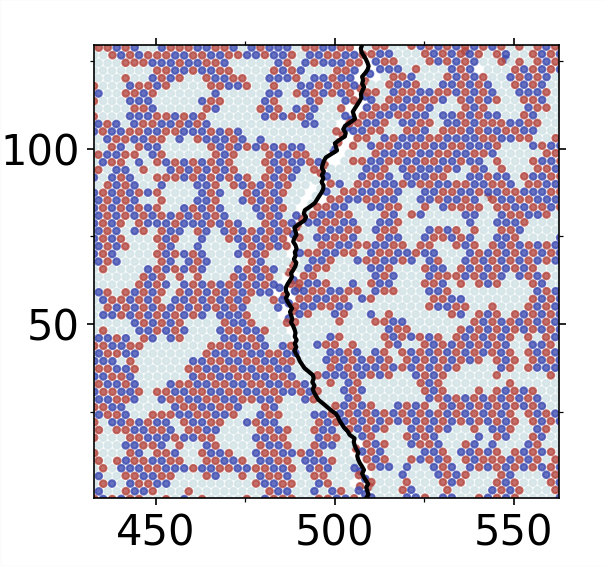}
        \LabelFig{19}{76}{$b)\scriptscriptstyle\sigma=600$\tiny(Mpa)}
        \Labelxy{50}{-6}{0}{$x$(\r{A})}
        \Labelxy{-6}{40}{90}{$z$(\r{A})}
    \end{overpic}
    \begin{overpic}[width=0.24\textwidth]{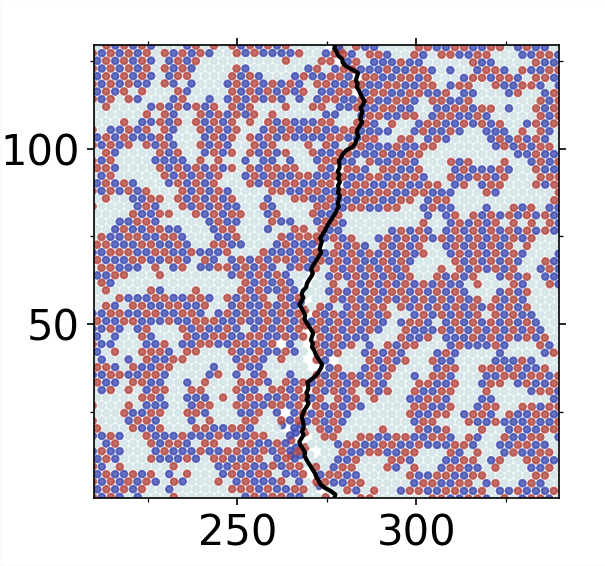}
        \LabelFig{19}{76}{$c)\scriptscriptstyle\sigma=650$\tiny(Mpa)}
        \Labelxy{50}{-6}{0}{$x$(\r{A})}
        \Labelxy{-6}{40}{90}{$z$(\r{A})}
    \end{overpic}
    \begin{overpic}[width=0.24\textwidth]{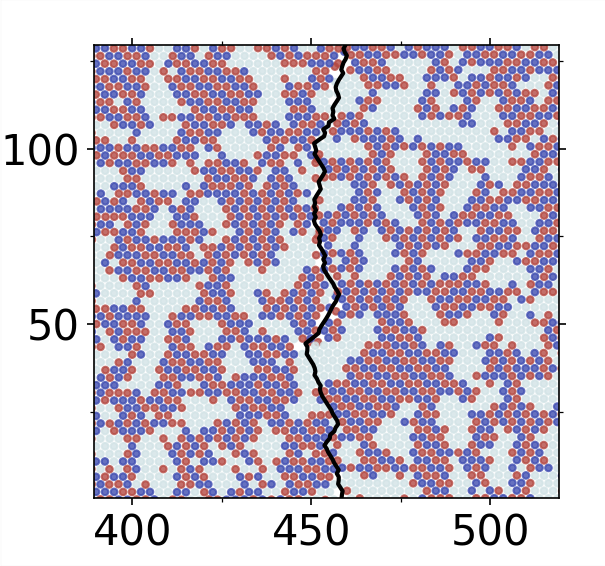}
        \LabelFig{19}{76}{$d)\scriptscriptstyle\sigma=750$\tiny(Mpa)}
        \Labelxy{50}{-6}{0}{$x$(\r{A})}
        \Labelxy{-6}{40}{90}{$z$(\r{A})}
    \end{overpic}
   \caption{Realizations of (immobile) partial edge dislocations in an annealed NiCoCr subject to the shear stress \textbf{a)} $\sigma=500$, \textbf{b)} $600$, \textbf{c)} $650$, and \textbf{d)} $750$ MPa. Here the two-dimensional stack denotes the $(111)$ glide plane.} 
   \label{fig:dislSnapshots}
\end{figure*}

\begin{figure*}
    \centering
    \begin{overpic}[width=0.24\textwidth]{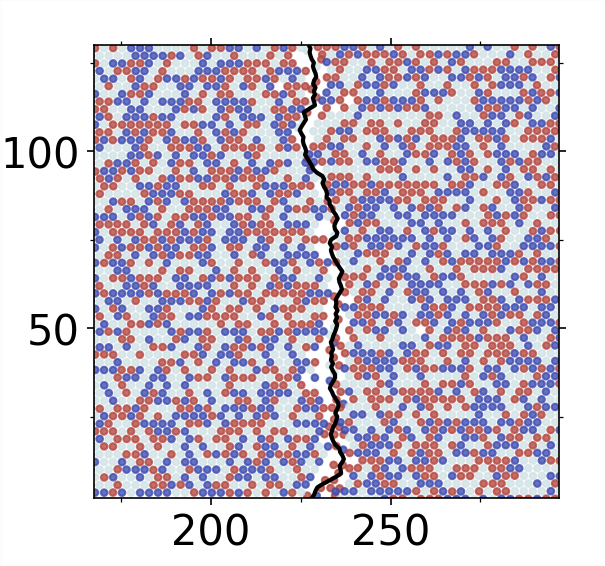}
        \LabelFig{19}{76}{$a)\scriptscriptstyle\sigma=400$\tiny(Mpa)}
        \Labelxy{50}{-6}{0}{$x$(\r{A})}
        \Labelxy{-6}{40}{90}{$z$(\r{A})}
    \end{overpic}
    \begin{overpic}[width=0.24\textwidth]{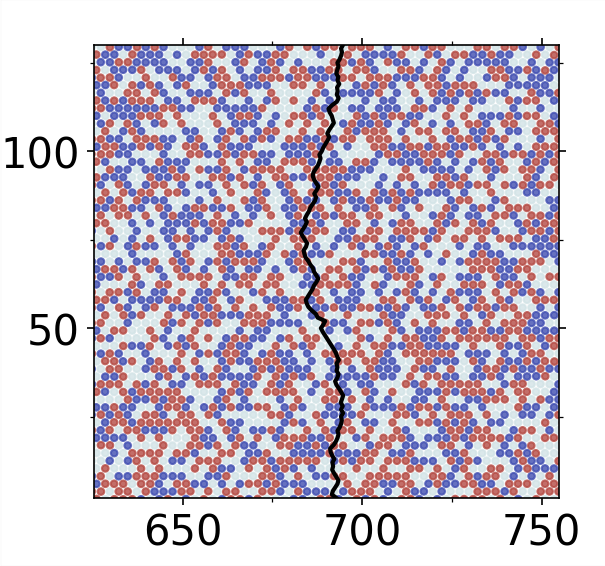}
        \LabelFig{19}{76}{$b)\scriptscriptstyle\sigma=500$\tiny(Mpa)}
        \Labelxy{50}{-6}{0}{$x$(\r{A})}
        \Labelxy{-6}{40}{90}{$z$(\r{A})}
    \end{overpic}
    \begin{overpic}[width=0.24\textwidth]{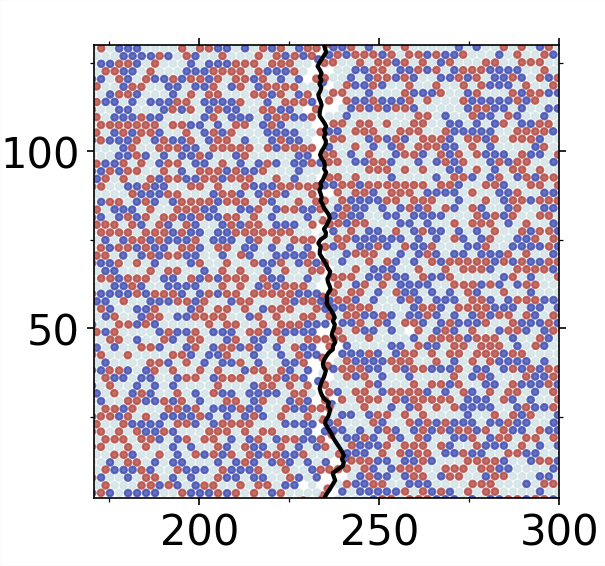}
        \LabelFig{19}{76}{$c)\scriptscriptstyle\sigma=550$\tiny(Mpa)}
        \Labelxy{50}{-6}{0}{$x$(\r{A})}
        \Labelxy{-6}{40}{90}{$z$(\r{A})}
    \end{overpic}
    \begin{overpic}[width=0.24\textwidth]{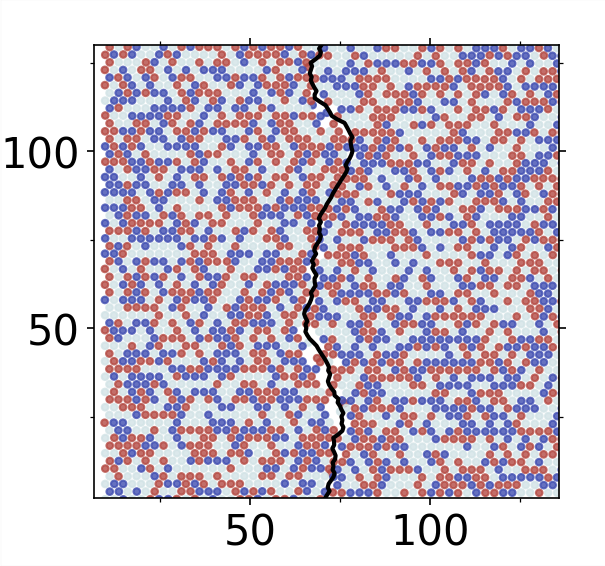}
        \LabelFig{19}{76}{$d)\scriptscriptstyle\sigma=600$\tiny(Mpa)}
        \Labelxy{50}{-6}{0}{$x$(\r{A})}
        \Labelxy{-6}{40}{90}{$z$(\r{A})}
    \end{overpic}
   \caption{Realizations of (immobile) partial dislocations in a NiCoCr RSA subject to the shear stress \textbf{a)} $\sigma=400$, \textbf{b)} $500$, \textbf{c)} $550$, and \textbf{d)} $600$ MPa.} 
   \label{fig:dislSnapshotsRsa}
\end{figure*}

The misfit volumes $\Delta V_b=\langle V_b\rangle-\langle V\rangle$ of Ni, Co, and Cr are determined as $+0.04$, $-0.03$, and $-0.01$ \r{A}$^3$, respectively.
The estimated atomic misfits appear to be at least one order of magnitude off from precise experiments \cite{Yin2020} but are, otherwise, reasonably close to ab initio-based estimates in \cite{oh2019engineering}.
We further explored the scaled fluctuation $\text{var}^{\frac{1}{2}}(V_b)/\langle V_b \rangle$ as a relevant measure of atomic distortions in Fig.~\ref{fig:voronoiRMS}(b) 
with $b$ denoting Ni, Co, and Cr.
$\text{var}^{\frac{1}{2}}(V_b)/\langle V_b \rangle$ shows a (fairly) monotonic increase for Ni as a function of $T_a$ till it saturates at a limiting value that appears to be lower than the one set by the RSA. 
This is in line with our SRO analysis that the abundance of Ni-Ni pairs within the first nearest neighbor distance and, therefore, ordering tends to curtail local atomic misfits or randomness in aged systems.
As for Co and Cr, we observe a non-monotonic evolution with a pronounced peak at $T_a\simeq 600$ K that, in the case of the former, even exceeds the associated RSA limit.
\add[KK]{Apart from the observed peaks, the relative variance for aged Co/Cr seem to always fall short of those of RSAs with a more dramatic decrease associated with Cr (the green diamonds).  
This, we conjecture, might be attributed to the favored formation of Cr-Co regions and less tendency for Cr-Ni as well as Cr-Cr bonding as evidenced by the behavior of the first-nearest-neighbor order parameters in Fig.~\ref{fig:sroSheng}.
We conclude this subsection by stating that short range order will have strong bearings on misfit volumes of NiCoCr as our data suggest direct correlations between the latter and the order parameters presented in Sec.~\ref{sec:sro_temp}.}
{One may infer a characteristic scale based on the rms fluctuations analysis presented in Fig.~\ref{fig:voronoiRMS}(b) which we interpret as the (mean) misfit size $\xi^\text{misfit}\simeq 1$ \r{A}. 
Together with nanoscopic SROs ($\xi^\text{sro}\simeq 10$ \r{A}), atomic-level misfit fluctuations will determine the dislocation glide resistance as discussed in Sec.~\ref{sec:sro_disl}.}
\note{PS: Ni is explained, but what about the dramatic decrease of the relative variance for Co/Cr? This is in even greater need for explanation. My guess is that by forming an ordered CoCr regions, the variance is diminished. The variance, most likely comes from the contact surface regions between Ni and CoCr precipitates. Also, one could look into the properties of equiatomic CoCr alloy (no Nickel) (e.g. \url{https://www.sciencedirect.com/science/article/pii/S0264127521002628}), to see how the local ordering of CoCr pairs in Nickel-free regions is similar/different from pure CoCr crystals.} \note[KK]{relevant discussions added.}


\subsection{Interplay between SROs and dislocations}\label{sec:sro_disl}
We follow two different approaches to address the dislocation-SRO interplay in NiCoCr CSAs: \romn{1}) study of dislocation effects on the nucleation of SROs in \emph{aging} alloys \romn{2}) investigations of strengthening mechanisms at play in \emph{as-aged} SRO-rich alloys driven out of equilibrium.
In \romn{1}), we aged samples with a dislocation allowing for both the {dislocation dissociation process} and spatio-temporal evolution of SROs while annealing. 
In \romn{2}), on the other hand, we embedded partial dislocations in as-annealed alloys and performed shear depining tests, with no appreciable change in the SRO microstructure.

In line with \romn{1}), Fig.~\ref{fig:stackingFault} compares the structure of SROs within the dislocation dissociation bounds and outside in aging NiCoCr at $T_a=600$ K. 
In Fig.~\ref{fig:stackingFault}(a), the denser population of SROs within the stacking fault is visually apparent in comparison with a dislocation-free two-dimensional stack at $y=10$~\r{A} illustrated in Fig.~\ref{fig:stackingFault}(b). 
We quantified the observed trend in Fig.~\ref{fig:stackingFault}(c) where the SRO parameter $p_\text{NiNi}$ associated with the former  reveals a shallower decay relative to that of atoms outside the fault plane. 
To our knowledge, the drastic increase in chemical ordering within the stacking fault region has not been previously reported in the literature.
As one possible mechanism at play, we speculate that the long-range stress field and mutual interactions between the two partials \cite{hull2001introduction} might favor the SRO nucleation and its growth within the dissociation zone. 
Figure~\ref{fig:dislStackingFault}(a) and (b) illustrates that such interactions at $T_a=600$ K lead to a notable reduction in the stacking fault width which implies the enhanced fault energy due to SROs \cite{Li2019,ding2018tunable}.
{It is expected that the fault dimension, and therefore the associated formation energy, will be strongly controlled by the annealing temperature as well.
We note that a detailed description of the SRO kinetics and dynamics of the dislocation dissociation as well as their (dynamical) interplay during the aging processs is outside the scope of our current study.}

Following approach \romn{2}), the notion of ``plastic flow" in solute strengthening theories directly links to the existence of the intrinsic friction stress $\sigma_c$ beyond which dislocations tend to glide rather smoothly at a non-negligible (mean) velocity $\langle v \rangle$.
Below this critical stress, by contrast, the migration of dislocations within CSAs (with a severely distorted energy landscape) typically occurs in a very intermittent manner with long periods of quiescent states (i.e. $\langle v \rangle\simeq 0$) interrupted by bursts of displacements \cite{osetsky2019two}.
In the absence of thermal activation, this depinning transition is phenomenologically described as \cite{zaiser2006scale} 
\begin{equation}\label{eq:depinning}
\langle v \rangle \propto (\sigma-\sigma_c)^{1/\beta},
\end{equation}
at $\sigma \ge \sigma_c$ and $\langle v \rangle=0$ otherwise, with $\beta \ge 1$ reflecting a marked discontinuity at $\sigma_c$ \cite{esfandiarpour2022edge}.
Here $\sigma$ is the applied shear stress resolved in the glide plane.

\begin{figure}
   \centering
    \begin{overpic}[width=0.27\textwidth]{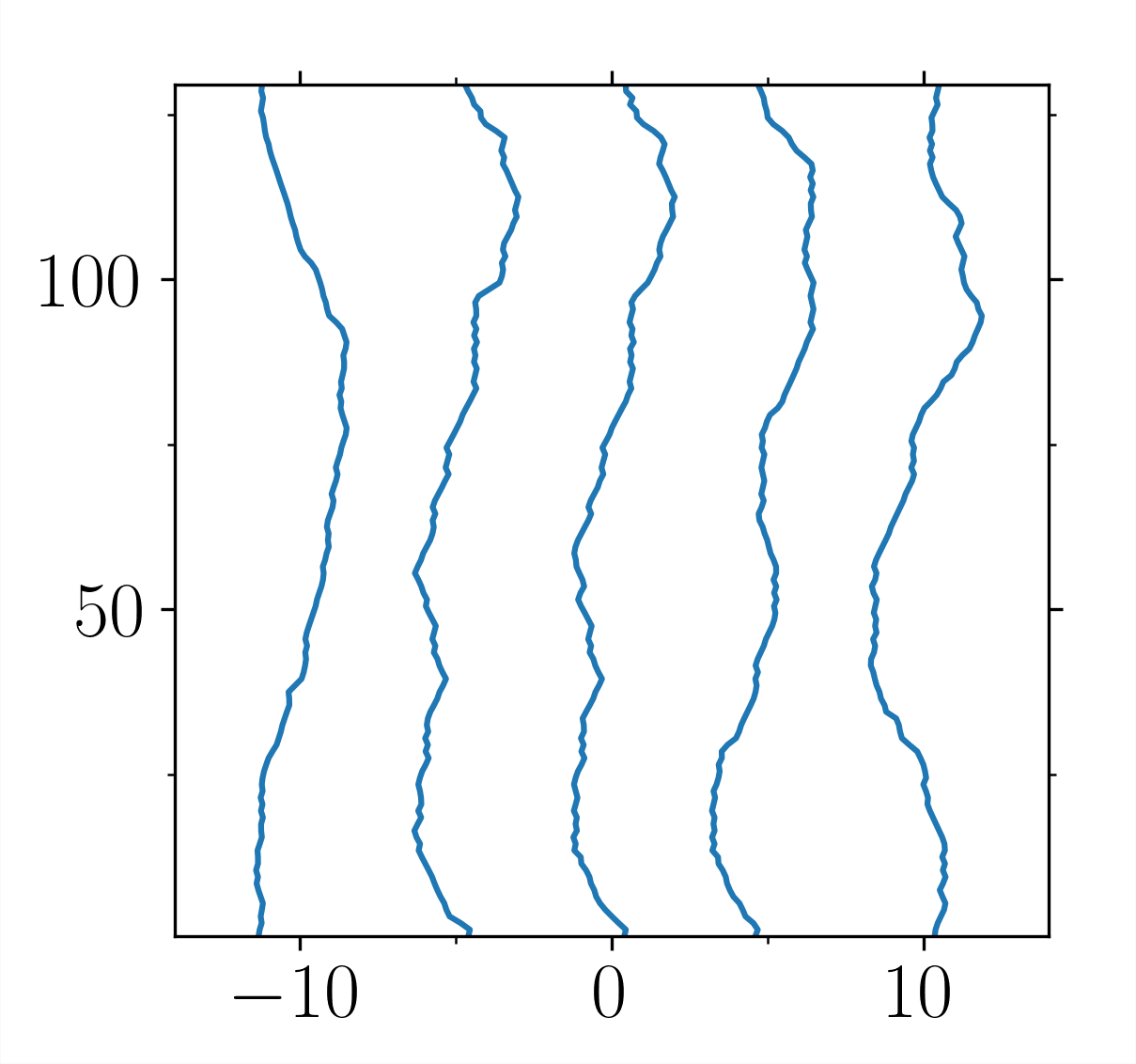}
        \Labelxy{44}{-3}{0}{$\hat{h}_x(z)$}
        \Labelxy{-6}{35}{90}{$z$(\r{A})}
        \LabelFig{17}{90}{$a)$ \scriptsize annealed}
        \Labelxy{17}{16}{90}{$\scriptstyle 500$\tiny (MPa)}
        \Labelxy{36}{16}{90}{$\scriptstyle 600$}
        \Labelxy{50}{16}{90}{$\scriptstyle 650$}
        \Labelxy{62}{16}{90}{$\scriptstyle 700$}
        \Labelxy{83}{16}{90}{$\scriptstyle 750$}
    \end{overpic}

    \begin{overpic}[width=0.27\textwidth]{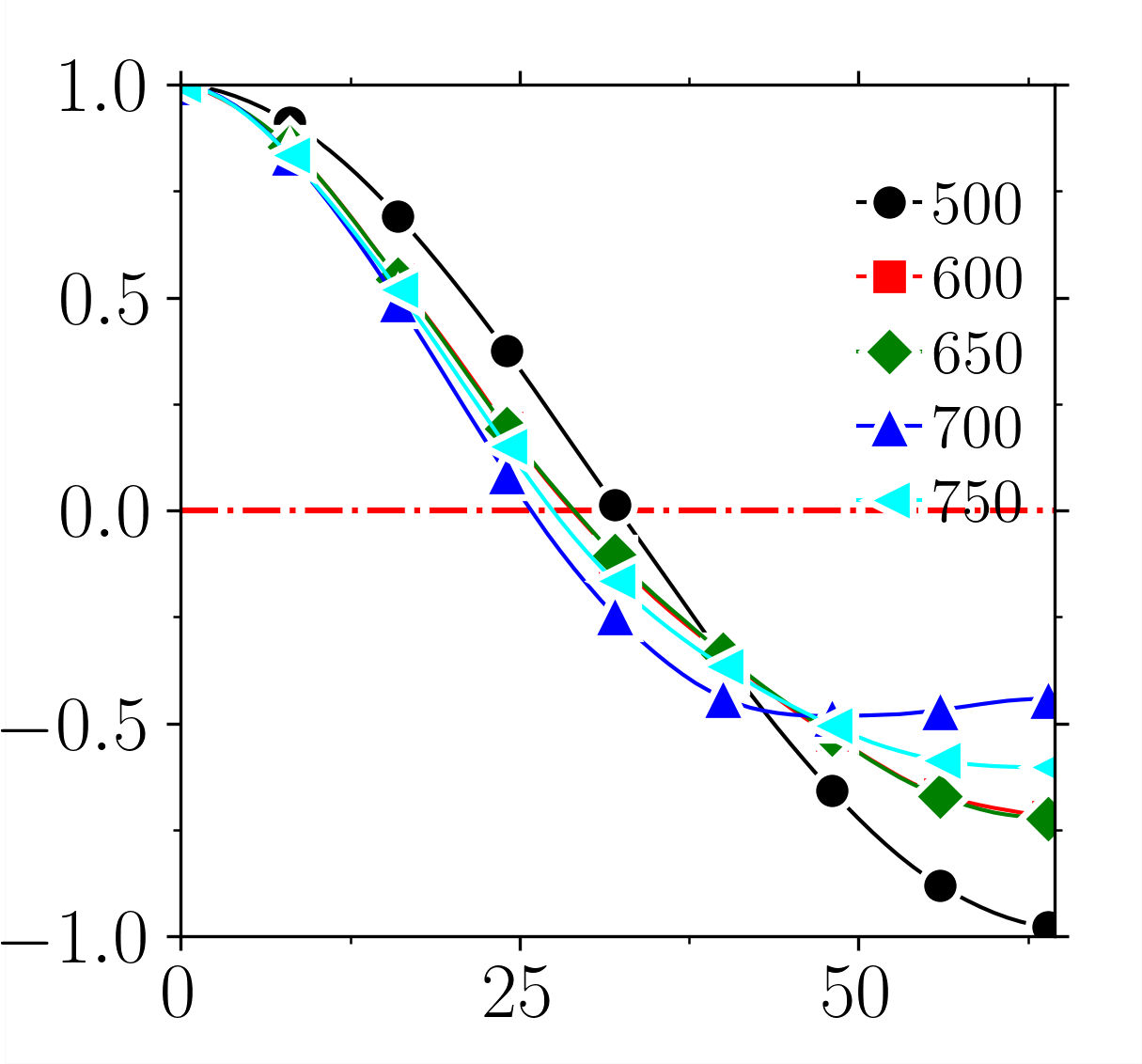}
        \Labelxy{44}{-3}{0}{$|z-z^\prime|$}
        \Labelxy{70}{80}{0}{$\sigma$\tiny (MPA)}
        \Labelxy{-6}{35}{90}{$c_h(|z-z^\prime|)$}
        \LabelFig{17}{15}{$b)$}
    \end{overpic}
    \caption{\textbf{a}) Configurations of (immobile) partial edge dislocations \textbf{b}) associated correlations $c_h(|z-z^\prime|)=\langle ~\hat{h}_x(z).\hat{h}_x(z^\prime)~\rangle$ as a function of distance $|z-z^\prime|$ in an annealed NiCoCr under different applied stresses (below the depinning transition) at $T_a=600$ K. This includes $\sigma=500$, $600$, $650$, $700$, and $750$ MPa. The shear tests were carried out at $5$ K. Here, the two-dimensional $x-z$ glide plane denotes a $(111)$ cross section. The dislocation configurations are shifted vertically for the clarity.}
    \label{fig:sroBelowStrsAnnealed}
\end{figure}

\begin{figure}
   \centering
    \begin{overpic}[width=0.27\textwidth]{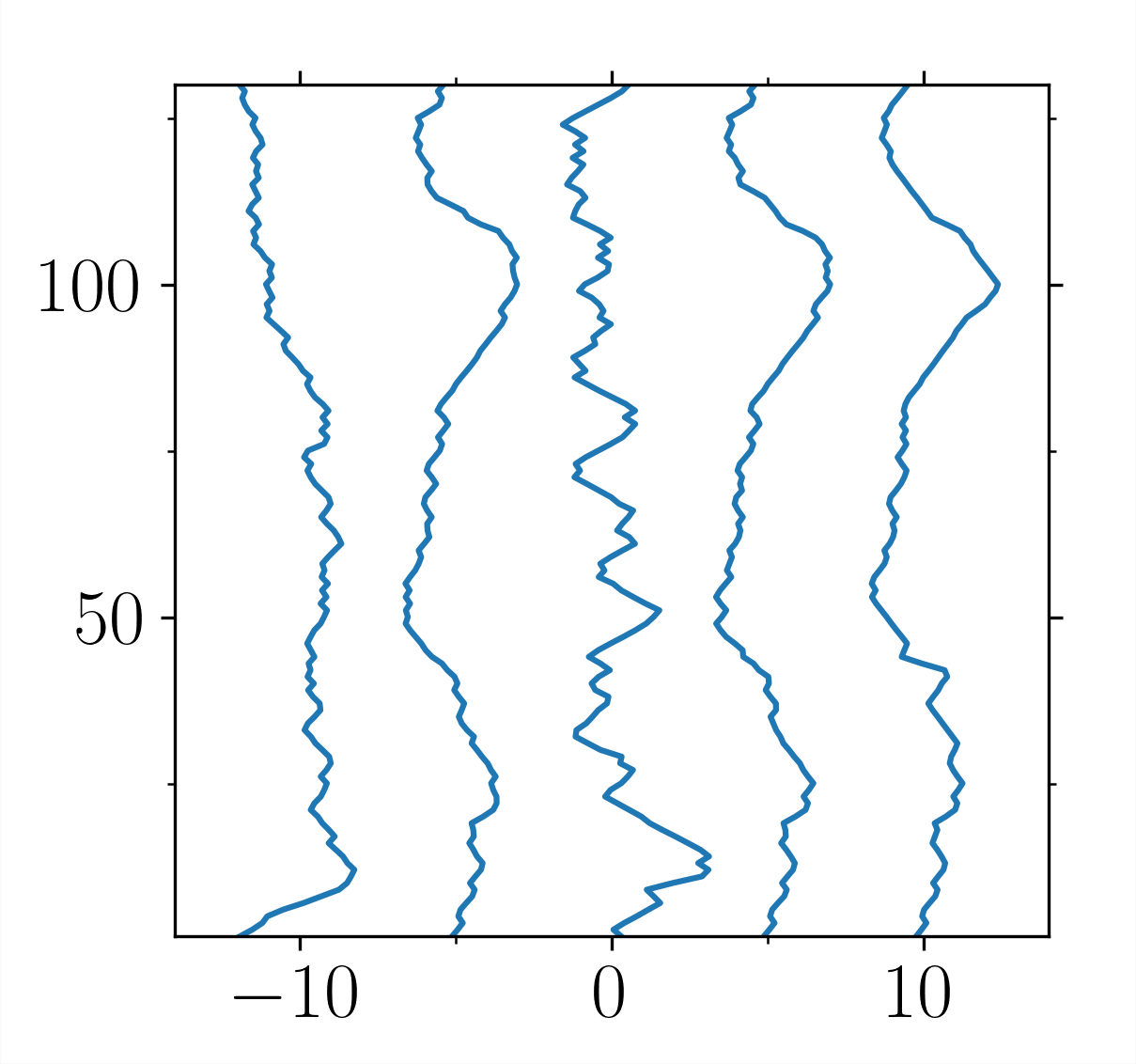}
        \Labelxy{44}{-3}{0}{$\hat{h}_x(z)$}
        \Labelxy{-6}{35}{90}{$z$(\r{A})}
        \LabelFig{17}{90}{$a)$ \scriptsize RSA}
        \Labelxy{19}{16}{90}{$\scriptstyle 400$\tiny (MPa)}
        \Labelxy{40}{16}{90}{$\scriptstyle 500$}
        \Labelxy{49}{16}{90}{$\scriptstyle 550$}
        \Labelxy{68}{16}{90}{$\scriptstyle 600$}
        \Labelxy{81}{16}{90}{$\scriptstyle 650$}
    \end{overpic}
    
    \begin{overpic}[width=0.27\textwidth]{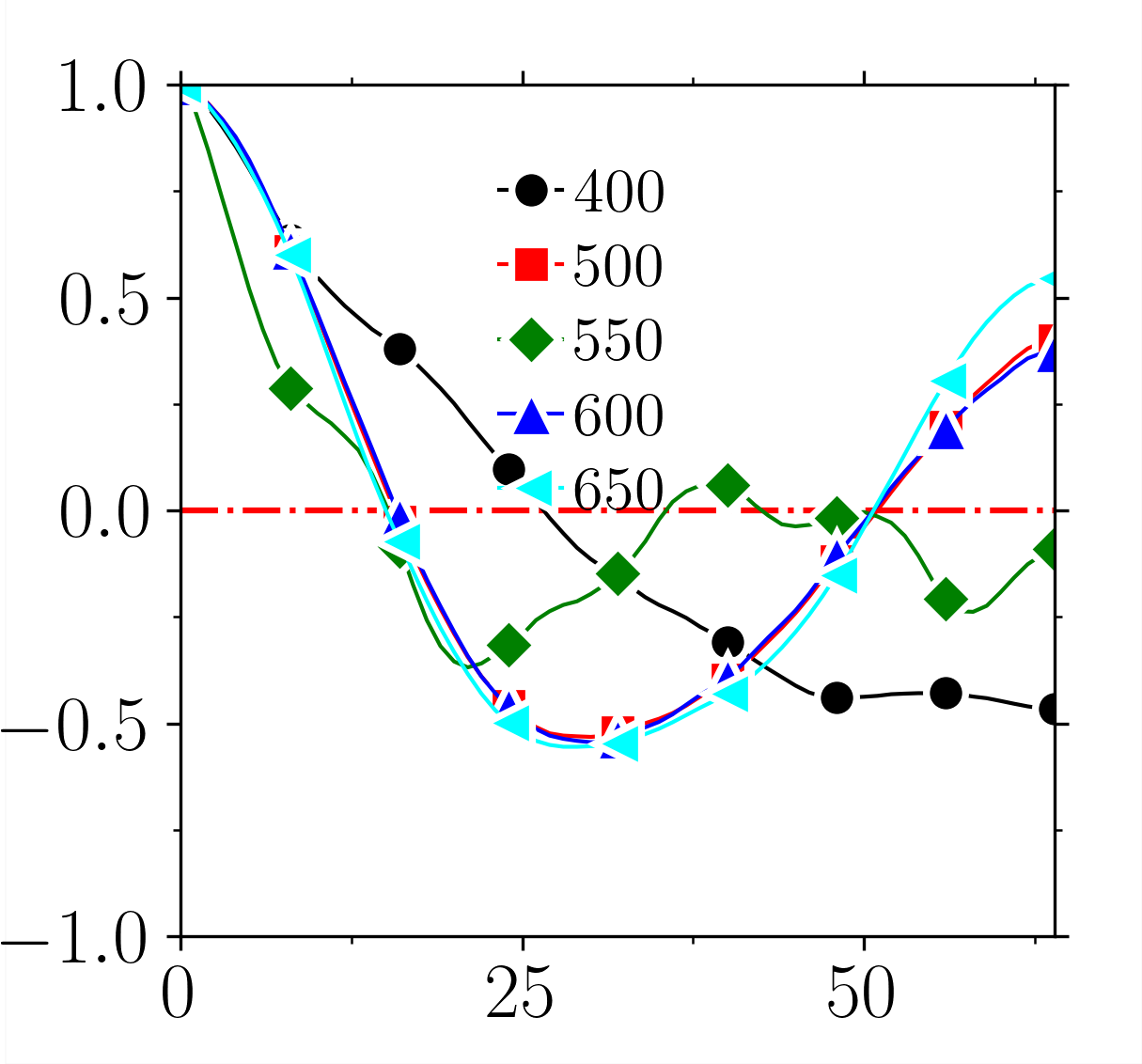}
        \Labelxy{44}{-3}{0}{$|z-z^\prime|$}
        \Labelxy{38}{80}{0}{$\sigma$\tiny (MPA)}
        \Labelxy{-6}{35}{90}{$c_h(|z-z^\prime|)$}
        \LabelFig{17}{15}{$b)$}
    \end{overpic}
    \caption{\textbf{a}) Configurations of (immobile) partial edge dislocations \textbf{b}) associated correlations $c_h(|z-z^\prime|)=\langle ~\hat{h}_x(z).\hat{h}_x(z^\prime)~\rangle$ as a function of distance $|z-z^\prime|$ in a NiCoCr RSA under different applied stresses below the depinning transition. This includes $\sigma=400$, $500$, $550$, $600$, and $650$ MPa}
    \label{fig:sroBelowStrsRsa}
\end{figure}

\begin{figure*}
    \centering
    \begin{overpic}[width=0.24\textwidth]{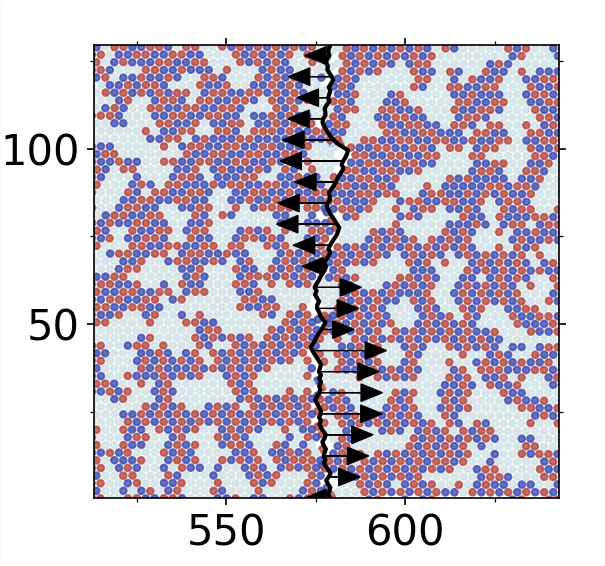}
        \LabelFig{19}{76}{$a)$}
        \Labelxy{50}{-6}{0}{$x$(\r{A})}
        \Labelxy{-6}{40}{90}{$z$(\r{A})}
    \end{overpic}
    \begin{overpic}[width=0.24\textwidth]{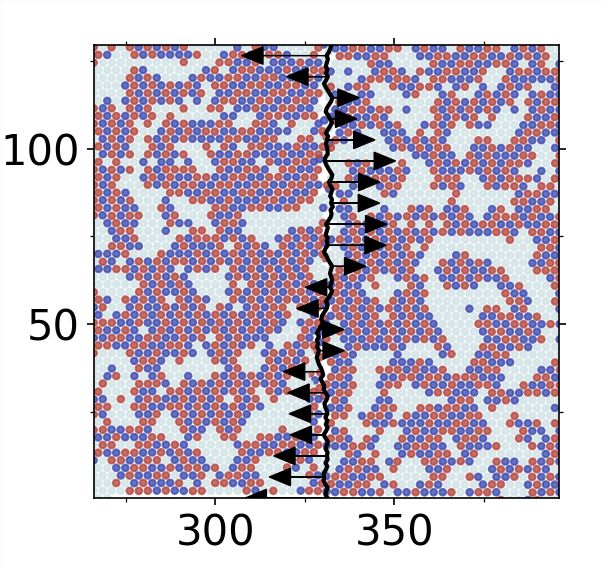}
        \LabelFig{19}{76}{$b)$}
        \Labelxy{50}{-6}{0}{$x$(\r{A})}
        \Labelxy{-6}{40}{90}{$z$(\r{A})}
    \end{overpic}
    \begin{overpic}[width=0.24\textwidth]{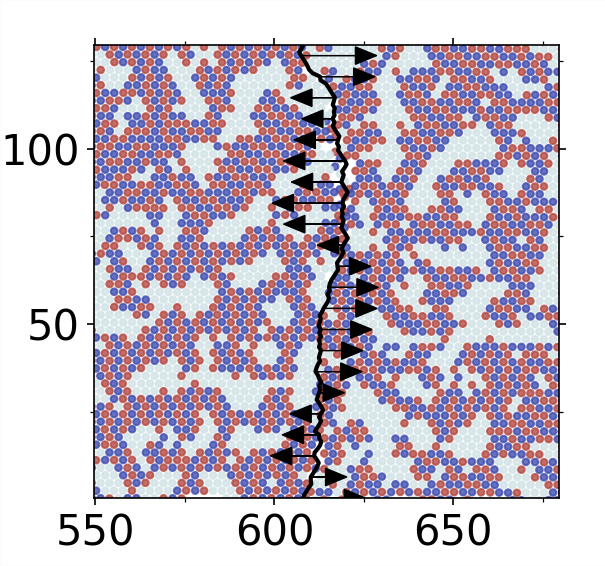}
        \LabelFig{19}{76}{$c)$}
        \Labelxy{50}{-6}{0}{$x$(\r{A})}
        \Labelxy{-6}{40}{90}{$z$(\r{A})}
    \end{overpic}
    \begin{overpic}[width=0.24\textwidth]{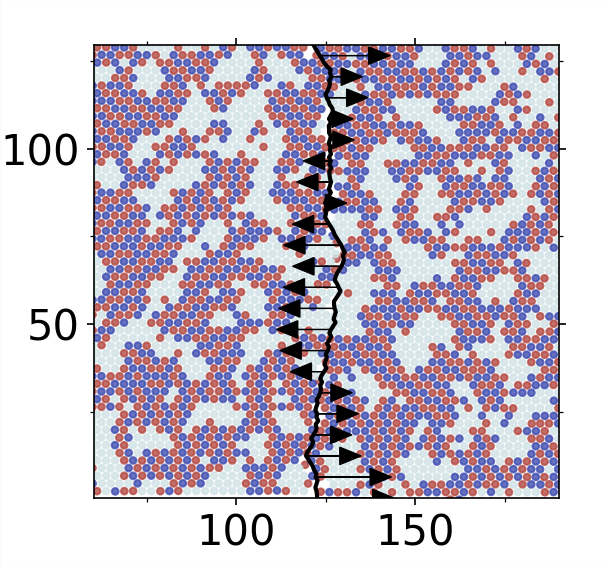}
        \LabelFig{19}{76}{$d)$}
        \Labelxy{50}{-6}{0}{$x$(\r{A})}
        \Labelxy{-6}{40}{90}{$z$(\r{A})}
    \end{overpic}
   \caption{ (z-scored) velocity of partial edge dislocations $\hat{v}_x(z)$, illustrated by the black arrows, in annealed NiCoCr at $T_a=600$ K and under the applied stress $\sigma =1200$ MPa (above $\sigma_c$). The shear tests were carried out at $5$ K. Here, the panels indicate different realizations associated with the gliding dislocations and the two-dimensional plane denotes a $(111)$ cross section. The arrows denote the velocity field.}
   \label{fig:dislVelSnapshots}
\end{figure*}

\begin{figure*}[t]
   \centering
    \begin{overpic}[width=0.27\textwidth]{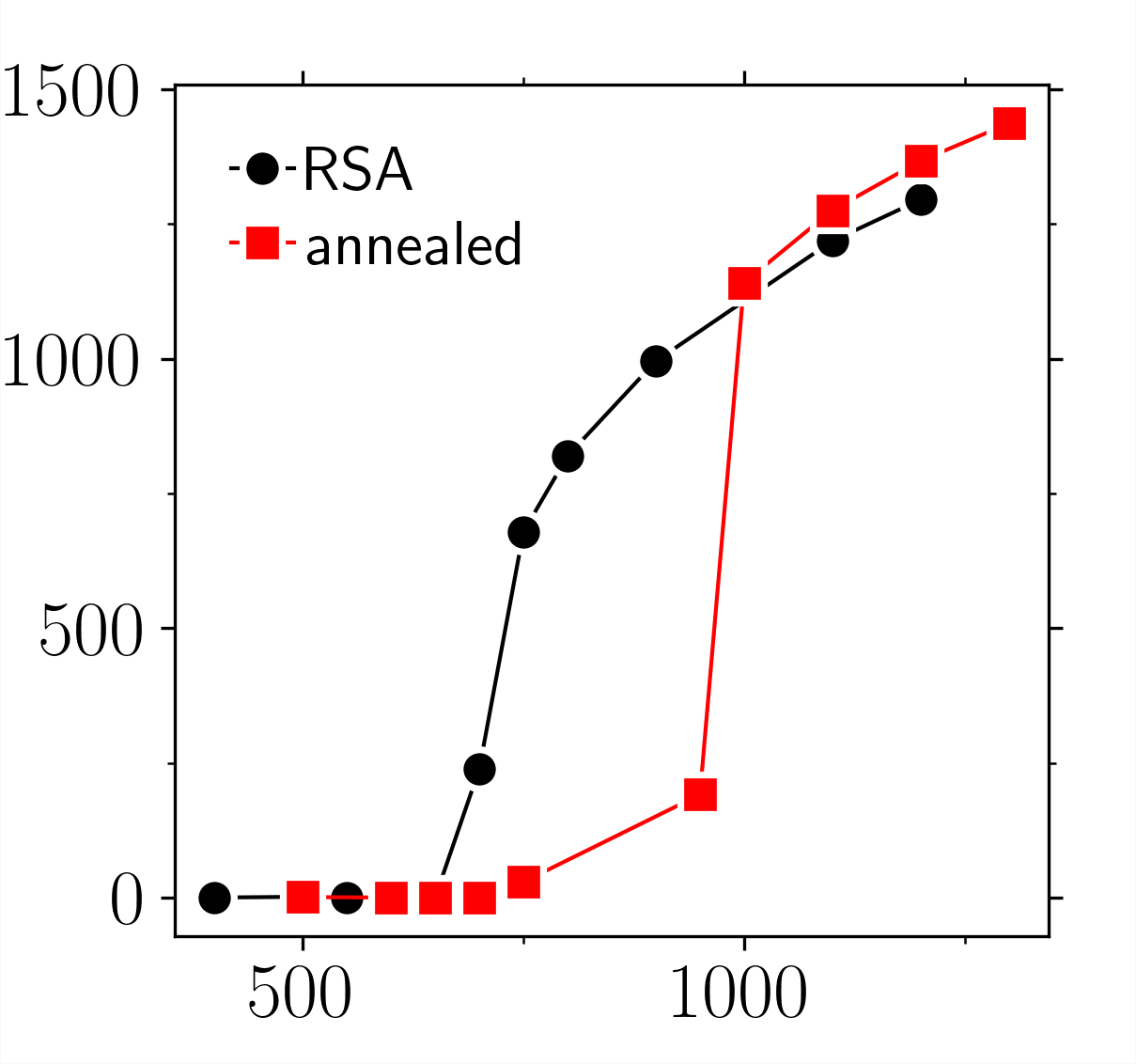}
        \Labelxy{44}{-3}{0}{$\sigma$~\scriptsize(MPa)}
        \Labelxy{-6}{35}{90}{$\langle v_x \rangle(\text{ms}^{-1})$}
        \LabelFig{17}{15}{$a)$}
    \end{overpic}
    \begin{overpic}[width=0.27\textwidth]{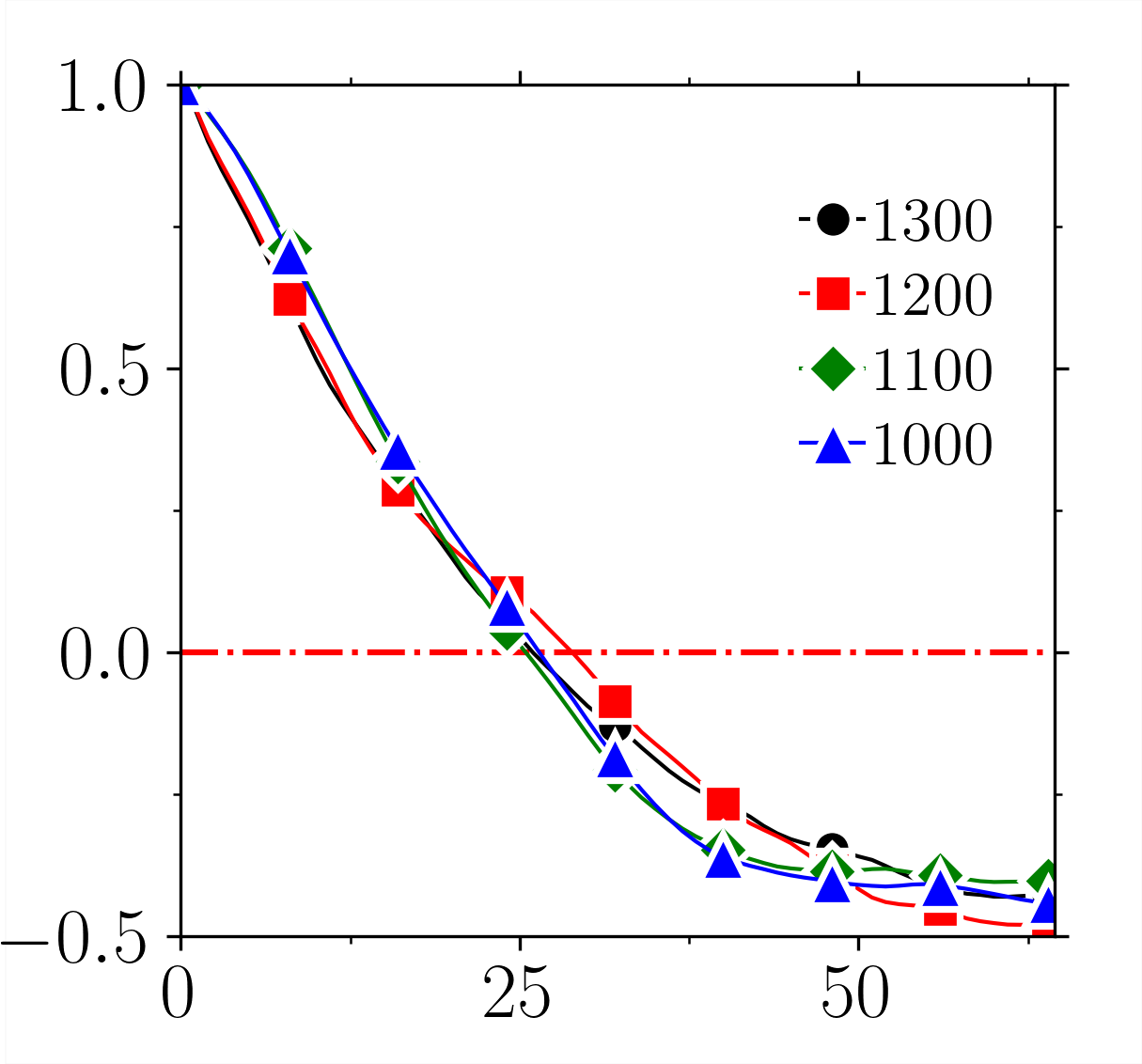}
        \Labelxy{44}{-3}{0}{$|z-z^\prime|$}
        \Labelxy{-6}{35}{90}{$\langle c_v(|z-z^\prime|) \rangle_\text{ens}$}
        \Labelxy{68}{80}{0}{$\sigma$\tiny (MPA)}
        \LabelFig{17}{15}{$b)$   \scriptsize annealed}
    \end{overpic}
    \begin{overpic}[width=0.27\textwidth]{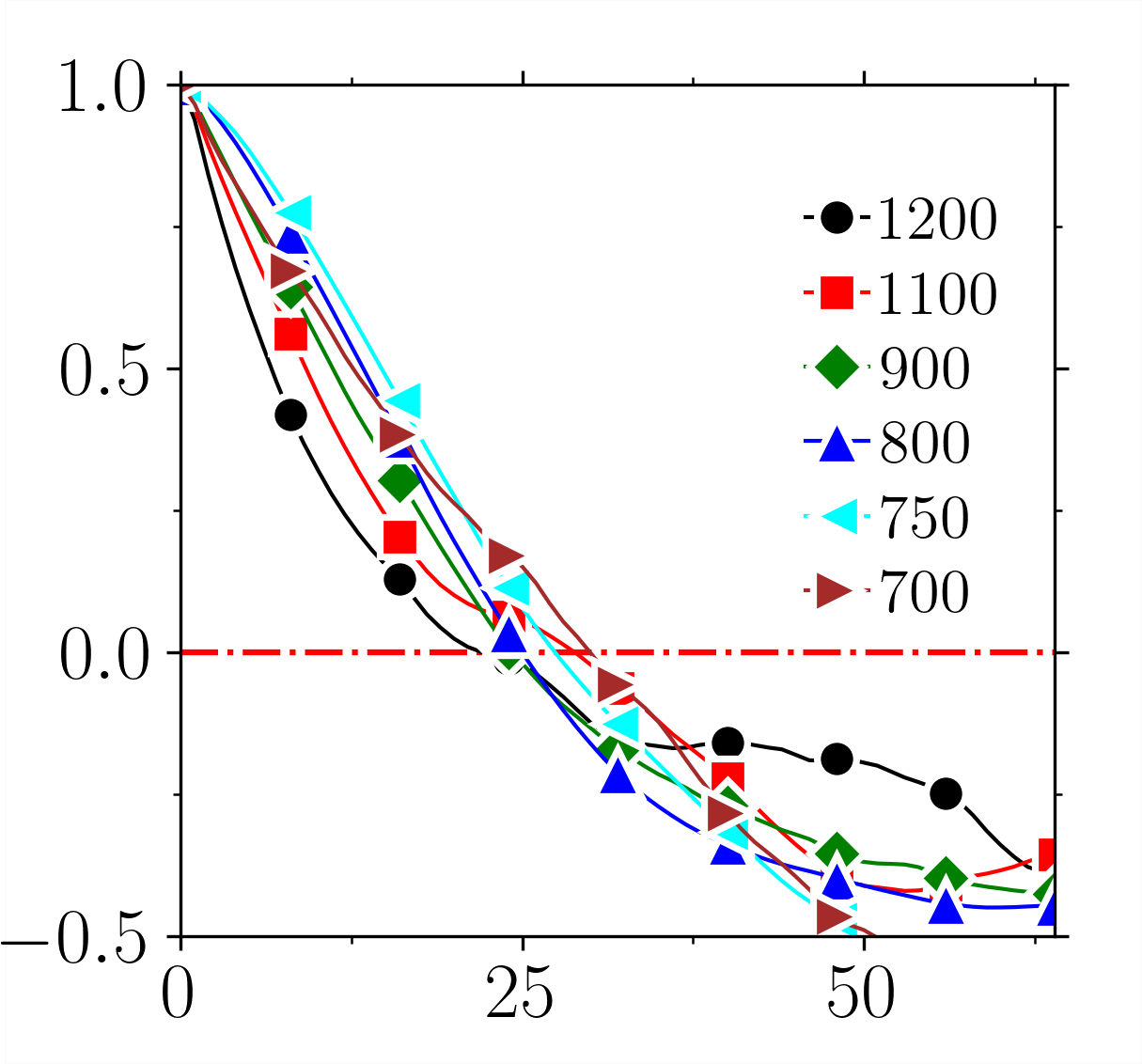}
        \Labelxy{44}{-3}{0}{$|z-z^\prime|$}
        \Labelxy{-6}{35}{90}{$\langle c_v(|z-z^\prime|) \rangle_\text{ens}$}
        \Labelxy{68}{80}{0}{$\sigma$\tiny(MPA)}
        \LabelFig{17}{15}{$c)$   \scriptsize RSA}
    \end{overpic}
    \caption{Stress-dependence of the mean dislocation velocity and associated fluctuations. \textbf{a}) Mobility rule describing the mean dislocation velocity $\langle v_x \rangle$ as a function of applied stress $\sigma$. Mean velocity auto correlations $\langle c_v(|z-z^\prime|) \rangle_\text{ens}$ as a function of distance $|z-z^\prime|$ in \textbf{b}) annealed NiCoCr \textbf{c}) NiCoCr RSA subject to multiple applied stresses $\sigma$ above $\sigma_y$. \note{i) how SROs affect mobility?}}
    \label{fig:velCrltn}
\end{figure*}

To estimate $\sigma_c$, we performed atomistic simulations of dislocation properties in fcc-based NiCoCr CSAs by studying dynamics of $\frac{1}{2}\langle{1}10\rangle\{111\}$ edge dislocations which, under an external drive, dissociate into two mixed partials and a stacking fault in between \cite{hull2001introduction}.
To measure the dislocation velocity and its spatial-temporal evolution, we first identified all dislocation line defects in the atomistic crystal, along  with their Burgers vectors, and output a line representation of the dislocations by using OVITO \cite{stukowski2012automated}.
Due to inherent lattice distortions, dislocation lines are not straight but show local fluctuations with respect to the average line direction along $z$.
We describe line fluctuations projected along the glide direction $x$ by the function $h_x(z)$ discretized via a fine grid of size $2$~\r{A} across the glide plane parallel to the $z$ direction.  
We obtain the dislocation velocity $v_x(z) = \delta h_x(z)/\delta t$ by considering successive dislocation snapshots that are apart by the time window $\delta t\simeq 4$ ps.
The latter is chosen to be at least three orders of magnitude longer than the discretization time $\Delta t$ yet short enough to resolve displacements down to atomistic scales.
The subsequent correlation analysis was performed on CSAs initially annealed at $T=600$ K and sheared, along with the RSAs, at $5$ K.

Figure~\ref{fig:dislSnapshots} and \ref{fig:dislSnapshotsRsa} illustrate configurations of (frozen) dislocations in an annealed NiCoCr as well as a NiCoCr RSA under different loads well below the depinning transition ($\sigma < \sigma_c$).
{The local curvatures associated with the dislocation segments in Fig.~\ref{fig:dislSnapshots} indicate fairly coherent pinning effects that somewhat correlate with the spatial locations of SROs. 
Such features might be also present in RSAs, as in Fig.~\ref{fig:dislSnapshotsRsa}, but to a very limited extent in space.}  
The local line curvature, and its positive sign with respect to the glide direction, should potentially indicate how effectively dislocations are pinned near SROs and/or due to local atomic misfits. 
In this context, line fluctuations associated with the aged alloy in Fig.~\ref{fig:sroBelowStrsAnnealed}(a) appear to be correlated over larger lengthscales than those of the random alloy in Fig.~\ref{fig:sroBelowStrsRsa}(a).
Similar trends could be also inferred from the associated correlation functions
\begin{equation}
c_h(|z-z^\prime|)=\langle~\hat{h}_x(z).\hat{h}_x(z^\prime)~\rangle,
\end{equation}
with the $z-$scored fluctuations $\hat{h}_x=(h_x-\langle h_x \rangle)/\text{var}^{\frac{1}{2}}(h_x)$.
Here the angular brackets $\langle .\rangle$ denote averaging in space.
Overall, the slower decay of correlations $c_h(|z-z^\prime|)$ in Fig.~\ref{fig:sroBelowStrsAnnealed}(b), in comparison with Fig.~\ref{fig:sroBelowStrsRsa}(b), may indicate additional SRO-induced pinning effects in annealed alloys.
We note that the absence of SROs does not necessarily rule out long-range fluctuation patterns in RSAs, as in Fig.~\ref{fig:sroBelowStrsRsa}(b), and, therefore, coherent pinning patterns due to atomic-scale distortions \cite{zhang4102468data,zaiser2021pinning}.

\begin{figure}[t]
  \centering
    \begin{overpic}[width=0.5\textwidth]{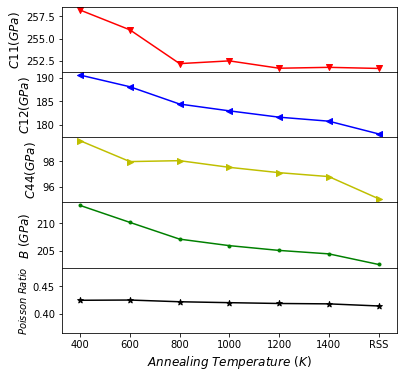}
    \Labelxy{91}{7.5}{0}{\scriptsize RSA}
    \end{overpic}
  \caption{Elastic constants and their dependence on the annealing temperature $T_a$. Measurements were carried out at $T=5$ K. The rightmost data point represents elastic properties of a random solid solution. Samples were generated based on the \potOne potential. \note{referenced in the text?, boot out the Poisson ratio since it is rather useless.}} 
  \label{fig:elasticConstants}
\end{figure}

\begin{figure}[b]
   \centering
    \begin{overpic}[width=0.27\textwidth]{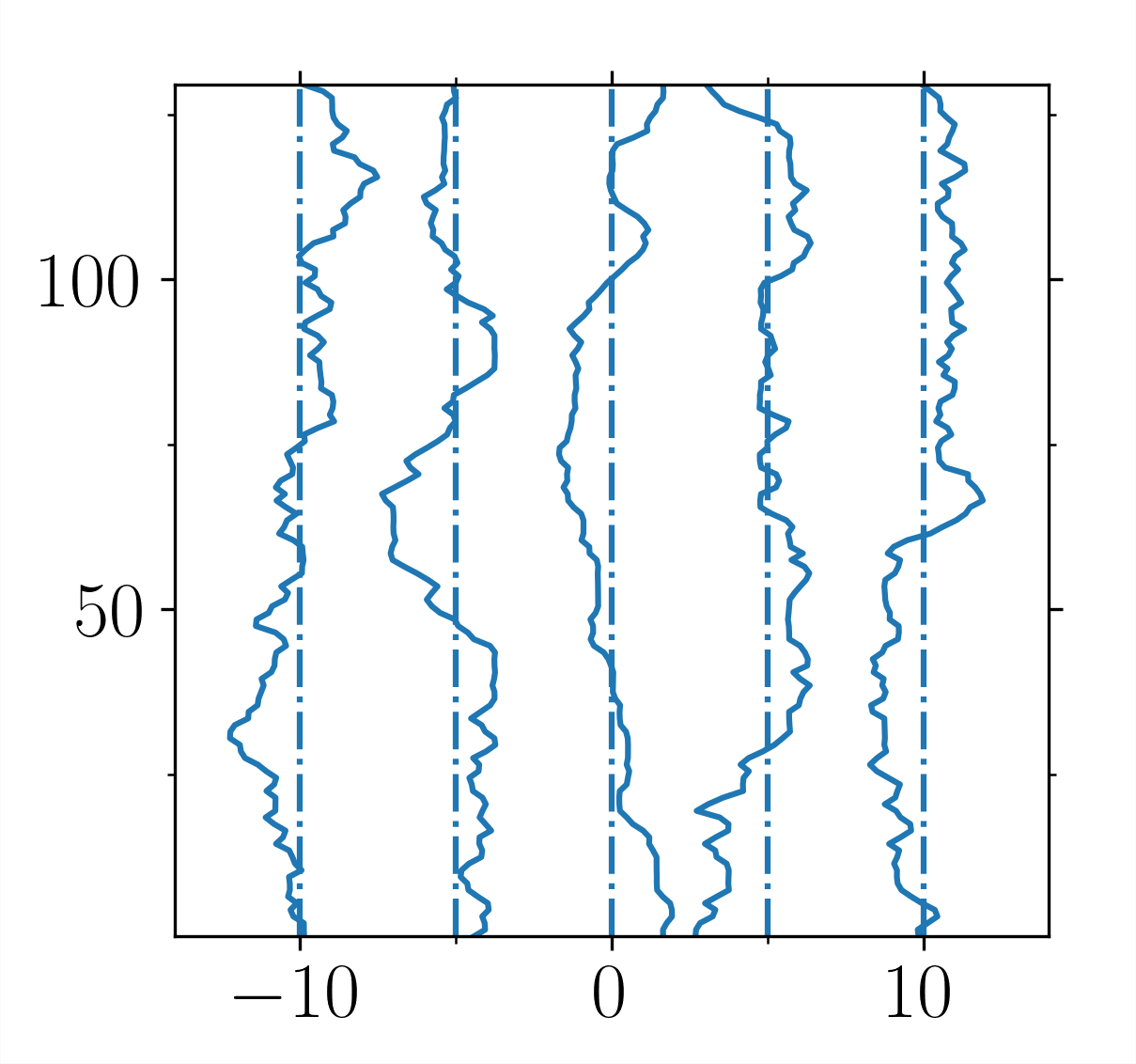}
        \Labelxy{44}{-3}{0}{$\hat{v}_x(z)$}
        \Labelxy{-6}{35}{90}{$z$(\r{A})}
        \LabelFig{17}{90}{$a)$ \scriptsize annealed}
        \Labelxy{18}{16}{90}{$\scriptstyle (1)$}
        \Labelxy{34}{16}{90}{$\scriptstyle (2)$}
        \Labelxy{48}{16}{90}{$\scriptstyle (3)$}
        \Labelxy{62}{16}{90}{$\scriptstyle (4)$}
        \Labelxy{76}{16}{90}{$\scriptstyle (5)$}
    \end{overpic}

    \begin{overpic}[width=0.27\textwidth]{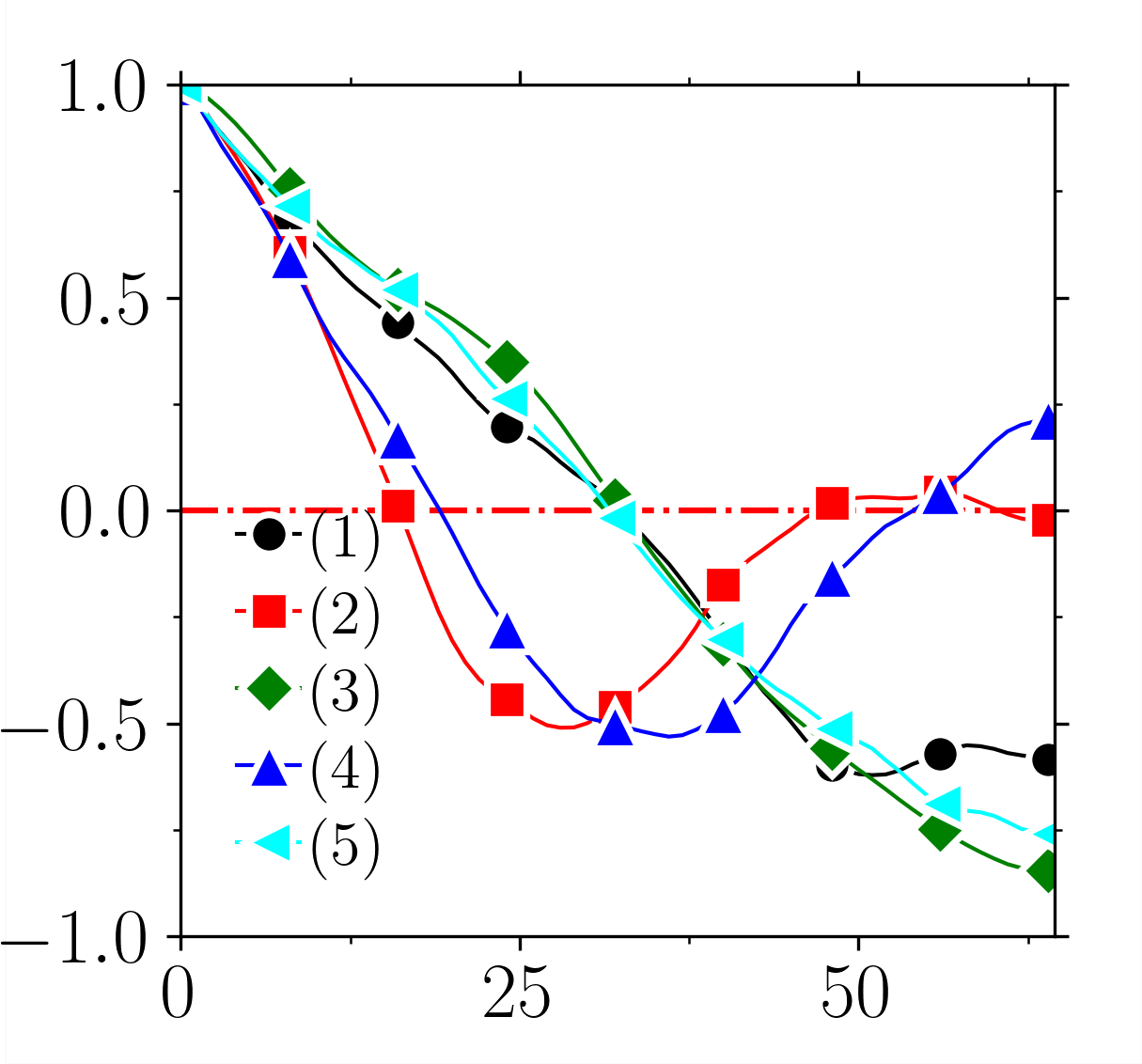}
        \Labelxy{44}{-3}{0}{$|z-z^\prime|$}
        \Labelxy{-6}{35}{90}{$c_v(|z-z^\prime|)$}
        \LabelFig{17}{15}{$b)$}
    \end{overpic}
    \caption{ \textbf{a}) (Scaled) velocity of partial edge dislocations $\hat{v}_x(z)$ \textbf{b}) associated correlations $c_v(|z-z^\prime|)=\langle ~\hat{v}_x(z).\hat{v}_x(z^\prime)~\rangle$ as a function of distance $|z-z^\prime|$ in an  annealed NiCoCr at $T_a=600$ K under the applied stress $\sigma =1200$ MPa (above $\sigma_c$). The shear tests were carried out at $5$ K. Here, the numbers indicate different realizations associated with gliding dislocations and the two-dimensional plane denotes a $(111)$ cross section. The velocity profiles are shifted vertically for the clarity.}
    \label{fig:sroAboveStrsAnnealed}
\end{figure}

\begin{figure}[t]
   \centering
    \begin{overpic}[width=0.27\textwidth]{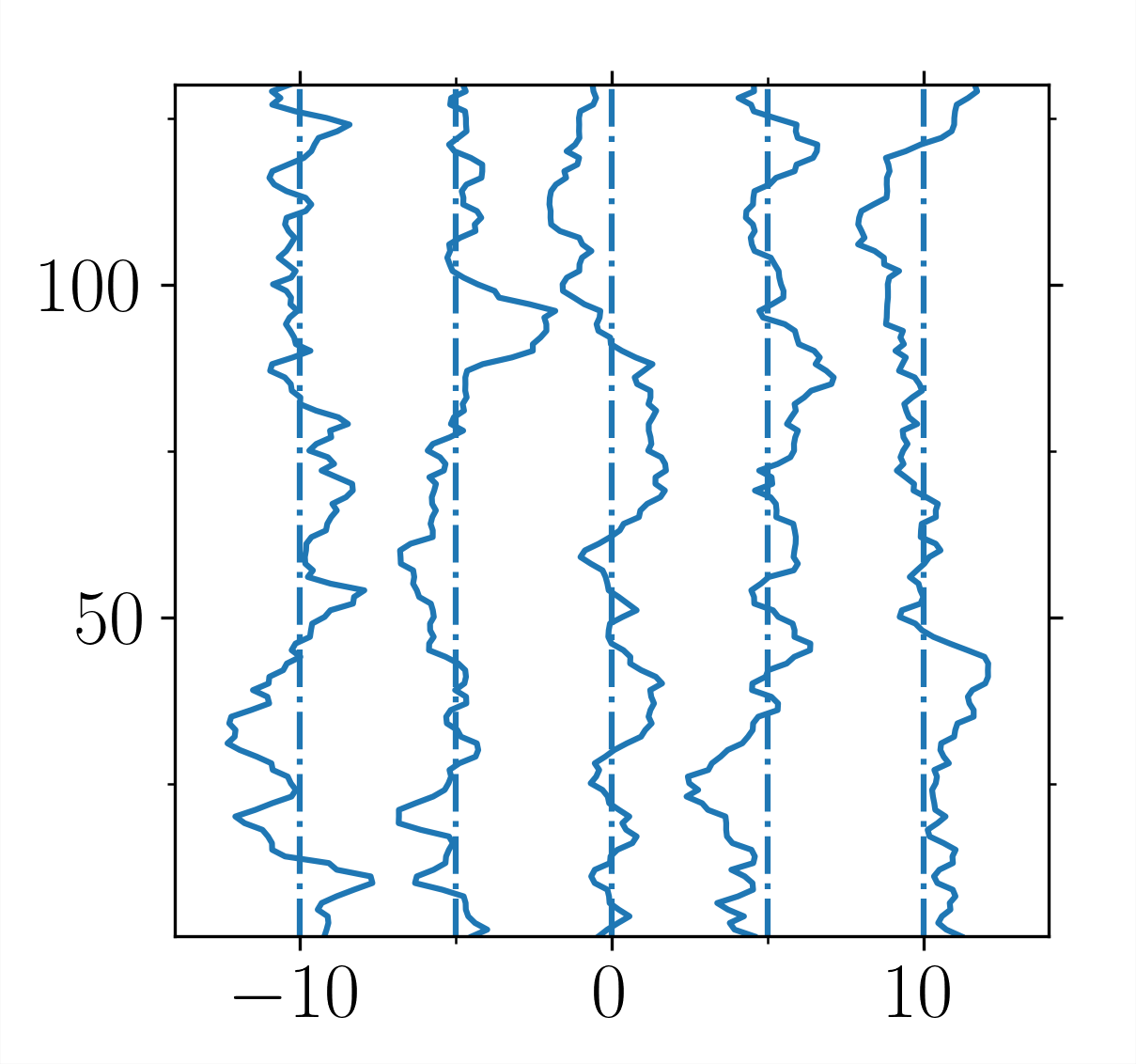}
        \Labelxy{44}{-3}{0}{$\hat{v}_x(z)$}
        \Labelxy{-6}{35}{90}{$z$(\r{A})}
        \LabelFig{17}{90}{$a)$ \scriptsize RSA}
        \Labelxy{18}{16}{90}{$\scriptstyle (1)$}
        \Labelxy{34}{16}{90}{$\scriptstyle (2)$}
        \Labelxy{48}{16}{90}{$\scriptstyle (3)$}
        \Labelxy{62}{16}{90}{$\scriptstyle (4)$}
        \Labelxy{76}{16}{90}{$\scriptstyle (5)$}
    \end{overpic}

    \begin{overpic}[width=0.27\textwidth]{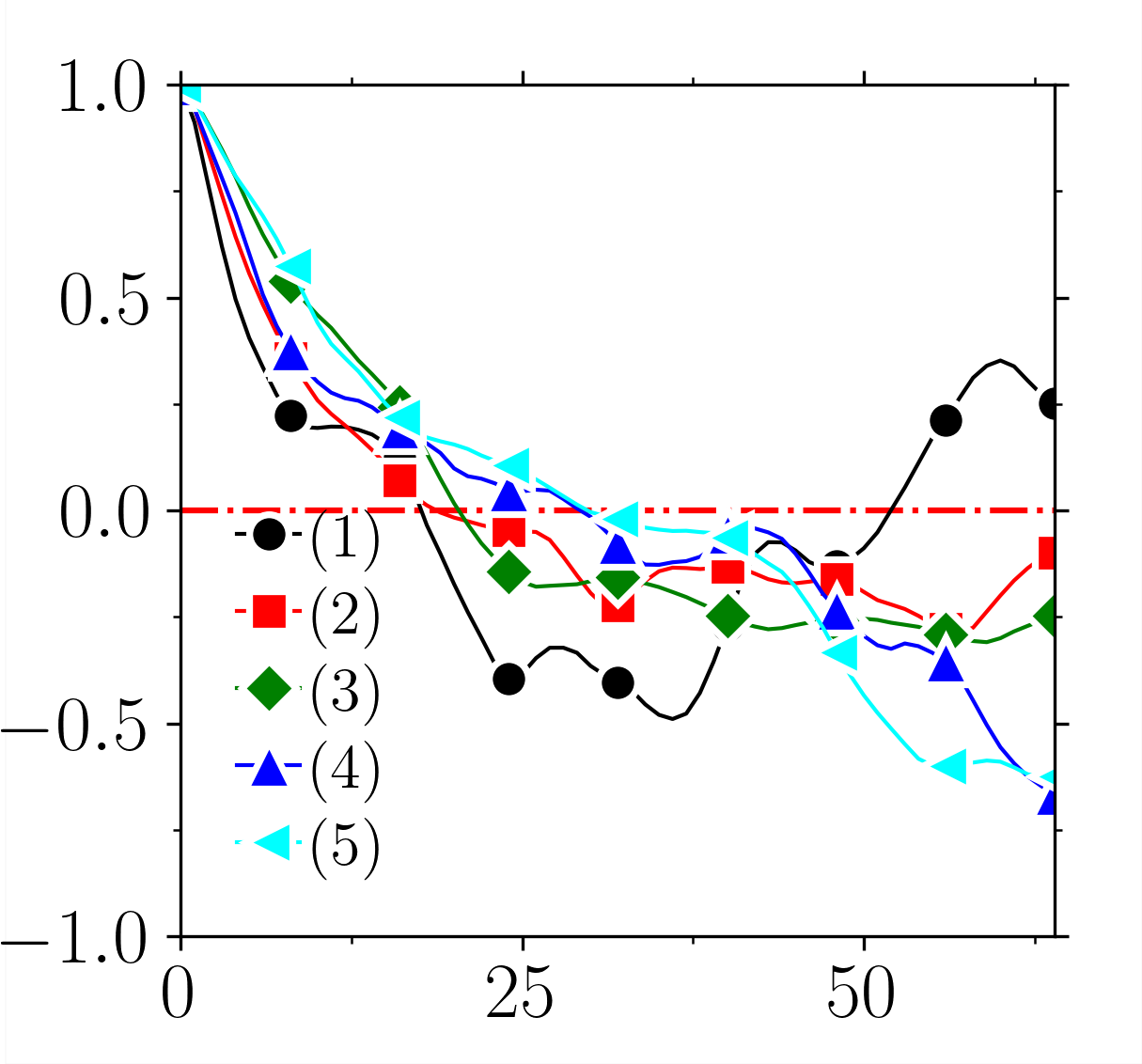}
        \Labelxy{44}{-3}{0}{$|z-z^\prime|$}
        \Labelxy{-6}{35}{90}{$c_v(|z-z^\prime|)$}
        \LabelFig{17}{15}{$b)$}
    \end{overpic}
    \caption{ \textbf{a}) (Scaled) velocity of partial edge dislocations $\hat{v}_x(z)$ \textbf{b}) associated correlations $c_v(|z-z^\prime|)=\langle ~\hat{v}_x(z).\hat{v}_x(z^\prime)~\rangle$ as a function of distance $|z-z^\prime|$ in a NiCoCr RSA under the applied stress $\sigma =1200$ MPa (above $\sigma_c$).}
    \label{fig:sroAboveStrsRsa}
\end{figure}

We repeated the above analysis by probing velocity fluctuations $v_x(z)$ associated with the gliding dislocations (at $\sigma>\sigma_c$) in annealed as well as random alloys (see Fig.~\ref{fig:dislVelSnapshots}). 
Figure~\ref{fig:velCrltn} illustrates the shear stress dependence of the mean dislocation velocity as well as (mean) velocity correlations (averaged over different configurations) for the aged and random alloys.
As shown in Fig.~\ref{fig:velCrltn}(a), the observed behavior of $\langle v_x\rangle$ versus $\sigma$ at $\sigma_c$ marks the dislocation pinning-to-depinning transition which is in agreement with the expected generic dependence around the transition. This seems to be fairly insensitive to annealing except for a meaningful shift of $\sigma_c$ to larger strengths.
The estimated critical shear stresses are $\sigma_c\simeq 950$ and $650$ MPa associated with aged and random samples, respectively.  
The (mean) velocity auto-correlations, averaged over different realizations, $\langle c_v(|z-z^\prime|) \rangle_\text{ens}$ (see the definition of $c_v(|z-z^\prime|)$ below) are shown at different stress levels beyond $\sigma_c$ in Fig.~\ref{fig:velCrltn}(b) and (c), both indicating a finite correlation length. 

{The marked increase of $\sigma_c$ is despite (relatively) insignificant variations of elastic properties (Fig.~\ref{fig:elasticConstants}) and, therefore, improving yielding properties against RSAs cannot be naively attributed to the enhancement in elasticity of aged alloys (see Sec.~\ref{sec:discussions}).
The elastic constants we probed in this study include $C_{11}$, $C_{12}$, and $C_{44}$ (based on the Voigt notation) as well as the bulk modulus $B$ and Poisson's ratio that were determined by using the \potOne interatomic potential.
Here the $x$, $y$, and $z$ dimensions are parallel to $[100]$, $[010]$, and $[001]$ crystal directions, respectively.
The overall trend we observe in Fig.~\ref{fig:elasticConstants} is consistent with the study of Li et al. \cite{Li2019} which reported the change of elastic properties with increasing SROs (upon decreasing $T_a$).
The elastic constants seem to develop features near $T_a\simeq 800$ where the dominant presence of chemical ordering is expected.}

We further investigate individual dislocation configurations and associated fluctuations of local velocities in Fig.~\ref{fig:sroAboveStrsAnnealed} and \ref{fig:sroAboveStrsRsa} where the dislocations move at an average speed $\langle v_x\rangle \simeq 1000 ~\text{ms}^{-1}$ in both systems subject to the applied shear stress of $\sigma=1200$ MPa, well above the corresponding depinning thresholds. 
The ($z$-scored) velocity profiles $\hat{v}_x(z)$ in Fig.~\ref{fig:sroAboveStrsAnnealed}(a) and \ref{fig:sroAboveStrsRsa}(a) correspond to five different snapshots of gliding dislocations that are shifted for a better view of variations across the (average) dislocation line direction $z$.
We remark that the regions to the left of the dashdotted lines indicate local velocities below the average speed $\langle v_x\rangle$, {as depicted by the left-headed arrows in Fig.~\ref{fig:dislVelSnapshots} }.
Statistically speaking, the segments with $v_x(z)<\langle v_x\rangle$ somewhat correlate with the positively-bent segments of dislocation lines which are mostly influenced by the existence of SROs and/or atomic misfits.
Nevertheless, velocity fluctuations quantified by the velocity auto correlations
\begin{equation}
c_v(|z-z^\prime|)=\langle~\hat{v}_x(z).\hat{v}_x(z^\prime)~\rangle,
\end{equation}
not appear to be statistically different in annealed and random alloys in Fig.~\ref{fig:sroAboveStrsAnnealed}(b) and \ref{fig:sroAboveStrsRsa}(b).


\note{expand}

%% file: sections/discussions.tex
\begin{table}[b]
\centering
\caption{Cubic elastic constants and depinning stress $\sigma_c$  at $5$ K for NiCoCr with and without annealing using the \potTwo potential. \note{values don't match those reported in the text!}}
\begin{tabular}{l c c c c c c c}
\hline
  &  $T_a=1100$(K) & $T_a=600$(K) & RSA & Experiment \\
\hline
$C_{11}$(GPa)    & 269.3 &  251.4 &  270.1 &  255.3 \cite{laplanche2020processing} \\
$C_{12}$(GPa)    & 146.8  & 149.1  & 147.9  & 159.4 \cite{laplanche2020processing} \\
$C_{44}$(GPa)    & 126.7 & 126.8 & 126.7 & 146.7 \cite{laplanche2020processing}\\
$\sigma_c$(GPa)    &  0.075 & 0.070 & 0.080 & 1.3 \cite{Gludovatz2016}\\
\end{tabular}
\label{tab:EC_Mo} 
\end{table}

\section{Discussions \& Conclusions}\label{sec:discussions}
Our atomistic simulations of \comp CSAs under special thermal treatments have revealed the formation of nanostructural local chemical ordering and enhanced dislocation glide resistance in close agreement with recent SRO-based studies in simulated and \emph{real} NiCoCr experiments \cite{Li2019,zhang2020short}.
On the ordering effects, we made use of the \potOne potential function that has been validated in terms of detailed and accurate modelling of  Ni, Co, and Cr interatomic interactions \cite{Li2019}.
Our direct measurements of local lattice strains agree very closely with a recent ab initio study \cite{oh2019engineering} but failed to fully reproduce experimental findings \cite{Yin2020}.
By using the \potTwo potential, we find very limited relevance to real annealed \comp alloys. 
The simulated alloys, in this context, exhibit no ordering (beyond statistical fluctuations), but also no notable improvement in yield strengths or elastic properties as reported in Table~\ref{tab:EC_Mo}.
We have interpreted the physical origin of such differences by using robust (experimentally-relevant) SRO descriptors in various thermal annealing scenarios. 
We find that the Li-Sheng-Ma potential, under the proper aging process, leads to an exceptional dislocation depinning strength with low stacking fault width that falls short of that of RSAs.
{The latter is associated with the enhanced stacking fault energy which might, in part, relate to improved elasticity as a result of SROs but, compared to the yield strength, the ordering effects appear to be less pronounced.} 
The intrinsic strengthening mechanism is mostly dominated by coherent SROs-induced pinning effects, but random spatial distributions of misfit volumes and the resulting roughening seem to be also at play. 

Our correlation analyses of the dislocation structure and its spatial-temporal evolution allow for inferring a characteristic pinning length $\xi_p$ and optimal displacement $w_p$ \cite{VARVENNE2016164}.
We interpret the latter as being the rms fluctuations in the dislocation height (with respect to the mean), i.e. $w_p=\langle h^2_x-\langle h_x\rangle^2\rangle^\frac{1}{2}$, whereas the former is determined as the shortest distance where the height correlations cross zero, e.g. $c_h(|z-z^\prime|)=0$ at $|z-z^\prime|=\xi_p$.
At $T_a=600$ K and $\sigma < \sigma_y$, it follows that $w_p=5.6-11.2$ \r{A} and $\xi_p=25-31$ \r{A} associated with the annealed alloy. 
{We conjecture that these two quantities should both correlate with the observed increase in the depinning stress. Based on our present data, however, we are not able to quantify such (anti-)correlations numerically.}
Both observables $w_p$ and $\xi_p$ \r{A} are also expected to show meaningful associations with the average SRO size $\xi^\text{sro}$ as well as the amplitude of misfit fluctuations (characterized by $\xi^\text{misfit}$) and are relevant ingredients in \emph{mean-field} solute models that make yield strength predictions based on dislocation line properties (e.g. line tension $\Gamma$ , length $L$, and Burgers vector $b$).
{In this mean-field picture, SROs introduce the characteristic scale $\xi^\text{sro}$ that \emph{effectively} decreases the pinning length $\xi_p$ leading to a reduction of the optimal displacement $w_p$ and, therefore, an extra strengthening.}
{Within these mean-field model frameworks, the depinning stress should scale with the line tension $\Gamma$, which itself is proportional to the shear modulus and, based on our findings, annealing is not expected to boost $\sigma_y$ simply because of such elasticity-based contributions but instead variations in the disorder strength and the pinning field are the key factors.}
To validate such theories in simulations, one must be vigilant to use appropriate mesoscopic lengths (beyond atomistic scales) where continuum-like concepts such as line tension and local curvature are well-defined \cite{VARVENNE2016164}.
To explore the full dislocation waviness in MD, it is also necessary that the dislocation length $L \gg \xi_p$ and associated deformation $w_p\ll \xi_p$ \cite{Priv_Comm_curtin}.  
However, the above separation of scales is typically a limit beyond atomistic modeling assumptions including the present study.

Complications might also arise in the application of strengthening theories (e.g. the VC model) due to SRO-induced correlations. 
The latter are at odds with the “randomness” hypothesis taken as granted based on solutes’ arrangements in RSAs.
Given that the VC theory is constructed exclusively on misfit information, an \emph{effective} treatment in the presence of spatial correlations is the assumption that SROs should alter dramatically local misfit distributions and, in that case, an alternative length $\xi^\text{misfit}_\text{eff}$ may be used to describe the strength of distortions. 
Alternatively, Zaiser and Wu (ZW) \cite{zaiser2021pinning, vaid2021pinning} formulated a more relevant approach based on the fact that pinning forces caused by obstacles are not fully random but rather correlated over some certain length $a$ \cite{geslin2021microelasticity} that we tend to interpret as the effective SRO size ($a\simeq\xi^\text{sro}$). 
In their formulation of dislocation dynamics, ZW introduced a order/disorder length that, along with the strength of misfit fluctuations, may describe the SRO- and misfit-induced noise field in a more accurate way than the VC methodology.
Similar efforts were made by Zhang et al. \cite{zhang2019effect} along these lines who developed a stochastic Peierls-Nabarro model to incorporate the role of both CSA randomness and short range ordering effects on glide dynamics of roughened dislocations.

{The existing literature reports the emergence of (varying degrees of) SROs as a rather generic feature across a broad range of high-entropy alloys (see \cite{wu2021short} and references therein). 
Nevertheless, the focus has been placed on different variants of NiCoCr-based alloys (including the well-studied Cantor alloy) and, owing to similar atomic size and electron negativity, such compositions might tend to favor SRO nucleation \cite{ding2019tuning}.
In terms of mechanical properties, the SRO-induced enhancement in the dislocation glide resistance may also constitute fairly universal mechanisms associated with it, i.e. coherent pinning and enhanced roughening, not specific to particular chemical compositions but their robustness over a broader range of compositionally complex solid solutions has yet to be fully explored.
}

There is a large multitude of results in this work, some of which including the dislocation roughening and SRO emergence could be potentially validated experimentally through the in/ex-situ electron microscopy analysis or other image-based characterization techniques.
{Furthermore, our finding will have important implications for Discrete Dislocation Dynamics (DDD) models and associated  mobility rules that additionally consider spatial correlations within the rough potential energy landscape \cite{salmenjoki2020plastic}.
This is conceptually similar to intrinsic Peierls stresses that are locally distributed in space but also correlated over certain microstructural scales.
Incorporating and tuning SROs' structural features as model ingredients will potentially lead to further improvements in DDD predictive capabilities and design-level hardening features in the context of NiCoCr-based CSAs with dense and complex networks of interacting dislocations.}

%% file: sections/conclusions.tex
\begin{acknowledgments}
This research was funded by the European Union Horizon 2020 research and innovation program under grant agreement no. 857470 and from the European Regional Development Fund via Foundation for Polish Science International Research Agenda PLUS program grant no. MAB PLUS/2018/8.
\end{acknowledgments}

%% file: sections/sm.tex
\clearpage
\renewcommand{\thefigure}{S\arabic{figure}}
\renewcommand{\thesection}{S~\Roman{section}}
\setcounter{figure}{0}    
\setcounter{section}{0}    
\section*{Supplementary Materials}
\begin{figure*}
   \centering
     \begin{overpic}[width=0.24\textwidth]{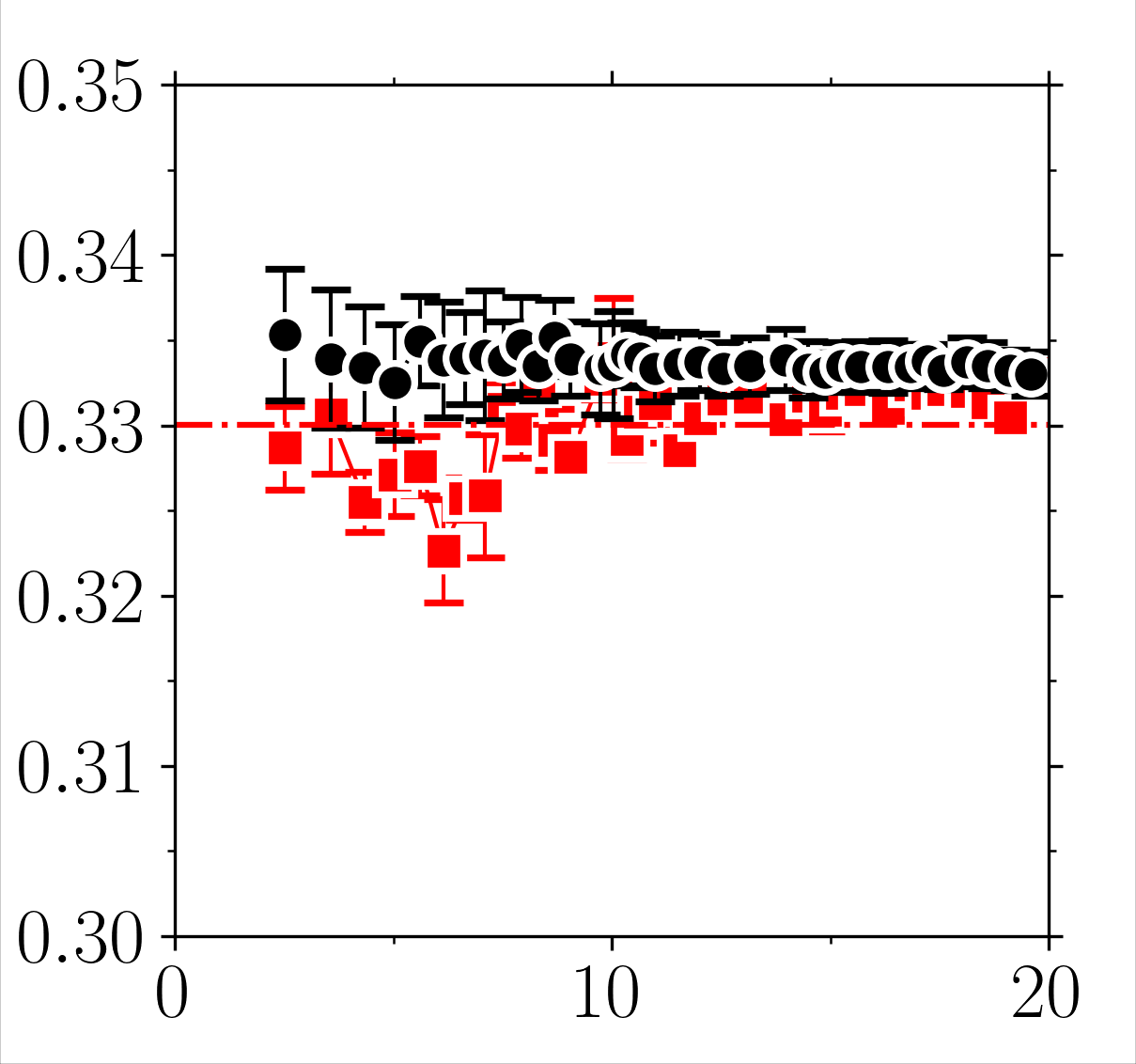}
        \Labelxy{50}{-3}{0}{$r$(\r{A})}
        \Labelxy{-6}{35}{90}{$p_\text{NiNi}$}
        \LabelFig{19}{76}{$a)$}
    \end{overpic}
    \begin{overpic}[width=0.24\textwidth]{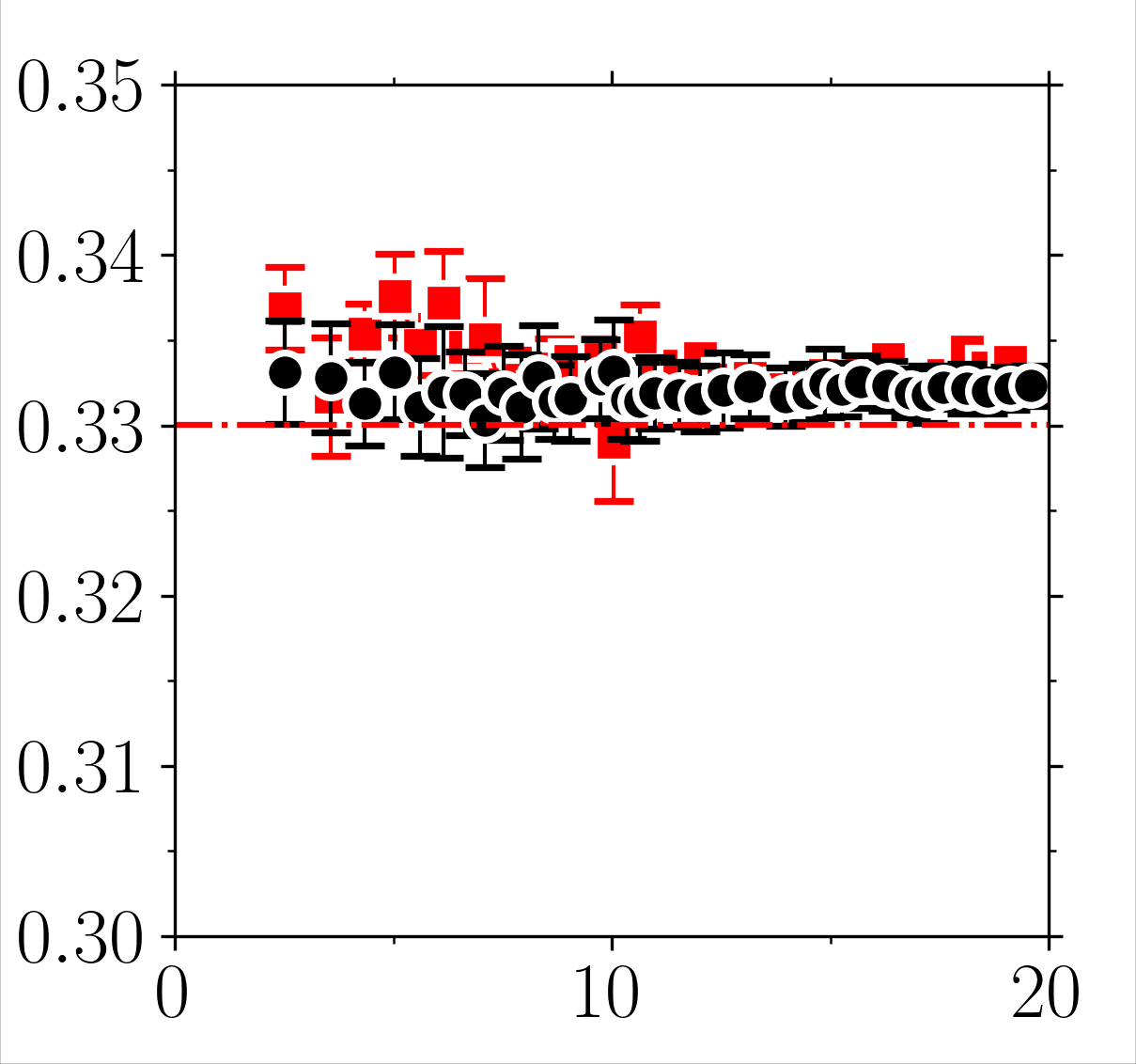}
        \Labelxy{50}{-3}{0}{$r$(\r{A})}
        \Labelxy{-6}{35}{90}{$p_\text{NiCo}$}
        \LabelFig{19}{76}{$b)$}
        \begin{tikzpicture}
            \legCirc{0.8}{1.1}{black}{\footnotesize annealed}{0.9}
            \legSq{0.8}{0.7}{red}{\footnotesize random CSA}{1.2}
        \end{tikzpicture}
    \end{overpic}
    \begin{overpic}[width=0.24\textwidth]{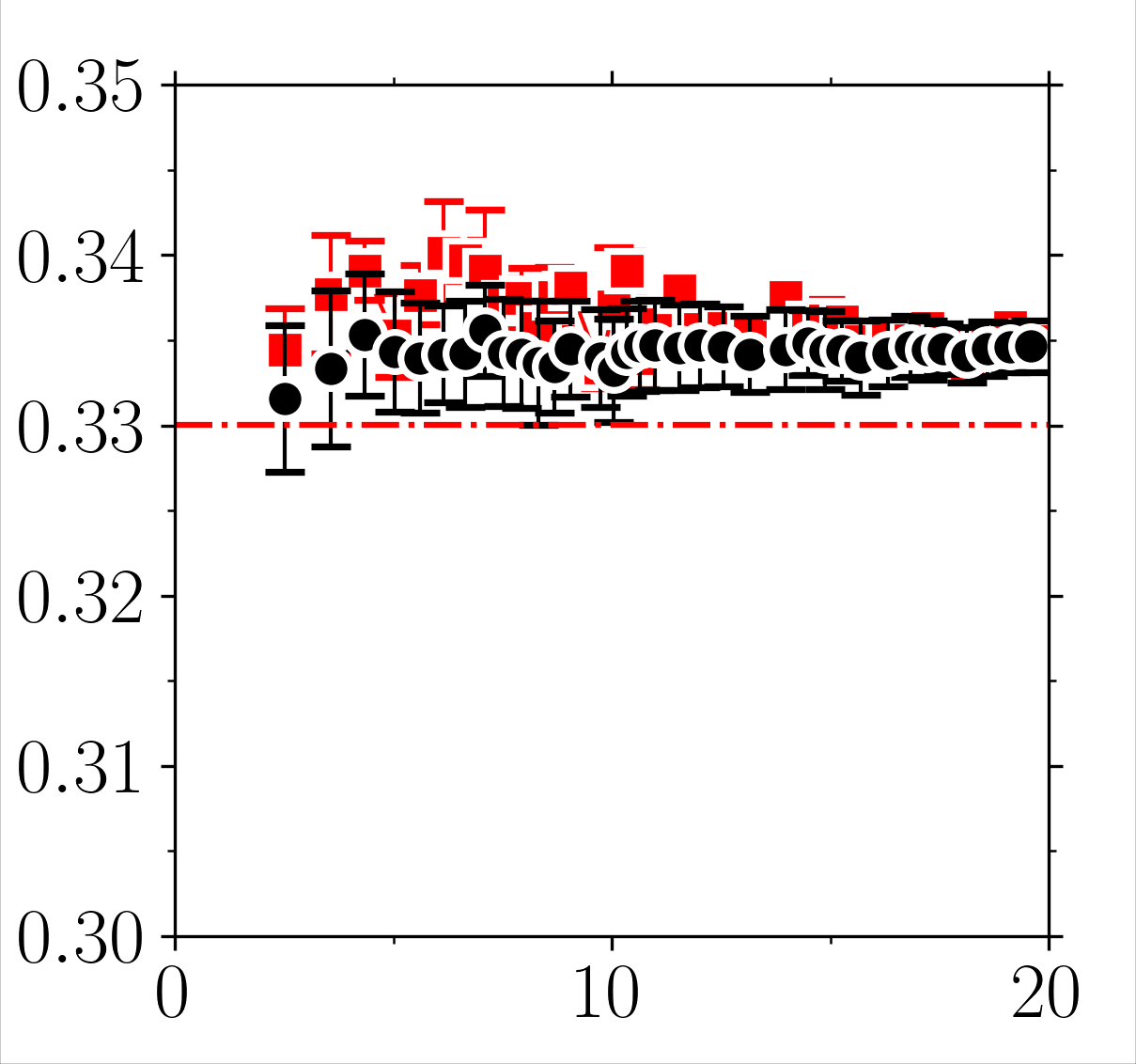}
        \Labelxy{50}{-3}{0}{$r$(\r{A})}
        \Labelxy{-6}{35}{90}{$p_\text{NiCr}$}
        \LabelFig{19}{76}{$c)$}
    \end{overpic}

    \begin{overpic}[width=0.24\textwidth]{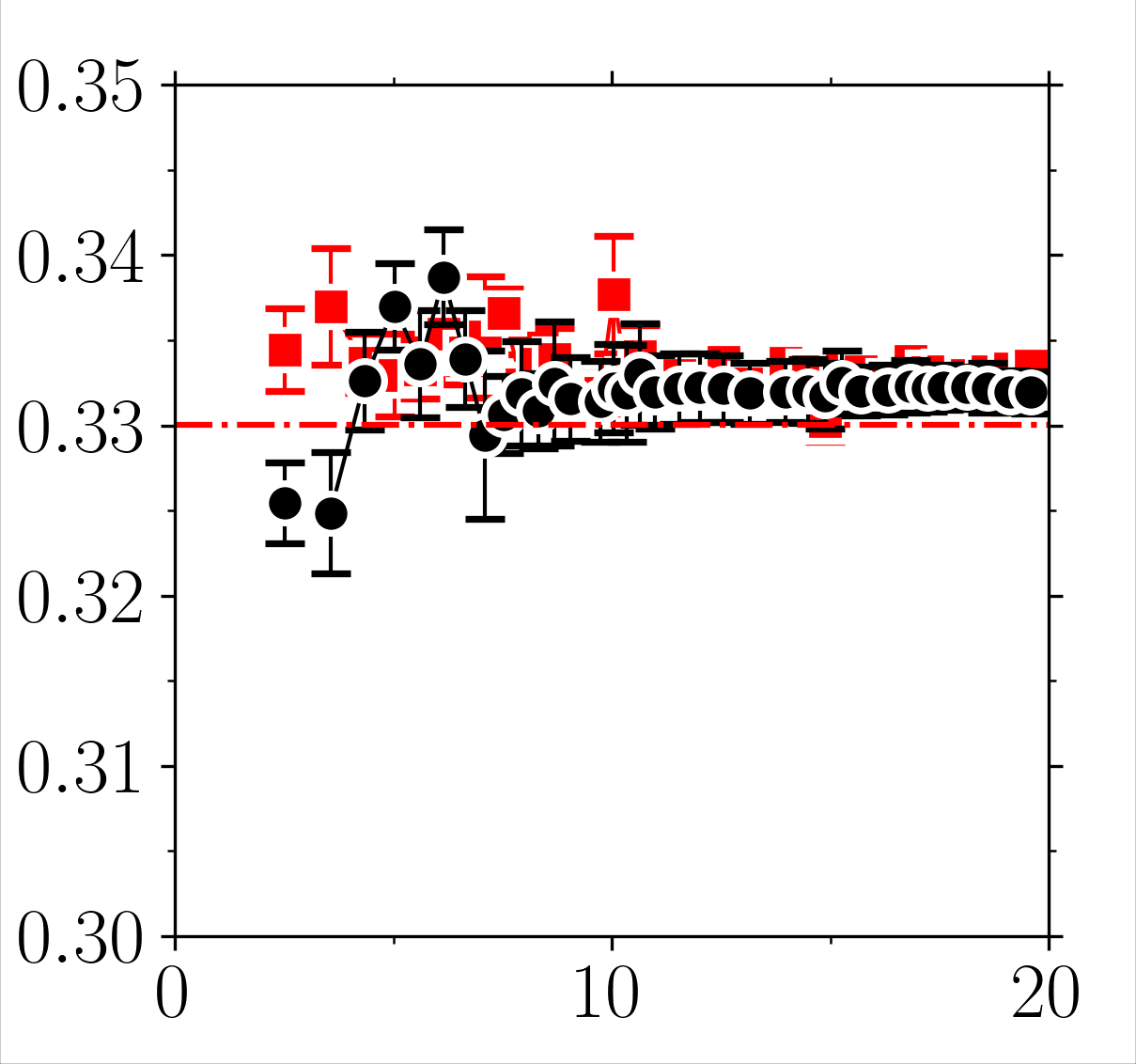}
        \Labelxy{50}{-3}{0}{$r$(\r{A})}
        \Labelxy{-6}{35}{90}{$p_\text{CoCo}$}
        \LabelFig{19}{76}{$d)$}
    \end{overpic}
    \begin{overpic}[width=0.24\textwidth]{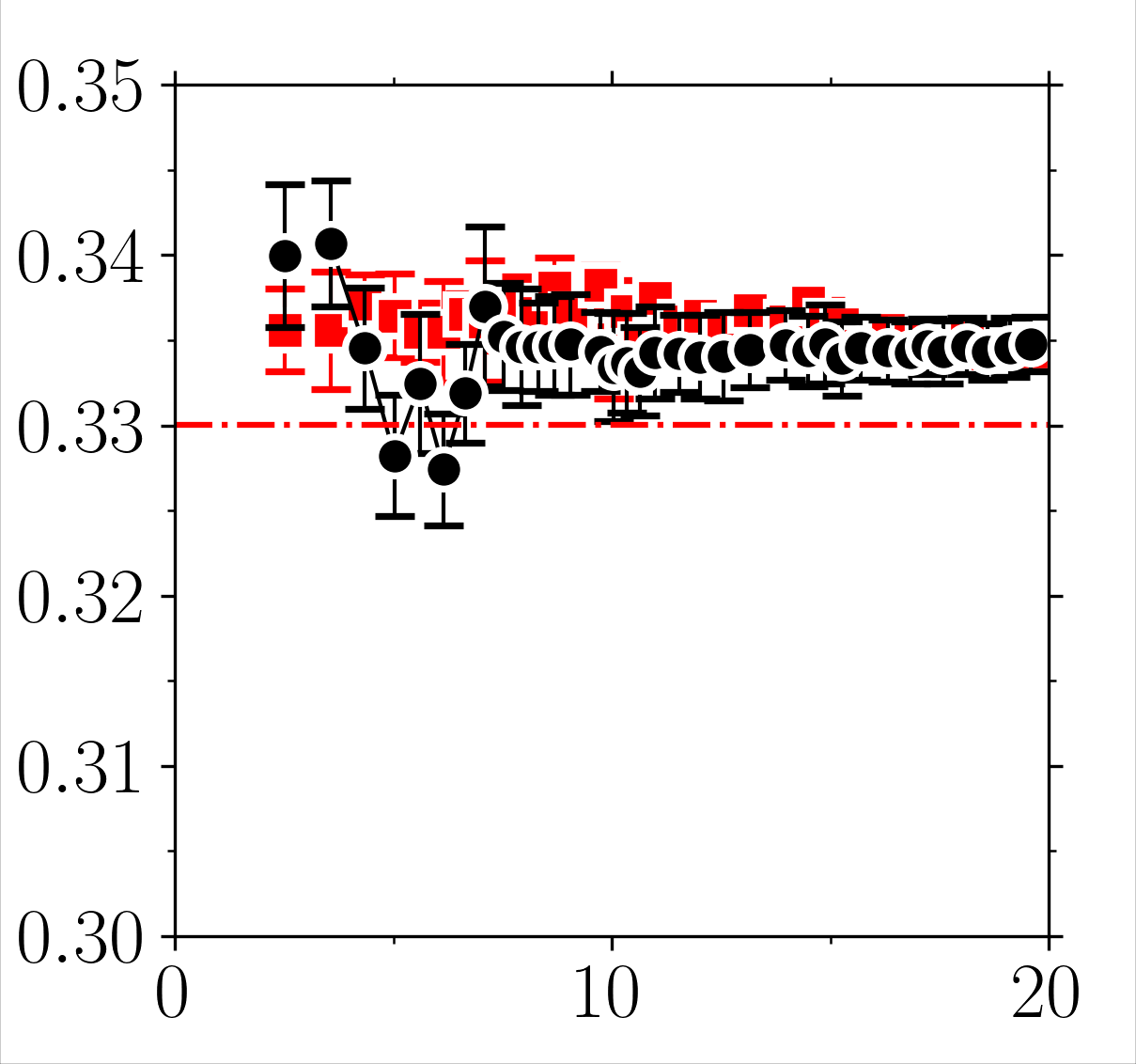}
        \Labelxy{50}{-3}{0}{$r$(\r{A})}
        \Labelxy{-6}{35}{90}{$p_\text{CoCr}$}
        \LabelFig{19}{76}{$e)$}
    \end{overpic}
    \begin{overpic}[width=0.24\textwidth]{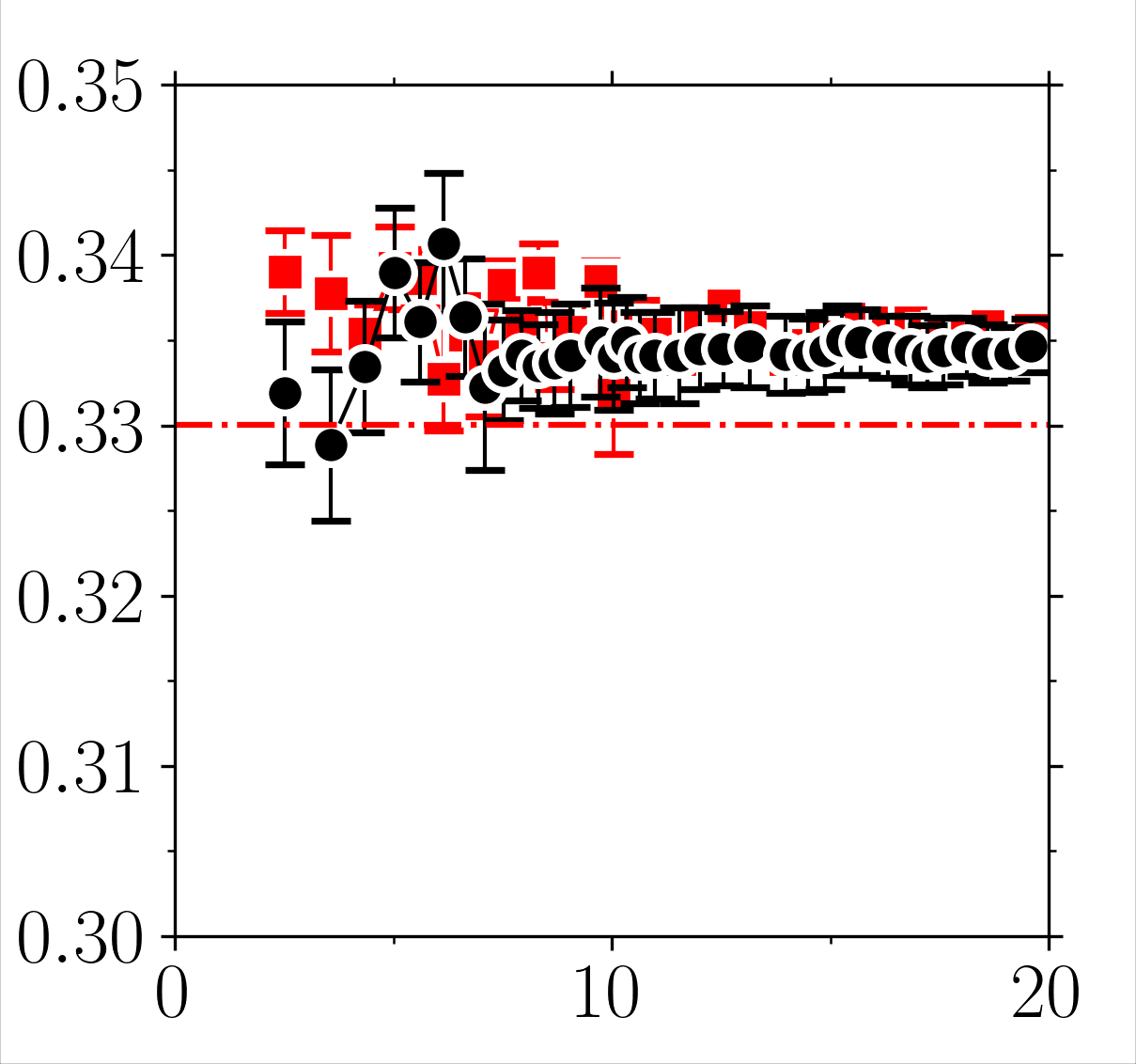}
        \Labelxy{50}{-3}{0}{$r$(\r{A})}
        \Labelxy{-6}{35}{90}{$p_\text{CrCr}$}
        \LabelFig{19}{76}{$f)$}
    \end{overpic}
  \caption{Short range ordering in annealed NiCoCr CSAs based on the \potTwo potential. Warren–Cowley SRO parameters including \textbf{a)} $p_\text{NiNi}$ \textbf{b)} $p_\text{NiCo}$ \textbf{c)} $p_\text{NiCr}$ \textbf{d)} $p_\text{CoCo}$ \textbf{e)} $p_\text{CoCr}$ \textbf{f)} $p_\text{CrCr}$ plotted against distance $r$ at $T_a=400$ K. The base (red) dashdotted line indicates the random concentration.} 
   \label{fig:sroFarkas}
\end{figure*}

In this Supplementary Materials we will present further discussions on local concentration fluctuations of model NiCoCr CSAs using the \potTwo potential function.
Figure~\ref{fig:sroFarkas}(a-f) illustrates that the the Warren–Cowley SRO parameters $p_{ab}$ associated with the annealed CSAs are statistically indistinguishable from $p_{ab}^{\text{rsa}}$ including the six (distinct) elemental pairs at $T_a=400$ K.
Therefore, the former can be essentially treated as RSAs with random distributions of the chemical compositions and associated fluctuations that exhibit the expected scale-dependence due to randomness (see Fig.~\ref{fig:std_concentrationFluctuations} (b,d)).
